\documentclass[sigplan,screen,balance=false]{acmart}
\AtBeginDocument{%
  }

\usepackage{amsmath}
\usepackage{textcomp}
\usepackage{pifont}
\usepackage{fontawesome5}

\usepackage{amssymb}
\usepackage{bm}
\usepackage{graphicx}
\usepackage{array}
\usepackage{longtable}
\usepackage{tabularx}
\usepackage{xltabular}
\usepackage{ragged2e}
\usepackage{nicefrac}
\usepackage{xcolor}
\usepackage{colortbl}
\usepackage{multirow}
\usepackage{xspace}
\usepackage{wrapfig}
\usepackage{mathpartir}
\usepackage{subcaption}
\usepackage{threeparttable}
\usepackage{listings}
\usepackage{outlines}
\usepackage{enumitem}       
\usepackage{pgf}
\usepackage{tikz}
\usetikzlibrary{calc, positioning, shapes.multipart, shapes, shapes.geometric, fit, backgrounds, arrows.meta}
\usepackage[symbol]{footmisc}
\usepackage{placeins}
\usepackage{cuted}
\usepackage{flushend}

\definecolor{gray}{RGB}{211,211,211}
\newcommand{\jbasicstyle}{\small\ttfamily} 

\newcommand{\jnumberstyle}{\scriptsize}

\lstdefinelanguage{pseudo}
{
  morekeywords={},
  keywordstyle=\bfseries,
  lineskip=-0.1em,
  numbers=left, 
  numberstyle=\jnumberstyle,
  numbersep=4pt,
  basicstyle=\jbasicstyle,
  breaklines=true,
  breakautoindent=true,
  tabsize=2,
  columns=fullflexible,
  morecomment=*[l][\textsl]{//},
  mathescape=true,
  xleftmargin=10pt,
}

\lstdefinelanguage{todo-comment}
{
  morekeywords={},
  keywordstyle=\bfseries,
  lineskip=-0.1em,
  numbers=none,
  basicstyle=\scriptsize\ttfamily,
  breaklines=true,
  breakautoindent=true,
  tabsize=2,
  columns=fullflexible,
  morecomment=*[l][\textsl]{//},
  mathescape=true,
  xleftmargin=0pt,
}

\definecolor{keywordcolor}{rgb}{0,0,1}      
\definecolor{modifiercolor}{rgb}{0.5,0,0.5} 
\definecolor{datatypecolor}{rgb}{0.82,0.16,0.46} 
\definecolor{methodcolor}{rgb}{0.25,0.5,0.35} 
\definecolor{byzantine}{rgb}{0.74, 0.2, 0.64}  
\definecolor{cadetblue}{rgb}{0.37, 0.62, 0.63}  
\definecolor{cadet}{rgb}{0.0, 0.42, 0.24}
\definecolor{brown(web)}{rgb}{0.65, 0.16, 0.16}  
\definecolor{bluegray}{rgb}{0.2, 0.2, 0.6}

\lstdefinelanguage{java-pretty}
{
  language=java,
  numbers=left,
  basicstyle=\scriptsize\ttfamily,
  numberstyle=\scriptsize,
  breaklines=true,
  columns=fullflexible,
  xleftmargin=18pt,
  tabsize=2,
  showstringspaces=false,
  deletekeywords={public, private, protected, static, final, class, interface, abstract, implements, extends, if, else, while, do, for, switch, case, default, break, continue, return, int, long, double, float, boolean, char, void, String,this},
  morekeywords=[1]{if, else, while, do, for, switch, case, default, break, continue, return},
  keywordstyle=[1]\color{byzantine}\bfseries,
  morekeywords=[2]{public, private, protected, static, final, class, interface, abstract, implements, extends},
  keywordstyle=[2]\color{bluegray}\bfseries,
  morekeywords=[3]{int, long, double, float, boolean, char, void, String, @Override, @Test},
  keywordstyle=[3]\color{cadet}\bfseries,
  morekeywords=[4]{class, interface, extends, implements, new, super, throw, throws, try, catch, finally},
  keywordstyle=[4]\color{methodcolor},
  morecomment=[l]{//},
  commentstyle=\color{cadet},
  stringstyle=\color{brown(web)},
}

\newcommand{\spanDel}[1]{%
\begingroup\setlength{\fboxsep}{1ex}\colorbox{red!10}{#1}\endgroup
}

\newcommand{\spanAdd}[1]{%
\begingroup\setlength{\fboxsep}{1ex}\colorbox{green!10}{#1}\endgroup
}

\lstdefinelanguage{imp-srp}{
	language=java,
	keywords=[2]{int, while, if, else, return, break, continue, halt},
        keywordstyle=[2]\bfseries,
	keywords=[3]{Order, Asc},
	keywordstyle=[3]\color{purple},
	basicstyle=\scriptsize\ttfamily,
        moredelim=**[is][\spanDel]{<<DEL>>}{<<END>>},
        moredelim=**[is][\spanAdd]{<<ADD>>}{<<END>>},
}

\lstdefinelanguage{imp-pretty-no-lines}
{
  language=imp-srp,
  numbers=none,
  basicstyle=\scriptsize\ttfamily,
  numberstyle=\scriptsize,
  breaklines=true,
  columns=fullflexible,
  aboveskip=0pt,
  belowskip=0pt,
  xleftmargin=0pt,
  tabsize=2,
  showstringspaces=false,
  deletekeywords={public, private, protected, static, final, class, interface, abstract, implements, extends, if, else, while, do, for, switch, case, default, break, continue, return, int, long, double, float, boolean, char, void, String,this},
  morekeywords=[1]{if, else, while, do, for, switch, case, default, break, continue, return},
  keywordstyle=[1]\color{byzantine}\bfseries,
  morekeywords=[2]{public, private, protected, static, final, class, interface, abstract, implements, extends},
  keywordstyle=[2]\color{bluegray}\bfseries,
  morekeywords=[3]{int, long, double, float, boolean, char, void, String, @Override, @Test},
  keywordstyle=[3]\color{cadet}\bfseries,
  morekeywords=[4]{class, interface, extends, implements, new, super, throw, throws, try, catch, finally},
  keywordstyle=[4]\color{methodcolor},
  morecomment=[l]{//},
  moredelim=**[is][\spanDel]{<<DEL>>}{<<END>>},
  moredelim=**[is][\spanAdd]{<<ADD>>}{<<END>>},
  commentstyle=\color{cadet},
  stringstyle=\color{brown(web)},
}

\lstdefinelanguage{java-pretty-small-framed}
{
  language=java,
  basicstyle=\tiny\ttfamily,
  breaklines=true,
  columns=fullflexible,
  tabsize=2,
  showstringspaces=false,
  deletekeywords={public, private, protected, static, final, class, interface, abstract, implements, extends, if, else, while, do, for, switch, case, default, break, continue, return, int, long, double, float, boolean, char, void, String,this},
  morekeywords=[1]{if, else, while, do, for, switch, case, default, break, continue, return},
  keywordstyle=[1]\color{byzantine}\bfseries,
  morekeywords=[2]{public, private, protected, static, final, class, interface, abstract, implements, extends},
  keywordstyle=[2]\color{bluegray}\bfseries,
  morekeywords=[3]{int, long, double, float, boolean, char, void, String, @Override, @Test},
  keywordstyle=[3]\color{cadet}\bfseries,
  morekeywords=[4]{class, interface, extends, implements, new, super, throw, throws, try, catch, finally},
  keywordstyle=[4]\color{methodcolor},
  morecomment=[l]{//},
  commentstyle=\color{cadet},
  stringstyle=\color{brown(web)},
}

\lstdefinelanguage{imp-grammar}{
	language=java,
	keywords=[2]{SYNTACTIC_ASSIGN, SYNTACTIC_IF, SYNTACTIC_ELSE, SYNTACTIC_WHILE, SYNTACTIC_ADD, SYNTACTIC_SUB, SYNTACTIC_MUL, SYNTACTIC_DIV, SYNTACTIC_MOD, SYNTACTIC_LT, SYNTACTIC_LTEQ, SYNTACTIC_GT, SYNTACTIC_GTEQ, SYNTACTIC_EQ, SYNTACTIC_NEQ, SYNTACTIC_NOT, SYNTACTIC_AND, SYNTACTIC_OR},
  keywordstyle=[2]\bfseries,
	keywords=[3]{Order, Asc},
	keywordstyle=[3]\color{purple},
	basicstyle=\scriptsize\ttfamily,
}

\lstdefinelanguage{operational-semantics}
{
  language=java,
  basicstyle=\tiny\ttfamily,
  breaklines=true,
  columns=fullflexible,
  tabsize=2,
  showstringspaces=false,
  deletekeywords={public, private, protected, static, final, class, interface, abstract, implements, extends, if, else, while, do, for, switch, case, default, break, continue, return, int, long, double, float, boolean, char, void, String,this},
  morekeywords=[1]{if, else, while, do, for, switch, case, default, break, continue, return},
  keywordstyle=[1]\color{byzantine}\bfseries,
  morekeywords=[2]{public, private, protected, static, final, class, interface, abstract, implements, extends},
  keywordstyle=[2]\color{bluegray}\bfseries,
  morekeywords=[3]{int, long, double, float, boolean, char, void, String, @Override, @Test},
  keywordstyle=[3]\color{cadet}\bfseries,
  morekeywords=[4]{class, interface, extends, implements, new, super, throw, throws, try, catch, finally},
  keywordstyle=[4]\color{methodcolor},
  morecomment=[l]{//},
  commentstyle=\color{cadet},
  stringstyle=\color{brown(web)},
}

\lstdefinelanguage{operational-semantics-small-framed}
{
  language=java,
  basicstyle=\tiny\ttfamily,
  breaklines=false,
  columns=fullflexible,
  tabsize=2,
  showstringspaces=false,
  deletekeywords={public, private, protected, static, final, class, interface, abstract, implements, extends, if, else, while, do, for, switch, case, default, break, continue, return, int, long, double, float, boolean, char, void, String,this},
  morekeywords=[1]{if, else, while, do, for, switch, case, default, break, continue, return},
  keywordstyle=[1]\color{byzantine}\bfseries,
  morekeywords=[2]{public, private, protected, static, final, class, interface, abstract, implements, extends},
  keywordstyle=[2]\color{bluegray}\bfseries,
  morekeywords=[3]{int, long, double, float, boolean, char, void, String, @Override, @Test},
  keywordstyle=[3]\color{cadet}\bfseries,
  morekeywords=[4]{class, interface, extends, implements, new, super, throw, throws, try, catch, finally},
  keywordstyle=[4]\color{methodcolor},
  morecomment=[l]{//},
  commentstyle=\color{cadet},
  stringstyle=\color{brown(web)},
}

\lstdefinelanguage{bnf-grammar}{
  language=java,
  basicstyle=\scriptsize\ttfamily,
  numbers=left,
  numberstyle=\scriptsize,
  xleftmargin=18pt,
  morekeywords=[1]{::=,|,program,stmt,stmt_list,decl_stmt,assign_stmt,id,type,exp,literal,letter,digit,type,decl,BOOL, MATHOP,LOGICALOP,LOGNOT,SP,ID,block,aexp,bexp,RELOP,id_list,ids,program, class_decls, class_decl, field_decls, field_decl, fields, field_list, method_decls, method_decl, params, param, param_list, var_decls, var_decl, vars, var_list, var_defs, var_def, stmt, assign_stmt, if_stmt, while_stmt, for_stmt, return_stmt, break_stmt, block, stmts, method_call_stmt, field_access_stmt, method_call_list, field_access_list, arg_list, args, expr, literal, unary_expr, binary_expr, method_call_expr, field_access_expr, array_index_expr, BOOL, ARRAY, CHAR, STRING, INTEGER, MATHOP, RELOP, NOT, LOGICALOP, ID, letter, digit, SP, alphanumeric, alphanumeric_list, mathop, relop, bool, lognot, logicalop
},
  keywordstyle=[1]\color{blue},
  commentstyle=\color{green},
  stringstyle=\color{red},
  breaklines=true,
  backgroundcolor=\color{white},
  showstringspaces=false,
  rulecolor=\color{black},
  morecomment=[l]{\#},
}

\lstdefinelanguage{bnf-grammar-tiny}{
  language=java,
  basicstyle=\tiny\ttfamily,
  numbers=left,
  numberstyle=\tiny,
  xleftmargin=18pt,
  morekeywords=[1]{::=,|,program,stmt,stmt_list,decl_stmt,assign_stmt,id,type,exp,literal,letter,digit,type,decl,BOOL, MATHOP,LOGICALOP,LOGNOT,SP,ID,block,aexp,bexp,RELOP,id_list,ids,program, class_decls, class_decl, field_decls, field_decl, fields, field_list, method_decls, method_decl, params, param, param_list, var_decls, var_decl, vars, var_list, var_defs, var_def, stmt, assign_stmt, if_stmt, while_stmt, for_stmt, return_stmt, break_stmt, block, stmts, method_call_stmt, field_access_stmt, method_call_list, field_access_list, arg_list, args, expr, literal, unary_expr, binary_expr, method_call_expr, field_access_expr, array_index_expr, BOOL, ARRAY, CHAR, STRING, INTEGER, MATHOP, RELOP, NOT, LOGICALOP, ID, letter, digit, SP, alphanumeric, alphanumeric_list
},
  keywordstyle=[1]\color{blue},
  commentstyle=\color{green},
  stringstyle=\color{red},
  breaklines=true,
  backgroundcolor=\color{white},
  showstringspaces=false,
  rulecolor=\color{black},
  morecomment=[l]{\#},
}

\lstdefinelanguage{k-framework}{
  language=java,
  basicstyle=\scriptsize\ttfamily,
  numbers=left,
  numberstyle=\scriptsize,
  xleftmargin=18pt,
  morekeywords=[1]{syntax, configuration, rule, imports, import},
  keywordstyle=[1]\textbf,
  morekeywords=[2]{module, endmodule},
  keywordstyle=[2]\textbf,
  breaklines=true,
  backgroundcolor=\color{white},
  showstringspaces=false,
  rulecolor=\color{black},
  morecomment=[l]{\#},
}

\lstdefinelanguage{imp}{
	language=java,
	keywords=[2]{int, while, if, else, return, break, halt},
        keywordstyle=[2]\bfseries,
	keywords=[3]{Order, Asc},
	keywordstyle=[3]\color{purple},
	basicstyle=\scriptsize\ttfamily,
}

\lstdefinelanguage{imp-unseen}{
	language=java,
	keywords=[2]{int, while, if, else, return, break, halt},
        keywordstyle=[2]\bfseries,
        inputencoding=utf8,
        extendedchars=true,
        literate=
             {𐕂}{{\char"10542}}1
             {𐕕}{{\char"10555}}1
             {𐕊}{{\char"1054A}}1
             {𐕃}{{\char"10543}}1
             {𐕐}{{\char"10550}}1
             {𐕏}{{\char"1054F}}1
             {𐔸}{{\char"10538}}1
             {𐕟}{{\char"1055F}}1,
	basicstyle=\scriptsize\ttfamily,
}

\newcommand{\ProgramLine}[2][]{%
  \if\relax\detokenize{#1}\relax
    \item \emph{#2}%
  \else
    \item[#1] \emph{#2}%
  \fi
}

\newcommand{\ProgramLineIndent}[2][]{%
  \if\relax\detokenize{#1}\relax
    \item \hspace{1.5em}\emph{#2}%
  \else
    \item[#1] \hspace{1.5em}\emph{#2}%
  \fi
}

\usepackage[skins]{tcolorbox}
\usetikzlibrary{tikzmark,decorations.pathreplacing,calc}

\definecolor{light-purple}{RGB}{151,156,171}
\definecolor{cherryblossompink}{rgb}{1.0, 0.72, 0.77}
\definecolor{blue-color}{RGB}{40,166,189}
\definecolor{pink-color}{RGB}{237,46,104}
\definecolor{dark-grey-color}{RGB}{79,91,102}
\definecolor{darkbyzantium}{rgb}{0.36, 0.22, 0.33}
\definecolor{bluebell}{rgb}{0.64, 0.64, 0.82}
\definecolor{airforceblue}{rgb}{0.36, 0.54, 0.66}
\definecolor{response}{RGB}{245,198,165}

\newtcolorbox{prompt}[1][]{
    colbacktitle=cherryblossompink,
    colframe=cherryblossompink,
    coltitle=darkbyzantium,
    fontupper=\footnotesize\ttfamily,
    boxsep=5pt,
    left=0pt,
    right=0pt,
    top=0pt,
    bottom=0pt,
    boxrule=1pt,
    enhanced,
    #1,
}

\newcommand{\DefMacro}[2]{\expandafter\def\csname rmk-#1\endcsname{#2}}
\newcommand{\UseMacro}[1]{\csname rmk-#1\endcsname}

\definecolor{SkyBlueAlpha}{HTML}{B0E5ED}
\definecolor{LavenderAlpha}{HTML}{FBC2E4}
\definecolor{annotatecolor}{rgb}{0.59,0,0.09}

\newcommand{\TableFont}{\scriptsize}
\newcommand{\PaperTitle}{Predicting Program Exit Code with LLMs and Programming Language Semantics}
\newcommand{\dataset}{\textsc{PLSemanticsBench}\xspace}
\newcommand{\humanwrit}{Human-Written\xspace}
\newcommand{\llmtrans}{LLM-Translated\xspace}
\newcommand{\fuzzgen}{Fuzzer-Generated\xspace}

\newcommand{\PEPTask}{Program Executability Prediction\xspace}
\newcommand{\pep}{PrEx\xspace}
\newcommand{\pcp}{PrEx\xspace}

\newcommand{\op}{POP\xspace}

\newcommand{\srp}{SRP\xspace}

\newcommand{\etp}{ETP\xspace}
\newcommand{\pretraining}{pretraining\xspace}

\newcommand{\COT}{CoT\xspace}

\newcommand{\ebnf}{EBNF\xspace}

\newcommand{\Code}[1]{{\ifmmode{\mathtt{#1}}\else$\mathtt{#1}$\fi}}
\newcommand{\CodeIn}[1]{\texttt{#1}}

\newcommand{\MyPara}[1]{\vspace{2pt}\noindent\textbf{#1}.}

\newcommand{\redcircled}[1]{H\xspace}

\newcommand{\delt}[1]{%
  \pgfmathtruncatemacro{\DeltaInt}{round(#1)}%
  \pgfmathtruncatemacro{\DeltaAbsInt}{abs(\DeltaInt)}%
  \ifnum\DeltaAbsInt<10
    \edef\DeltaFmt{0\DeltaAbsInt}%
  \else
    \edef\DeltaFmt{\DeltaAbsInt}%
  \fi
  \ifnum\DeltaInt>0
    \textcolor{green!45!black}{\scriptsize(+\DeltaFmt)}%
  \else\ifnum\DeltaInt<0
    \textcolor{red!60!black}{\scriptsize(\texttt{-}\DeltaFmt)}%
  \else
    \textcolor{black!55}{\scriptsize(000)}%
  \fi\fi
}

\newcommand{\accdrobust}[2]{%
  \pgfmathtruncatemacro{\DeltaTmp}{round((#1) - (#2))}%
  #1\,\delt{\DeltaTmp}%
}
\newcommand{\accdrobustbf}[2]{%
  \pgfmathtruncatemacro{\DeltaTmp}{round((#1) - (#2))}%
  \textbf{#1}\,\delt{\DeltaTmp}%
}

\newcommand{\KTool}{K-framework\xspace}
\newcommand{\Kos}{\ensuremath{\mathbb{K}}\xspace}
\newcommand{\Sos}{\ensuremath{\mathbb{S}}-semantics\xspace}

\newcommand{\C}{C$^*$\xspace}
\newcommand{\IMP}{C$^*$\xspace}

\newcommand{\LLM}{LLM\xspace}
\newcommand{\LLMs}{LLMs\xspace}
\newcommand{\LLMses}{LLMs'\xspace}

\newcommand{\K}{\ensuremath{\mathbb{K}}\xspace}

\renewcommand{\S}{\ensuremath{\mathbb{S}}\xspace}

\newcommand{\OriginalValidProgs}{492\xspace}
\newcommand{\BaseIMPDataset}{491\xspace}
\newcommand{\InvalidIMPDataset}{2455\xspace}

\newcommand{\IMPPCPDataset}{2946\xspace}

\newcommand{\datspts}{Dataset Splits\xspace}

\newcommand{\modbyzero}{modulo-by-zero\xspace}
\newcommand{\divbyzero}{divide-by-zero\xspace}
\newcommand{\varusebdec}{variable-use-before-declare\xspace}
\newcommand{\contloop}{continue-outside-loop\xspace}
\newcommand{\breakloop}{break-outside-loop\xspace}
\newcommand{\transfs}{invalid transformations\xspace}
\newcommand{\transformationprocess}{transformation process\xspace}
\newcommand{\transformation}{transformation\xspace}

\newcommand{\keywordMut}{KeywordSwap\xspace}
\newcommand{\kswap}{KeywordSwap\xspace}
\newcommand{\kobf}{KeywordObf\xspace}
\newcommand{\KeywordSwap}{\keywordMut}

\newcommand{\keywordObf}{KeywordObf\xspace}
\newcommand{\KeywordObf}{\keywordObf}

\newcommand{\standardSem}{standard\xspace}
\newcommand{\nstandardSem}{nonstandard\xspace}

\newcommand{\standardSemCap}{Standard\xspace}

\newcommand{\Program}{$p$}

\newcommand{\MutatedProgramKS}{$p'_{ks}$}
\newcommand{\MutatedProgramKO}{$p'_{ko}$}
\newcommand{\OperationalSemantics}{$s$}

\newcommand{\KeywordMutOperationalSemantics}{${s'_{ks}}$}
\newcommand{\KeywordObfOperationalSemantics}{${s'_{ko}}$}

\newcommand{\uk}{\boldmath(\OperationalSemantics,\Program)}

\newcommand{\ksMk}{\boldmath(\KeywordMutOperationalSemantics, \MutatedProgramKS)}
\newcommand{\koMk}{\boldmath(\KeywordObfOperationalSemantics, \MutatedProgramKO)}
\newcommand{\nk}{\boldmath\Program}

\newcommand{\KOADD}{\raisebox{-0.05ex}{\large$\star$}}
\newcommand{\KOSUB}{$\blacktriangle$}
\newcommand{\KOMUL}{\texttt{\$}}
\newcommand{\KODIV}{$\blacksquare$}
\newcommand{\KOMOD}{$\blacktriangledown$}
\newcommand{\KOASSIGN}{\texteuro}
\newcommand{\KOLT}{$\bullet$}
\newcommand{\KOGT}{$\circledcirc$}
\newcommand{\KOLTEQ}{$\bigtriangledown$}
\newcommand{\KOGTEQ}{\textcolor{red}{$\heartsuit$}}
\newcommand{\KOEQ}{$\blacklozenge$}
\newcommand{\KONEQ}{$\blacktriangleright$}
\newcommand{\KOAND}{$\Diamond$}
\newcommand{\KOOR}{$\bigtriangleup$}
\newcommand{\KONOT}{$\bigcirc$}
\newcommand{\KOBREAK}{$\checkmark$}
\newcommand{\KOIF}{$\triangleright$}
\newcommand{\KOELSE}{\textyen}
\newcommand{\KOWHILE}{\faShekelSign}
\newcommand{\KOHALT}{\faRupeeSign}
\newcommand{\KOCONTINUE}{\ding{51}}

\newcommand{\llamaBig}{\textsc{Llama-3.3 70B}\xspace}
\newcommand{\gemini}{\textsc{Gemini-2.5-pro}\xspace}
\newcommand{\qwenCoderGeneral}{\textsc{Qwen2.5-Coder}\xspace}
\newcommand{\qwenCoder}[1]{\textsc{Qwen2.5-Coder #1B\xspace}}
\newcommand{\qwenCoderCoT}[1]{\qwenCoder{#1}-\COT}
\newcommand{\ministral}[1]{\textsc{Ministral 3 #1B\xspace}}
\newcommand{\ministralCoT}[1]{\ministral{#1}-\COT}
\newcommand{\dpskQwen}[1]{\textsc{DeepSeek-Qwen #1B\xspace}}
\newcommand{\qwq}{\textsc{QwQ 32B}\xspace}
\newcommand{\dpskLlama}[1]{\textsc{DeepSeek-Llama #1B\xspace}}
\newcommand{\gptfo}{\textsc{GPT-4o}\xspace}
\newcommand{\othree}{\textsc{o3-mini}\xspace}

\newcommand{\FigPCPQualitativeExamplesCaption}{Minimal \humanwrit \pcp failures under \Sos (seed~\UseMacro{pcp-qual-seed}). Both programs are \UseMacro{pcp-qual-example-loc} lines long. Highlighted lines introduce the semantic error. Even on these small examples, top models may predict the correct invalidity but cite the wrong rule.}
\newcommand{\FigPCPErrorTypeRadarCaption}{Per-error-type accuracy on invalid \C programs across all evaluated models under \Sos. The top row shows \humanwrit programs and the bottom row shows \fuzzgen programs (Standard, \keywordMut, and \keywordObf configurations). Radar axes are models and each colored polygon is a semantic error type. Radial grid lines mark accuracy percentages, using the same rule-level criterion as Tables~\ref{tab:pcp-IMP-K-IMP-SOS-accuracy-qwen-coder-by-split-human-written} and~\ref{tab:pcp-IMP-K-IMP-SOS-accuracy-qwen-coder-by-split-fuzzer-generated}.\label{figure:pcp-error-type-radar}}

\newcommand{\FigPCPSOSUKRadarCaption}{\standardSemCap, \Sos}
\newcommand{\FigPCPSOSSwapRadarCaption}{\keywordMut, \Sos}
\newcommand{\FigPCPSOSObfRadarCaption}{\keywordObf, \Sos}

\newcommand{\FigValidInvalidExampleCaption}{Example of a valid program and semantically invalid programs with their corresponding error categories.\label{figure:valid-invalid-example}}
\newcommand{\FigPEPTaskExampleCaption}{Valid and invalid \pep task examples.\label{figure:pep-task-example}}
\newcommand{\FigPEPTaskValidCaption}{Valid program\label{figure:pep-task-valid}}
\newcommand{\FigPEPTaskInvalidCaption}{Invalid program\label{figure:pep-task-invalid}}
\newcommand{\PepTaskExampleDivByZeroLine}{8}

\newcommand{\FigIMPEBNFCaption}{Complete syntax of \IMP in EBNF (\texttt{<letter>} and \texttt{<digit>} rules are omitted and have their usual definitions).\label{figure:imp-syntax}}

\newcommand{\FigPCPPromptOverviewCaption}{Overview of the \pcp prompt: the model receives the \C{} grammar, semantic rules (\S{} or \K), and a program, and must predict executability.\label{figure:pcp-prompt-overview}}
\newcommand{\FigIMPSOSRulesSubsetCaption}{Some \Sos rules of the \C language.\label{table:imp-sos-rules}}
\newcommand{\FigIMPKRulesSubsetCaption}{Some \Kos-semantics rules of the \C language.\label{table:imp-k-rules}}

\DefMacro{DataExample}{One example of the original \CodeIn{C++} solution and the corresponding re-written \IMP program.}
\DefMacro{cProgramExample}{The \texttt{C++} solution to the problem ``MBCPP/962'' in BabelCode MBPP and one public test case. The public test we use is \CodeIn{sumEven(3, 8)==18}.}
\DefMacro{impProgramExample}{The \IMP program re-written from the C++ solution.}

\DefMacro{TCap-benchmark-dataset-stats}{Statistics of \dataset.\vspace{-0.1cm}}
\DefMacro{TCap-models-srp-etp-IMP-SOS}{Models' exact-match accuracies on \srp and \etp tasks. \label{tab:srp-etp}}
\DefMacro{TCap-models-pcp-op-IMP-SOS}{Models' accuracies on \pep and \op tasks. \label{tab:pep-pop}}

\newcommand{\TableNonStandardRulesCaption}{Transformations applied to the \standardSem
semantics to derive the \nstandardSem \keywordMut
and \keywordObf.
\label{tab:mutation-rules-combined}}

\DefMacro{dataset-imp-dataset-avg-loc}{15.9}
\DefMacro{dataset-imp-dataset-avg-rules-per-state}{5.0}
\DefMacro{dataset-imp-dataset-avg-tokens}{87.1}
\DefMacro{dataset-imp-dataset-avg-trace-length}{72.1}
\DefMacro{dataset-imp-dataset-max-rules-per-state}{26}
\DefMacro{dataset-imp-dataset-max-trace-length}{2,407}
\DefMacro{dataset-imp-dataset-min-rules-per-state}{1}
\DefMacro{dataset-imp-dataset-min-trace-length}{3}
\DefMacro{dataset-imp-dataset-num-programs}{162}

\DefMacro{res-Qwen-QwQ-32B-da-op-nk-IMP-SOS-acc}{99}
\DefMacro{res-Qwen-QwQ-32B-da-op-nk-IMP-SOS-malformed-count}{0}
\DefMacro{res-o3-mini-da-op-nk-IMP-SOS-acc}{99}
\DefMacro{res-o3-mini-da-op-nk-IMP-SOS-malformed-count}{0}
\DefMacro{res-gpt-4o-mini-cot-op-nk-IMP-SOS-acc}{77}
\DefMacro{res-gpt-4o-mini-cot-op-nk-IMP-SOS-malformed-count}{100}
\DefMacro{res-gemini-2.5-pro-preview-05-06-da-op-nk-IMP-SOS-acc}{85}
\DefMacro{res-gemini-2.5-pro-preview-05-06-da-op-nk-IMP-SOS-malformed-count}{0}
\DefMacro{res-Qwen-Qwen2.5-Coder-3B-Instruct-da-op-nk-IMP-SOS-acc}{29}
\DefMacro{res-Qwen-Qwen2.5-Coder-3B-Instruct-da-op-nk-IMP-SOS-malformed-count}{0}
\DefMacro{res-Qwen-Qwen2.5-Coder-3B-Instruct-cot-op-nk-IMP-SOS-acc}{61}
\DefMacro{res-Qwen-Qwen2.5-Coder-3B-Instruct-cot-op-nk-IMP-SOS-malformed-count}{11}
\DefMacro{res-Qwen-Qwen2.5-Coder-7B-Instruct-da-op-nk-IMP-SOS-acc}{35}
\DefMacro{res-Qwen-Qwen2.5-Coder-7B-Instruct-da-op-nk-IMP-SOS-malformed-count}{0}
\DefMacro{res-Qwen-Qwen2.5-Coder-7B-Instruct-cot-op-nk-IMP-SOS-acc}{69}
\DefMacro{res-Qwen-Qwen2.5-Coder-7B-Instruct-cot-op-nk-IMP-SOS-malformed-count}{8}
\DefMacro{res-Qwen-Qwen2.5-Coder-14B-Instruct-da-op-nk-IMP-SOS-acc}{46}
\DefMacro{res-Qwen-Qwen2.5-Coder-14B-Instruct-da-op-nk-IMP-SOS-malformed-count}{0}
\DefMacro{res-Qwen-Qwen2.5-Coder-14B-Instruct-cot-op-nk-IMP-SOS-acc}{81}
\DefMacro{res-Qwen-Qwen2.5-Coder-14B-Instruct-cot-op-nk-IMP-SOS-malformed-count}{100}
\DefMacro{res-Qwen-Qwen2.5-Coder-32B-Instruct-da-op-nk-IMP-SOS-acc}{47}
\DefMacro{res-Qwen-Qwen2.5-Coder-32B-Instruct-da-op-nk-IMP-SOS-malformed-count}{0}
\DefMacro{res-Qwen-Qwen2.5-Coder-32B-Instruct-cot-op-nk-IMP-SOS-acc}{86}
\DefMacro{res-Qwen-Qwen2.5-Coder-32B-Instruct-cot-op-nk-IMP-SOS-malformed-count}{0}
\DefMacro{res-meta-llama-Llama-3.3-70B-Instruct-da-op-nk-IMP-SOS-acc}{44}
\DefMacro{res-meta-llama-Llama-3.3-70B-Instruct-da-op-nk-IMP-SOS-malformed-count}{0}
\DefMacro{res-meta-llama-Llama-3.3-70B-Instruct-cot-op-nk-IMP-SOS-acc}{89}
\DefMacro{res-meta-llama-Llama-3.3-70B-Instruct-cot-op-nk-IMP-SOS-malformed-count}{0}
\DefMacro{res-deepseek-ai-DeepSeek-R1-Distill-Llama-70B-da-op-nk-IMP-SOS-acc}{83}
\DefMacro{res-deepseek-ai-DeepSeek-R1-Distill-Llama-70B-da-op-nk-IMP-SOS-malformed-count}{13}
\DefMacro{res-deepseek-ai-DeepSeek-R1-Distill-Qwen-14B-da-op-nk-IMP-SOS-acc}{90}
\DefMacro{res-deepseek-ai-DeepSeek-R1-Distill-Qwen-14B-da-op-nk-IMP-SOS-malformed-count}{0}
\DefMacro{res-deepseek-ai-DeepSeek-R1-Distill-Qwen-32B-da-op-nk-IMP-SOS-acc}{88}
\DefMacro{res-deepseek-ai-DeepSeek-R1-Distill-Qwen-32B-da-op-nk-IMP-SOS-malformed-count}{0}
\DefMacro{res-o3-mini-da-op-uk-IMP-SOS-acc}{\cellcolor{green!10}100}
\DefMacro{res-o3-mini-da-op-uk-IMP-SOS-malformed-count}{0}
\DefMacro{res-gpt-5-mini-da-op-uk-IMP-SOS-acc}{100}
\DefMacro{res-gpt-5-mini-da-op-uk-IMP-SOS-malformed-count}{0}
\DefMacro{res-o3-mini-da-op-mk-IMP-SOS-addSub_mulDiv_negateRelation-acc}{63}
\DefMacro{res-o3-mini-da-op-mk-IMP-SOS-unseen-acc}{95}
\DefMacro{res-o3-mini-da-op-mk-IMP-SOS-malformed-count}{7.3}
\DefMacro{res-gpt-5-mini-da-op-mk-IMP-SOS-addSub_mulDiv_negateRelation-acc}{79}
\DefMacro{res-gpt-5-mini-da-op-mk-IMP-SOS-unseen-acc}{99}
\DefMacro{res-gpt-5-mini-da-op-mk-IMP-SOS-malformed-count}{5.7}

\DefMacro{dataset-mk-IMP-SOS-mutation-rate-addSub}{85.8}
\DefMacro{dataset-mk-IMP-SOS-mutation-rate-addSub-mulDiv}{92.0}
\DefMacro{dataset-mk-IMP-SOS-mutation-rate-addSub-mulDiv-negateRelation}{100}
\DefMacro{dataset-mk-IMP-SOS-mutation-rate-addSub-negateRelation}{96.9}
\DefMacro{dataset-mk-IMP-SOS-mutation-rate-mulDiv}{70.4}
\DefMacro{dataset-mk-IMP-SOS-mutation-rate-mulDiv-negateRelation}{98.8}
\DefMacro{dataset-mk-IMP-SOS-mutation-rate-negateRelation}{92.0}
\DefMacro{dataset-mk-IMP-SOS-num-mutated-programs-addSub}{139}
\DefMacro{dataset-mk-IMP-SOS-num-mutated-programs-addSub-mulDiv}{149}
\DefMacro{dataset-mk-IMP-SOS-num-mutated-programs-addSub-mulDiv-negateRelation}{162}
\DefMacro{dataset-mk-IMP-SOS-num-mutated-programs-addSub-negateRelation}{157}
\DefMacro{dataset-mk-IMP-SOS-num-mutated-programs-mulDiv}{114}
\DefMacro{dataset-mk-IMP-SOS-num-mutated-programs-mulDiv-negateRelation}{160}
\DefMacro{dataset-mk-IMP-SOS-num-mutated-programs-negateRelation}{149}

\DefMacro{dataset-benchmark-imp-num-programs}{162}
\DefMacro{dataset-benchmark-imp-mean-loc}{15.9}
\DefMacro{dataset-benchmark-imp-max-loc}{52}
\DefMacro{dataset-benchmark-imp-min-loc}{3}
\DefMacro{dataset-benchmark-imp-mean-tokens}{87.3}
\DefMacro{dataset-benchmark-imp-max-tokens}{308}
\DefMacro{dataset-benchmark-imp-min-tokens}{17}
\DefMacro{dataset-benchmark-imp-mean-vars}{4.2}
\DefMacro{dataset-benchmark-imp-max-vars}{15}
\DefMacro{dataset-benchmark-imp-min-vars}{2}
\DefMacro{dataset-benchmark-etp-mean-trace-length}{23.1}
\DefMacro{dataset-benchmark-etp-max-trace-length}{107}
\DefMacro{dataset-benchmark-etp-min-trace-length}{3}
\DefMacro{dataset-benchmark-srp-mean-selected-stmts}{7.1}
\DefMacro{dataset-benchmark-srp-max-selected-stmts}{10}
\DefMacro{dataset-benchmark-srp-min-selected-stmts}{3}
\DefMacro{dataset-benchmark-srp-mean-rules-per-stmt}{4.7}
\DefMacro{dataset-benchmark-srp-max-rules-per-stmt}{26}
\DefMacro{dataset-benchmark-srp-min-rules-per-stmt}{1}

\DefMacro{dataset-program-executability-total-programs}{2946}
\DefMacro{dataset-program-executability-success-count}{491}
\DefMacro{dataset-program-executability-success-percent}{16.7\%}
\DefMacro{dataset-program-executability-break-outside-loop-count}{491}
\DefMacro{dataset-program-executability-break-outside-loop-percent}{16.7\%}
\DefMacro{dataset-program-executability-continue-outside-loop-count}{491}
\DefMacro{dataset-program-executability-continue-outside-loop-percent}{16.7\%}
\DefMacro{dataset-program-executability-divide-by-zero-count}{491}
\DefMacro{dataset-program-executability-divide-by-zero-percent}{16.7\%}
\DefMacro{dataset-program-executability-modulo-zero-count}{491}
\DefMacro{dataset-program-executability-modulo-zero-percent}{16.7\%}
\DefMacro{dataset-program-executability-var-use-before-declare-count}{491}
\DefMacro{dataset-program-executability-var-use-before-declare-percent}{16.7\%}

\DefMacro{dataset-imp-split-loc-human-written-num-programs}{162}
\DefMacro{dataset-imp-split-loc-human-written-min-loc}{4}
\DefMacro{dataset-imp-split-loc-human-written-median-loc}{19}
\DefMacro{dataset-imp-split-loc-human-written-max-loc}{80}
\DefMacro{dataset-imp-split-loc-human-written-min-tokens}{18}
\DefMacro{dataset-imp-split-loc-human-written-median-tokens}{81}
\DefMacro{dataset-imp-split-loc-human-written-max-tokens}{351}
\DefMacro{dataset-imp-split-loc-synthetic-cpp-num-programs}{165}
\DefMacro{dataset-imp-split-loc-synthetic-cpp-min-loc}{8}
\DefMacro{dataset-imp-split-loc-synthetic-cpp-median-loc}{106}
\DefMacro{dataset-imp-split-loc-synthetic-cpp-max-loc}{597}
\DefMacro{dataset-imp-split-loc-synthetic-cpp-min-tokens}{33}
\DefMacro{dataset-imp-split-loc-synthetic-cpp-median-tokens}{538}
\DefMacro{dataset-imp-split-loc-synthetic-cpp-max-tokens}{2,092}
\DefMacro{dataset-imp-split-loc-fuzzer-generated-num-programs}{164}
\DefMacro{dataset-imp-split-loc-fuzzer-generated-min-loc}{85}
\DefMacro{dataset-imp-split-loc-fuzzer-generated-median-loc}{786}
\DefMacro{dataset-imp-split-loc-fuzzer-generated-max-loc}{1,961}
\DefMacro{dataset-imp-split-loc-fuzzer-generated-min-tokens}{982}
\DefMacro{dataset-imp-split-loc-fuzzer-generated-median-tokens}{9,081}
\DefMacro{dataset-imp-split-loc-fuzzer-generated-max-tokens}{22,239}

\DefMacro{dataset-imp-split-complexity-human-written-num-programs}{162}
\DefMacro{dataset-imp-split-complexity-human-written-median-cc}{3}
\DefMacro{dataset-imp-split-complexity-human-written-median-max-nested-if}{1}
\DefMacro{dataset-imp-split-complexity-human-written-median-max-nested-while}{1}
\DefMacro{dataset-imp-split-complexity-synthetic-cpp-num-programs}{165}
\DefMacro{dataset-imp-split-complexity-synthetic-cpp-median-cc}{9}
\DefMacro{dataset-imp-split-complexity-synthetic-cpp-median-max-nested-if}{1}
\DefMacro{dataset-imp-split-complexity-synthetic-cpp-median-max-nested-while}{1}
\DefMacro{dataset-imp-split-complexity-fuzzer-generated-num-programs}{164}
\DefMacro{dataset-imp-split-complexity-fuzzer-generated-median-cc}{100}
\DefMacro{dataset-imp-split-complexity-fuzzer-generated-median-max-nested-if}{7}
\DefMacro{dataset-imp-split-complexity-fuzzer-generated-median-max-nested-while}{6}

\DefMacro{pcp-split-drop-capable-hw-threshold}{45}
\DefMacro{pcp-split-drop-capable-mean-fuzzgen-min-pp}{13}
\DefMacro{pcp-split-drop-capable-mean-fuzzgen-max-pp}{40}
\DefMacro{pcp-split-drop-capable-mean-fuzzgen-median-pp}{24}

\DefMacro{pcp-split-drop-dpsk32-mean-hw}{88.7}
\DefMacro{pcp-split-drop-dpsk32-uk-k-fuzzgen-pp}{11}
\DefMacro{pcp-split-drop-dpsk32-uk-k-fuzzgen-pct}{10.8\%}
\DefMacro{pcp-split-drop-dpsk32-mk-k-swap-fuzzgen-pp}{19}
\DefMacro{pcp-split-drop-dpsk32-mk-k-swap-fuzzgen-pct}{24.9\%}
\DefMacro{pcp-split-drop-dpsk32-mk-k-obf-fuzzgen-pp}{29}
\DefMacro{pcp-split-drop-dpsk32-mk-k-obf-fuzzgen-pct}{30.1\%}
\DefMacro{pcp-split-drop-dpsk32-uk-sos-fuzzgen-pp}{9}
\DefMacro{pcp-split-drop-dpsk32-uk-sos-fuzzgen-pct}{8.7\%}
\DefMacro{pcp-split-drop-dpsk32-mk-sos-swap-fuzzgen-pp}{15}
\DefMacro{pcp-split-drop-dpsk32-mk-sos-swap-fuzzgen-pct}{20.5\%}
\DefMacro{pcp-split-drop-dpsk32-mk-sos-obf-fuzzgen-pp}{32}
\DefMacro{pcp-split-drop-dpsk32-mk-sos-obf-fuzzgen-pct}{35.8\%}
\DefMacro{pcp-split-drop-dpsk32-uk-k-llmtrans-pp}{1}
\DefMacro{pcp-split-drop-dpsk32-uk-k-llmtrans-pct}{1.3\%}
\DefMacro{pcp-split-drop-dpsk32-mk-k-swap-llmtrans-pp}{2}
\DefMacro{pcp-split-drop-dpsk32-mk-k-swap-llmtrans-pct}{2.7\%}
\DefMacro{pcp-split-drop-dpsk32-mk-k-obf-llmtrans-pp}{6}
\DefMacro{pcp-split-drop-dpsk32-mk-k-obf-llmtrans-pct}{6.3\%}
\DefMacro{pcp-split-drop-dpsk32-uk-sos-llmtrans-pp}{2}
\DefMacro{pcp-split-drop-dpsk32-uk-sos-llmtrans-pct}{1.6\%}
\DefMacro{pcp-split-drop-dpsk32-mk-sos-swap-llmtrans-pp}{1}
\DefMacro{pcp-split-drop-dpsk32-mk-sos-swap-llmtrans-pct}{1.3\%}
\DefMacro{pcp-split-drop-dpsk32-mk-sos-obf-llmtrans-pp}{5}
\DefMacro{pcp-split-drop-dpsk32-mk-sos-obf-llmtrans-pct}{6\%}

\DefMacro{pcp-split-drop-dpsk32-mean-llmtrans-pp}{2.9}
\DefMacro{pcp-split-drop-dpsk32-mean-fuzzgen-pp}{19}
\DefMacro{pcp-split-drop-dpsk32-mean-fuzzgen-pct}{21.8\%}
\DefMacro{pcp-split-drop-dpsk32-max-fuzzgen-pp}{32}
\DefMacro{pcp-split-drop-dpsk32-max-fuzzgen-pct}{35.8\%}
\DefMacro{pcp-split-drop-dpsk32-mean-llmtrans-pct}{3.2\%}
\DefMacro{pcp-split-drop-dpsk32-max-llmtrans-pp}{6}
\DefMacro{pcp-split-drop-dpsk32-max-llmtrans-pct}{6.3\%}

\DefMacro{pcp-split-drop-min14cot-mean-hw}{89.1}
\DefMacro{pcp-split-drop-min14cot-uk-k-fuzzgen-pp}{11}
\DefMacro{pcp-split-drop-min14cot-uk-k-fuzzgen-pct}{11.3\%}
\DefMacro{pcp-split-drop-min14cot-mk-k-swap-fuzzgen-pp}{25}
\DefMacro{pcp-split-drop-min14cot-mk-k-swap-fuzzgen-pct}{31.7\%}
\DefMacro{pcp-split-drop-min14cot-mk-k-obf-fuzzgen-pp}{25}
\DefMacro{pcp-split-drop-min14cot-mk-k-obf-fuzzgen-pct}{27.4\%}
\DefMacro{pcp-split-drop-min14cot-uk-sos-fuzzgen-pp}{23}
\DefMacro{pcp-split-drop-min14cot-uk-sos-fuzzgen-pct}{23.3\%}
\DefMacro{pcp-split-drop-min14cot-mk-sos-swap-fuzzgen-pp}{28}
\DefMacro{pcp-split-drop-min14cot-mk-sos-swap-fuzzgen-pct}{35.2\%}
\DefMacro{pcp-split-drop-min14cot-mk-sos-obf-fuzzgen-pp}{37}
\DefMacro{pcp-split-drop-min14cot-mk-sos-obf-fuzzgen-pct}{41.6\%}
\DefMacro{pcp-split-drop-min14cot-uk-k-llmtrans-pp}{2}
\DefMacro{pcp-split-drop-min14cot-uk-k-llmtrans-pct}{2\%}
\DefMacro{pcp-split-drop-min14cot-mk-k-swap-llmtrans-pp}{2}
\DefMacro{pcp-split-drop-min14cot-mk-k-swap-llmtrans-pct}{2\%}
\DefMacro{pcp-split-drop-min14cot-mk-k-obf-llmtrans-pp}{10}
\DefMacro{pcp-split-drop-min14cot-mk-k-obf-llmtrans-pct}{11.1\%}
\DefMacro{pcp-split-drop-min14cot-uk-sos-llmtrans-pp}{10}
\DefMacro{pcp-split-drop-min14cot-uk-sos-llmtrans-pct}{10.4\%}
\DefMacro{pcp-split-drop-min14cot-mk-sos-swap-llmtrans-pp}{7}
\DefMacro{pcp-split-drop-min14cot-mk-sos-swap-llmtrans-pct}{9.3\%}
\DefMacro{pcp-split-drop-min14cot-mk-sos-obf-llmtrans-pp}{21}
\DefMacro{pcp-split-drop-min14cot-mk-sos-obf-llmtrans-pct}{23.7\%}

\DefMacro{pcp-split-drop-min14cot-mean-llmtrans-pp}{8.7}
\DefMacro{pcp-split-drop-min14cot-mean-fuzzgen-pp}{24.8}
\DefMacro{pcp-split-drop-min14cot-mean-fuzzgen-pct}{28.4\%}
\DefMacro{pcp-split-drop-min14cot-max-fuzzgen-pp}{37}
\DefMacro{pcp-split-drop-min14cot-max-fuzzgen-pct}{41.6\%}
\DefMacro{pcp-split-drop-min14cot-mean-llmtrans-pct}{9.8\%}
\DefMacro{pcp-split-drop-min14cot-max-llmtrans-pp}{21}
\DefMacro{pcp-split-drop-min14cot-max-llmtrans-pct}{23.7\%}

\DefMacro{pcp-split-drop-qwen32cot-mean-hw}{82.5}
\DefMacro{pcp-split-drop-qwen32cot-uk-k-fuzzgen-pp}{26}
\DefMacro{pcp-split-drop-qwen32cot-uk-k-fuzzgen-pct}{26.2\%}
\DefMacro{pcp-split-drop-qwen32cot-mk-k-swap-fuzzgen-pp}{32}
\DefMacro{pcp-split-drop-qwen32cot-mk-k-swap-fuzzgen-pct}{39.9\%}
\DefMacro{pcp-split-drop-qwen32cot-mk-k-obf-fuzzgen-pp}{48}
\DefMacro{pcp-split-drop-qwen32cot-mk-k-obf-fuzzgen-pct}{66.2\%}
\DefMacro{pcp-split-drop-qwen32cot-uk-sos-fuzzgen-pp}{21}
\DefMacro{pcp-split-drop-qwen32cot-uk-sos-fuzzgen-pct}{21.2\%}
\DefMacro{pcp-split-drop-qwen32cot-mk-sos-swap-fuzzgen-pp}{35}
\DefMacro{pcp-split-drop-qwen32cot-mk-sos-swap-fuzzgen-pct}{44.1\%}
\DefMacro{pcp-split-drop-qwen32cot-mk-sos-obf-fuzzgen-pp}{39}
\DefMacro{pcp-split-drop-qwen32cot-mk-sos-obf-fuzzgen-pct}{57\%}
\DefMacro{pcp-split-drop-qwen32cot-uk-k-llmtrans-pp}{4}
\DefMacro{pcp-split-drop-qwen32cot-uk-k-llmtrans-pct}{3.9\%}
\DefMacro{pcp-split-drop-qwen32cot-mk-k-swap-llmtrans-pp}{5}
\DefMacro{pcp-split-drop-qwen32cot-mk-k-swap-llmtrans-pct}{6.8\%}
\DefMacro{pcp-split-drop-qwen32cot-mk-k-obf-llmtrans-pp}{20}
\DefMacro{pcp-split-drop-qwen32cot-mk-k-obf-llmtrans-pct}{28.3\%}
\DefMacro{pcp-split-drop-qwen32cot-uk-sos-llmtrans-pp}{2}
\DefMacro{pcp-split-drop-qwen32cot-uk-sos-llmtrans-pct}{2.1\%}
\DefMacro{pcp-split-drop-qwen32cot-mk-sos-swap-llmtrans-pp}{1}
\DefMacro{pcp-split-drop-qwen32cot-mk-sos-swap-llmtrans-pct}{1.4\%}
\DefMacro{pcp-split-drop-qwen32cot-mk-sos-obf-llmtrans-pp}{19}
\DefMacro{pcp-split-drop-qwen32cot-mk-sos-obf-llmtrans-pct}{27.9\%}

\DefMacro{pcp-split-drop-qwen32cot-mean-llmtrans-pp}{8.6}
\DefMacro{pcp-split-drop-qwen32cot-mean-fuzzgen-pp}{33.3}
\DefMacro{pcp-split-drop-qwen32cot-mean-fuzzgen-pct}{42.4\%}
\DefMacro{pcp-split-drop-qwen32cot-max-fuzzgen-pp}{48}
\DefMacro{pcp-split-drop-qwen32cot-max-fuzzgen-pct}{66.2\%}
\DefMacro{pcp-split-drop-qwen32cot-mean-llmtrans-pct}{11.7\%}
\DefMacro{pcp-split-drop-qwen32cot-max-llmtrans-pp}{20}
\DefMacro{pcp-split-drop-qwen32cot-max-llmtrans-pct}{28.3\%}

\DefMacro{pcp-split-drop-dpsk14-mean-hw}{77.5}
\DefMacro{pcp-split-drop-dpsk14-uk-k-fuzzgen-pp}{13}
\DefMacro{pcp-split-drop-dpsk14-uk-k-fuzzgen-pct}{14.4\%}
\DefMacro{pcp-split-drop-dpsk14-mk-k-swap-fuzzgen-pp}{26}
\DefMacro{pcp-split-drop-dpsk14-mk-k-swap-fuzzgen-pct}{38.1\%}
\DefMacro{pcp-split-drop-dpsk14-mk-k-obf-fuzzgen-pp}{29}
\DefMacro{pcp-split-drop-dpsk14-mk-k-obf-fuzzgen-pct}{36.4\%}
\DefMacro{pcp-split-drop-dpsk14-uk-sos-fuzzgen-pp}{14}
\DefMacro{pcp-split-drop-dpsk14-uk-sos-fuzzgen-pct}{14.8\%}
\DefMacro{pcp-split-drop-dpsk14-mk-sos-swap-fuzzgen-pp}{29}
\DefMacro{pcp-split-drop-dpsk14-mk-sos-swap-fuzzgen-pct}{40.5\%}
\DefMacro{pcp-split-drop-dpsk14-mk-sos-obf-fuzzgen-pp}{24}
\DefMacro{pcp-split-drop-dpsk14-mk-sos-obf-fuzzgen-pct}{39.2\%}
\DefMacro{pcp-split-drop-dpsk14-uk-k-llmtrans-pp}{1}
\DefMacro{pcp-split-drop-dpsk14-uk-k-llmtrans-pct}{1.1\%}
\DefMacro{pcp-split-drop-dpsk14-mk-k-swap-llmtrans-pp}{-2}
\DefMacro{pcp-split-drop-dpsk14-mk-k-swap-llmtrans-pct}{-2.9\%}
\DefMacro{pcp-split-drop-dpsk14-mk-k-obf-llmtrans-pp}{3}
\DefMacro{pcp-split-drop-dpsk14-mk-k-obf-llmtrans-pct}{4.3\%}
\DefMacro{pcp-split-drop-dpsk14-uk-sos-llmtrans-pp}{4}
\DefMacro{pcp-split-drop-dpsk14-uk-sos-llmtrans-pct}{4.6\%}
\DefMacro{pcp-split-drop-dpsk14-mk-sos-swap-llmtrans-pp}{2}
\DefMacro{pcp-split-drop-dpsk14-mk-sos-swap-llmtrans-pct}{2.6\%}
\DefMacro{pcp-split-drop-dpsk14-mk-sos-obf-llmtrans-pp}{1}
\DefMacro{pcp-split-drop-dpsk14-mk-sos-obf-llmtrans-pct}{1.9\%}

\DefMacro{pcp-split-drop-dpsk14-mean-llmtrans-pp}{1.6}
\DefMacro{pcp-split-drop-dpsk14-mean-fuzzgen-pp}{22.5}
\DefMacro{pcp-split-drop-dpsk14-mean-fuzzgen-pct}{30.6\%}
\DefMacro{pcp-split-drop-dpsk14-max-fuzzgen-pp}{29}
\DefMacro{pcp-split-drop-dpsk14-max-fuzzgen-pct}{40.5\%}
\DefMacro{pcp-split-drop-dpsk14-mean-llmtrans-pct}{1.9\%}
\DefMacro{pcp-split-drop-dpsk14-max-llmtrans-pp}{4}
\DefMacro{pcp-split-drop-dpsk14-max-llmtrans-pct}{4.6\%}

\DefMacro{pcp-split-drop-qwen14da-mean-hw}{53.8}
\DefMacro{pcp-split-drop-qwen14da-uk-k-fuzzgen-pp}{6}
\DefMacro{pcp-split-drop-qwen14da-uk-k-fuzzgen-pct}{9.4\%}
\DefMacro{pcp-split-drop-qwen14da-mk-k-swap-fuzzgen-pp}{11}
\DefMacro{pcp-split-drop-qwen14da-mk-k-swap-fuzzgen-pct}{27.2\%}
\DefMacro{pcp-split-drop-qwen14da-mk-k-obf-fuzzgen-pp}{1}
\DefMacro{pcp-split-drop-qwen14da-mk-k-obf-fuzzgen-pct}{3.6\%}
\DefMacro{pcp-split-drop-qwen14da-uk-sos-fuzzgen-pp}{18}
\DefMacro{pcp-split-drop-qwen14da-uk-sos-fuzzgen-pct}{21.3\%}
\DefMacro{pcp-split-drop-qwen14da-mk-sos-swap-fuzzgen-pp}{35}
\DefMacro{pcp-split-drop-qwen14da-mk-sos-swap-fuzzgen-pct}{47.6\%}
\DefMacro{pcp-split-drop-qwen14da-mk-sos-obf-fuzzgen-pp}{10}
\DefMacro{pcp-split-drop-qwen14da-mk-sos-obf-fuzzgen-pct}{27.9\%}
\DefMacro{pcp-split-drop-qwen14da-uk-k-llmtrans-pp}{9}
\DefMacro{pcp-split-drop-qwen14da-uk-k-llmtrans-pct}{13.4\%}
\DefMacro{pcp-split-drop-qwen14da-mk-k-swap-llmtrans-pp}{2}
\DefMacro{pcp-split-drop-qwen14da-mk-k-swap-llmtrans-pct}{4.8\%}
\DefMacro{pcp-split-drop-qwen14da-mk-k-obf-llmtrans-pp}{5}
\DefMacro{pcp-split-drop-qwen14da-mk-k-obf-llmtrans-pct}{21.2\%}
\DefMacro{pcp-split-drop-qwen14da-uk-sos-llmtrans-pp}{8}
\DefMacro{pcp-split-drop-qwen14da-uk-sos-llmtrans-pct}{9.9\%}
\DefMacro{pcp-split-drop-qwen14da-mk-sos-swap-llmtrans-pp}{4}
\DefMacro{pcp-split-drop-qwen14da-mk-sos-swap-llmtrans-pct}{5.9\%}
\DefMacro{pcp-split-drop-qwen14da-mk-sos-obf-llmtrans-pp}{7}
\DefMacro{pcp-split-drop-qwen14da-mk-sos-obf-llmtrans-pct}{20\%}

\DefMacro{pcp-split-drop-qwen14da-mean-llmtrans-pp}{6}
\DefMacro{pcp-split-drop-qwen14da-mean-fuzzgen-pp}{13.5}
\DefMacro{pcp-split-drop-qwen14da-mean-fuzzgen-pct}{22.8\%}
\DefMacro{pcp-split-drop-qwen14da-max-fuzzgen-pp}{35}
\DefMacro{pcp-split-drop-qwen14da-max-fuzzgen-pct}{47.6\%}
\DefMacro{pcp-split-drop-qwen14da-mean-llmtrans-pct}{12.5\%}
\DefMacro{pcp-split-drop-qwen14da-max-llmtrans-pp}{9}
\DefMacro{pcp-split-drop-qwen14da-max-llmtrans-pct}{21.2\%}

\DefMacro{pcp-split-drop-overall-max-fuzzgen-pp}{55}
\DefMacro{pcp-split-drop-overall-max-fuzzgen-pct}{70.1\%}
\DefMacro{pcp-split-drop-top3-max-fuzzgen-pp}{48}
\DefMacro{pcp-split-drop-top3-max-fuzzgen-pct}{66.2\%}
\DefMacro{pcp-split-drop-dpsk14-vs-top3-mean-fuzzgen-gap-pp}{3.2}

\DefMacro{pcp-sem-drop-capable-count}{14}
\DefMacro{pcp-sem-drop-all-swap-median-pp}{19}
\DefMacro{pcp-sem-drop-all-swap-mean-pp}{16.8}
\DefMacro{pcp-sem-drop-all-obf-median-pp}{32}
\DefMacro{pcp-sem-drop-all-obf-mean-pp}{30.4}
\DefMacro{pcp-sem-drop-hw-obf-minus-swap-mean-pp}{13.6}

\DefMacro{pcp-sem-drop-capable-mean-min-pp}{15}
\DefMacro{pcp-sem-drop-capable-mean-max-pp}{32}
\DefMacro{pcp-sem-drop-capable-mean-median-pp}{25}
\DefMacro{pcp-sem-drop-capable-mean-pp}{25.1}
\DefMacro{pcp-sem-drop-capable-swap-median-pp}{20}
\DefMacro{pcp-sem-drop-capable-swap-mean-pp}{18.2}
\DefMacro{pcp-sem-drop-capable-obf-median-pp}{34}
\DefMacro{pcp-sem-drop-capable-obf-mean-pp}{32}

\DefMacro{pcp-sem-drop-dpsk32-mean-hw-pp}{15.2}
\DefMacro{pcp-sem-drop-dpsk32-mean-swap-hw-pp}{24.6}
\DefMacro{pcp-sem-drop-dpsk32-mean-obf-hw-pp}{5.7}
\DefMacro{pcp-sem-drop-dpsk32-k-obf-hw-pp}{2}
\DefMacro{pcp-sem-drop-dpsk32-sos-obf-hw-pp}{10}
\DefMacro{pcp-sem-drop-min14cot-mean-hw-pp}{14.8}
\DefMacro{pcp-sem-drop-min14cot-mean-swap-hw-pp}{20.2}
\DefMacro{pcp-sem-drop-min14cot-mean-obf-hw-pp}{9.3}
\DefMacro{pcp-sem-drop-min14cot-k-obf-hw-pp}{9}
\DefMacro{pcp-sem-drop-min14cot-sos-obf-hw-pp}{10}
\DefMacro{pcp-sem-drop-qwen32cot-mean-hw-pp}{23.3}
\DefMacro{pcp-sem-drop-qwen32cot-mean-swap-hw-pp}{18.5}
\DefMacro{pcp-sem-drop-qwen32cot-mean-obf-hw-pp}{28}
\DefMacro{pcp-sem-drop-qwen32cot-k-obf-hw-pp}{27}
\DefMacro{pcp-sem-drop-qwen32cot-sos-obf-hw-pp}{29}
\DefMacro{pcp-sem-drop-top3-mean-hw-min-pp}{14.8}
\DefMacro{pcp-sem-drop-top3-mean-hw-max-pp}{23.3}

\DefMacro{pcp-sem-drop-max-hw-pp}{59}

\DefMacro{pcp-sem-drop-qwen32cot-mean-fuzzgen-pp}{38.3}
\DefMacro{pcp-sem-drop-fuzzgen-swap-mean-pp}{21}
\DefMacro{pcp-sem-drop-fuzzgen-obf-mean-pp}{26.3}
\DefMacro{pcp-sem-drop-fuzzgen-obf-minus-swap-mean-pp}{5.3}

\DefMacro{pcp-sem-drop-qwen14da-mean-hw-pp}{29}
\DefMacro{pcp-sem-drop-qwen14cot-mean-hw-pp}{25.7}
\DefMacro{pcp-sem-drop-qwen14cot-vs-da-mean-hw-gap-pp}{3.3}
\DefMacro{pcp-sem-drop-qwen32da-mean-hw-pp}{27.8}
\DefMacro{pcp-sem-drop-qwen32cot-vs-da-mean-hw-gap-pp}{4.5}

\DefMacro{pcp-qual-seed}{42}
\DefMacro{pcp-qual-failure-mode-count}{4}
\DefMacro{pcp-qual-config-count}{6}
\DefMacro{pcp-qual-top-n-failures}{2}
\DefMacro{pcp-qual-example-count}{2}
\DefMacro{pcp-qual-example-loc}{5}

\DefMacro{pcp-qual-sos-rule-continue}{76}
\DefMacro{pcp-qual-sos-rule-break}{73}
\DefMacro{pcp-qual-sos-rule-modbyzero}{23}
\DefMacro{pcp-qual-sos-rule-modbyzero-neighbor}{24}
\DefMacro{pcp-qual-sos-rule-varusebdec}{2}

\DefMacro{pcp-qual-hw-programs-total}{5{,}832}
\DefMacro{pcp-qual-hw-failures-min}{661}
\DefMacro{pcp-qual-hw-failures-max}{1028}
\DefMacro{pcp-qual-hw-false-success-min}{509}
\DefMacro{pcp-qual-hw-false-success-max}{779}
\DefMacro{pcp-qual-hw-wrong-rule-min}{63}
\DefMacro{pcp-qual-hw-wrong-rule-max}{169}
\DefMacro{pcp-qual-hw-false-error-min}{45}
\DefMacro{pcp-qual-hw-false-error-max}{61}
\DefMacro{pcp-qual-hw-failures-dpsk32}{669}
\DefMacro{pcp-qual-hw-false-success-dpsk32}{549}
\DefMacro{pcp-qual-hw-wrong-rule-dpsk32}{75}
\DefMacro{pcp-qual-hw-false-error-dpsk32}{45}
\DefMacro{pcp-qual-hw-failures-qwen32cot}{1028}
\DefMacro{pcp-qual-hw-false-success-qwen32cot}{779}
\DefMacro{pcp-qual-hw-wrong-rule-qwen32cot}{169}
\DefMacro{pcp-qual-hw-false-error-qwen32cot}{61}
\DefMacro{pcp-qual-hw-failures-min14cot}{661}
\DefMacro{pcp-qual-hw-false-success-min14cot}{509}
\DefMacro{pcp-qual-hw-wrong-rule-min14cot}{63}
\DefMacro{pcp-qual-hw-false-error-min14cot}{45}

\DefMacro{pcp-qual-fuzzgen-programs-total}{5{,}904}
\DefMacro{pcp-qual-fuzzgen-failures-min}{1762}
\DefMacro{pcp-qual-fuzzgen-failures-max}{2991}
\DefMacro{pcp-qual-fuzzgen-false-success-min}{372}
\DefMacro{pcp-qual-fuzzgen-false-success-max}{1484}
\DefMacro{pcp-qual-fuzzgen-wrong-rule-min}{597}
\DefMacro{pcp-qual-fuzzgen-wrong-rule-max}{1003}
\DefMacro{pcp-qual-fuzzgen-false-error-min}{341}
\DefMacro{pcp-qual-fuzzgen-false-error-max}{453}
\DefMacro{pcp-qual-fuzzgen-failures-dpsk32}{1762}
\DefMacro{pcp-qual-fuzzgen-false-success-dpsk32}{372}
\DefMacro{pcp-qual-fuzzgen-wrong-rule-dpsk32}{833}
\DefMacro{pcp-qual-fuzzgen-false-error-dpsk32}{453}
\DefMacro{pcp-qual-fuzzgen-failures-qwen32cot}{2991}
\DefMacro{pcp-qual-fuzzgen-false-success-qwen32cot}{1484}
\DefMacro{pcp-qual-fuzzgen-wrong-rule-qwen32cot}{1003}
\DefMacro{pcp-qual-fuzzgen-false-error-qwen32cot}{369}
\DefMacro{pcp-qual-fuzzgen-failures-min14cot}{2803}
\DefMacro{pcp-qual-fuzzgen-false-success-min14cot}{432}
\DefMacro{pcp-qual-fuzzgen-wrong-rule-min14cot}{597}
\DefMacro{pcp-qual-fuzzgen-false-error-min14cot}{341}

\DefMacro{pcp-qual-llmtrans-programs-total}{5{,}940}
\DefMacro{pcp-qual-llmtrans-failures-min}{822}
\DefMacro{pcp-qual-llmtrans-failures-max}{1749}
\DefMacro{pcp-qual-llmtrans-false-success-min}{543}
\DefMacro{pcp-qual-llmtrans-false-success-max}{1251}
\DefMacro{pcp-qual-llmtrans-wrong-rule-min}{64}
\DefMacro{pcp-qual-llmtrans-wrong-rule-max}{211}
\DefMacro{pcp-qual-llmtrans-false-error-min}{26}
\DefMacro{pcp-qual-llmtrans-false-error-max}{59}
\DefMacro{pcp-qual-llmtrans-failures-dpsk32}{822}
\DefMacro{pcp-qual-llmtrans-false-success-dpsk32}{695}
\DefMacro{pcp-qual-llmtrans-wrong-rule-dpsk32}{88}
\DefMacro{pcp-qual-llmtrans-false-error-dpsk32}{38}
\DefMacro{pcp-qual-llmtrans-failures-qwen32cot}{1543}
\DefMacro{pcp-qual-llmtrans-false-success-qwen32cot}{1251}
\DefMacro{pcp-qual-llmtrans-wrong-rule-qwen32cot}{211}
\DefMacro{pcp-qual-llmtrans-false-error-qwen32cot}{59}
\DefMacro{pcp-qual-llmtrans-failures-min14cot}{1749}
\DefMacro{pcp-qual-llmtrans-false-success-min14cot}{543}
\DefMacro{pcp-qual-llmtrans-wrong-rule-min14cot}{64}
\DefMacro{pcp-qual-llmtrans-false-error-min14cot}{26}

\DefMacro{exectime-n}{100}
\DefMacro{exectime-k-startup}{2.88}
\DefMacro{exectime-k-e2e-median}{3.1}
\DefMacro{exectime-k-e2e-mean}{3.2}
\DefMacro{exectime-k-e2e-max}{4.5}
\DefMacro{exectime-k-pure-median}{0.23}
\DefMacro{exectime-k-pure-mean}{0.29}
\DefMacro{exectime-k-pure-p90}{0.55}
\DefMacro{exectime-k-pure-max}{1.59}
\DefMacro{exectime-llm-median}{0.80}
\DefMacro{exectime-llm-mean}{2.6}
\DefMacro{exectime-llm-p90}{4.2}
\DefMacro{exectime-llm-max}{32.2}
\DefMacro{exectime-llm-min}{0.49}
\DefMacro{exectime-llm-out-median}{17}
\DefMacro{exectime-llm-out-max}{916}
\DefMacro{exectime-llm-prompt-median}{3800}
\DefMacro{exectime-llm-prompt-max}{24882}
\DefMacro{exectime-llm-toks-per-sec}{20.3}
\DefMacro{exectime-llm-slower-pure}{97}
\DefMacro{exectime-llm-faster-e2e}{81}
\DefMacro{exectime-r-tok}{0.95}
\DefMacro{exectime-r-size-llm}{0.48}
\DefMacro{exectime-speedup-pure-median}{7.1}

\DefMacro{TCap-imp-dataset-dataset-stats}{IMP Dataset statistics. All programs are executable by K-framework. Len(tr) is the length of the logged execution trace. We report the \# semantic rules per statement.}
\DefMacro{THead-dataset}{dataset}
\DefMacro{THead-num-programs}{\# programs}
\DefMacro{THead-avg-loc}{avg. loc}
\DefMacro{THead-avg-tokens}{avg. \# tokens}
\DefMacro{THead-avg-trace-length}{avg. len(tr)}
\DefMacro{THead-max-trace-length}{max. len(tr)}
\DefMacro{THead-min-trace-length}{min. len(tr)}
\DefMacro{THead-avg-rules-per-state}{avg. \# rules}
\DefMacro{THead-max-rules-per-state}{max. \# rules}
\DefMacro{THead-min-rules-per-state}{min. \# rules}
\DefMacro{THead-imp-dataset}{IMP}

\DefMacro{TCap-models-pcp-IMP-SOS}{Results of the models.}
\DefMacro{THead-pcp-IMP-SOS-models}{Models}
\DefMacro{THead-pcp-IMP-SOS-pcp-uk-IMP-SOS-accuracy}{\uk}
\DefMacro{THead-pcp-IMP-SOS-pcp-mk-IMP-SOS-addSub_mulDiv_negateRelation-accuracy}{\ksMk}
\DefMacro{THead-pcp-IMP-SOS-pcp-mk-IMP-SOS-unseen-accuracy}{\koMk}
\DefMacro{THead-pcp-IMP-SOS-Qwen-Qwen2.5-Coder-14B-Instruct-da}{\qwenCoder{14}}
\DefMacro{THead-pcp-IMP-SOS-Qwen-Qwen2.5-Coder-14B-Instruct-cot}{\qwenCoderCoT{14}}
\DefMacro{THead-pcp-IMP-SOS-Qwen-Qwen2.5-Coder-32B-Instruct-da}{\qwenCoder{32}}
\DefMacro{THead-pcp-IMP-SOS-Qwen-Qwen2.5-Coder-32B-Instruct-cot}{\qwenCoderCoT{32}}
\DefMacro{THead-pcp-IMP-SOS-meta-llama-Llama-3.3-70B-Instruct-da}{\llamaBig}
\DefMacro{THead-pcp-IMP-SOS-meta-llama-Llama-3.3-70B-Instruct-cot}{\llamaBig-\COT}
\DefMacro{THead-pcp-IMP-SOS-gpt-4o-mini-da}{\gptfo}
\DefMacro{THead-pcp-IMP-SOS-gpt-4o-mini-cot}{\gptfo-\COT}
\DefMacro{THead-pcp-IMP-SOS-deepseek-ai-DeepSeek-R1-Distill-Qwen-14B-da}{\dpskQwen{14}}
\DefMacro{THead-pcp-IMP-SOS-deepseek-ai-DeepSeek-R1-Distill-Qwen-32B-da}{\dpskQwen{32}}
\DefMacro{THead-pcp-IMP-SOS-deepseek-ai-DeepSeek-R1-Distill-Llama-70B-da}{\dpskLlama{70}}
\DefMacro{THead-pcp-IMP-SOS-Qwen-QwQ-32B-da}{\qwq}
\DefMacro{THead-pcp-IMP-SOS-o3-mini-da}{\othree}
\DefMacro{THead-pcp-IMP-SOS-gemini-2.5-pro-da}{\gemini}

\DefMacro{TCap-mk-IMP-SOS-dataset-stats}{Mutated IMP Dataset statistics.}
\DefMacro{THead-mk-IMP-SOS-dataset}{Mutated IMP dataset}
\DefMacro{THead-mk-IMP-SOS-mutation-rate}{\% Mutated Program}
\DefMacro{THead-mk-IMP-SOS-num-mutated-programs}{\# Mutated Program}
\DefMacro{THead-addSub}{addSub}
\DefMacro{THead-mulDiv}{mulDiv}
\DefMacro{THead-negateRelation}{negRel}
\DefMacro{THead-addSub-mulDiv}{addSub+mulDiv}
\DefMacro{THead-addSub-negateRelation}{addSub+negRel}
\DefMacro{THead-mulDiv-negateRelation}{mulDiv+negRel}
\DefMacro{THead-addSub-mulDiv-negateRelation}{addSub+mulDiv+negRel}
\DefMacro{THead-mutation-rate}{mutation-rate}
\DefMacro{THead-num-mutated-programs}{\# mutated programs}

\DefMacro{title1}{\textbf{Rule}}
\DefMacro{title2}{\textbf{Operator}}
\DefMacro{title3}{\textbf{Operator}}
\DefMacro{addSubRule}{addSub}
\DefMacro{mulDivRule}{mulDiv}
\DefMacro{negateRule}{negRel}
\DefMacro{addSubRuleOp1}{\texttt{+}}
\DefMacro{addSubRuleOp2}{\texttt{-}}
\DefMacro{mulDivRuleOp1}{\texttt{*}}
\DefMacro{mulDivRuleOp2}{\texttt{/}}
\DefMacro{negateRuleOp1}{\texttt{<}}
\DefMacro{negateRuleOp2}{\texttt{>}}
\DefMacro{negateRuleOp3}{\texttt{<=}}
\DefMacro{negateRuleOp4}{\texttt{>=}}
\DefMacro{negateRuleOp5}{\texttt{!=}}
\DefMacro{negateRuleOp6}{\texttt{==}}
\DefMacro{negateRuleOp7}{\texttt{\&\&}}
\DefMacro{negateRuleOp8}{\texttt{||}}

\DefMacro{Caucasiantitle1}{\textbf{Type}}
\DefMacro{Caucasiantitle2}{\textbf{\standardSemCap}}
\DefMacro{Caucasiantitle3}{\textbf{\keywordObf}}
\DefMacro{Caucasiantitle4}{\textbf{\keywordMut}}
\DefMacro{CaucasianArithmetic}{Arithmetic}
\DefMacro{CaucasianAssignment}{Assignment}
\DefMacro{CaucasianRelational}{Relational}
\DefMacro{CaucasianLogical}{Logical}
\DefMacro{CaucasianKeyword}{Keyword}

\DefMacro{THead-pcp-op-IMP-SOS-models}{Models}
\DefMacro{THead-pcp-op-IMP-SOS-pcp-uk-IMP-SOS-accuracy}{\uk}
\DefMacro{THead-pcp-op-IMP-SOS-pcp-mk-IMP-SOS-addSub_mulDiv_negateRelation-accuracy}{\ksMk}
\DefMacro{THead-pcp-op-IMP-SOS-pcp-mk-IMP-SOS-unseen-accuracy}{\koMk}
\DefMacro{THead-pcp-op-IMP-SOS-op-nk-IMP-SOS-acc}{\nk}
\DefMacro{THead-pcp-op-IMP-SOS-op-uk-IMP-SOS-acc}{\uk}
\DefMacro{THead-pcp-op-IMP-SOS-op-mk-IMP-SOS-addSub_mulDiv_negateRelation-acc}{\ksMk}
\DefMacro{THead-pcp-op-IMP-SOS-op-mk-IMP-SOS-unseen-acc}{\koMk}
\DefMacro{THead-pcp-op-IMP-SOS-meta-llama-Llama-3.3-70B-Instruct-da}{\llamaBig}
\DefMacro{THead-pcp-op-IMP-SOS-meta-llama-Llama-3.3-70B-Instruct-cot}{\llamaBig-\COT}
\DefMacro{THead-pcp-op-IMP-SOS-Qwen-Qwen2.5-Coder-14B-Instruct-da}{\qwenCoder{14}}
\DefMacro{THead-pcp-op-IMP-SOS-Qwen-Qwen2.5-Coder-14B-Instruct-cot}{\qwenCoderCoT{14}}
\DefMacro{THead-pcp-op-IMP-SOS-Qwen-Qwen2.5-Coder-32B-Instruct-da}{\qwenCoder{32}}
\DefMacro{THead-pcp-op-IMP-SOS-Qwen-Qwen2.5-Coder-32B-Instruct-cot}{\qwenCoderCoT{32}}
\DefMacro{THead-pcp-op-IMP-SOS-gpt-4o-mini-da}{\gptfo}
\DefMacro{THead-pcp-op-IMP-SOS-gpt-4o-mini-cot}{\gptfo-\COT}
\DefMacro{THead-pcp-op-IMP-SOS-deepseek-ai-DeepSeek-R1-Distill-Qwen-14B-da}{\dpskQwen{14}}
\DefMacro{THead-pcp-op-IMP-SOS-deepseek-ai-DeepSeek-R1-Distill-Qwen-32B-da}{\dpskQwen{32}}
\DefMacro{THead-pcp-op-IMP-SOS-deepseek-ai-DeepSeek-R1-Distill-Llama-70B-da}{\dpskLlama{70}}
\DefMacro{THead-pcp-op-IMP-SOS-gemini-2.5-pro-da}{\gemini}
\DefMacro{THead-pcp-op-IMP-SOS-o3-mini-da}{\othree}
\DefMacro{THead-pcp-op-IMP-SOS-Qwen-QwQ-32B-da}{\qwq}

\DefMacro{TCap-models-pcp-IMP-K-IMP-SOS-accuracy-qwen-coder}{Results of the models.}
\DefMacro{THead-pcp-IMP-K-IMP-SOS-accuracy-qwen-coder-models}{Models}
\DefMacro{THead-pcp-IMP-K-IMP-SOS-accuracy-qwen-coder-pcp-uk-IMP-K-accuracy}{\uk}
\DefMacro{THead-pcp-IMP-K-IMP-SOS-accuracy-qwen-coder-pcp-mk-IMP-K-KeywordSwap-accuracy}{\KeywordSwap}
\DefMacro{THead-pcp-IMP-K-IMP-SOS-accuracy-qwen-coder-pcp-mk-IMP-K-KeywordObf-accuracy}{\KeywordObf}
\DefMacro{THead-pcp-IMP-K-IMP-SOS-accuracy-qwen-coder-pcp-uk-IMP-SOS-accuracy}{\uk}
\DefMacro{THead-pcp-IMP-K-IMP-SOS-accuracy-qwen-coder-pcp-mk-IMP-SOS-KeywordSwap-accuracy}{\KeywordSwap}
\DefMacro{THead-pcp-IMP-K-IMP-SOS-accuracy-qwen-coder-pcp-mk-IMP-SOS-KeywordObf-accuracy}{\KeywordObf}
\DefMacro{THead-pcp-IMP-K-IMP-SOS-accuracy-qwen-coder-Qwen-Qwen2.5-Coder-3B-Instruct-da}{\qwenCoder{3}}
\DefMacro{THead-pcp-IMP-K-IMP-SOS-accuracy-qwen-coder-Qwen-Qwen2.5-Coder-3B-Instruct-cot}{\qwenCoderCoT{3}}
\DefMacro{THead-pcp-IMP-K-IMP-SOS-accuracy-qwen-coder-Qwen-Qwen2.5-Coder-7B-Instruct-da}{\qwenCoder{7}}
\DefMacro{THead-pcp-IMP-K-IMP-SOS-accuracy-qwen-coder-Qwen-Qwen2.5-Coder-7B-Instruct-cot}{\qwenCoderCoT{7}}
\DefMacro{THead-pcp-IMP-K-IMP-SOS-accuracy-qwen-coder-Qwen-Qwen2.5-Coder-14B-Instruct-da}{\qwenCoder{14}}
\DefMacro{THead-pcp-IMP-K-IMP-SOS-accuracy-qwen-coder-Qwen-Qwen2.5-Coder-14B-Instruct-cot}{\qwenCoderCoT{14}}
\DefMacro{THead-pcp-IMP-K-IMP-SOS-accuracy-qwen-coder-Qwen-Qwen2.5-Coder-32B-Instruct-da}{\qwenCoder{32}}
\DefMacro{THead-pcp-IMP-K-IMP-SOS-accuracy-qwen-coder-Qwen-Qwen2.5-Coder-32B-Instruct-cot}{\qwenCoderCoT{32}}
\DefMacro{THead-pcp-IMP-K-IMP-SOS-accuracy-qwen-coder-random-guesser-dataset-distribution-da}{Random}

\DefMacro{TCap-models-pcp-IMP-K-IMP-SOS-accuracy-qwen-coder-by-split-human-written}{\pcp accuracy on \humanwrit programs.}
\DefMacro{THead-pcp-IMP-K-IMP-SOS-accuracy-qwen-coder-by-split-human-written-models}{Models}
\DefMacro{THead-pcp-IMP-K-IMP-SOS-accuracy-qwen-coder-by-split-human-written-pcp-uk-IMP-K-accuracy}{\standardSemCap}
\DefMacro{THead-pcp-IMP-K-IMP-SOS-accuracy-qwen-coder-by-split-human-written-pcp-mk-IMP-K-KeywordSwap-accuracy}{\KeywordSwap}
\DefMacro{THead-pcp-IMP-K-IMP-SOS-accuracy-qwen-coder-by-split-human-written-pcp-mk-IMP-K-KeywordObf-accuracy}{\KeywordObf}
\DefMacro{THead-pcp-IMP-K-IMP-SOS-accuracy-qwen-coder-by-split-human-written-pcp-uk-IMP-SOS-accuracy}{\standardSemCap}
\DefMacro{THead-pcp-IMP-K-IMP-SOS-accuracy-qwen-coder-by-split-human-written-pcp-mk-IMP-SOS-KeywordSwap-accuracy}{\KeywordSwap}
\DefMacro{THead-pcp-IMP-K-IMP-SOS-accuracy-qwen-coder-by-split-human-written-pcp-mk-IMP-SOS-KeywordObf-accuracy}{\KeywordObf}
\DefMacro{THead-pcp-IMP-K-IMP-SOS-accuracy-qwen-coder-by-split-human-written-random-guesser-dataset-distribution-da}{Random}
\DefMacro{THead-pcp-IMP-K-IMP-SOS-accuracy-qwen-coder-by-split-human-written-mistralai-Ministral-3-3B-Instruct-2512-BF16-da}{\ministral{3}}
\DefMacro{THead-pcp-IMP-K-IMP-SOS-accuracy-qwen-coder-by-split-human-written-mistralai-Ministral-3-8B-Instruct-2512-BF16-da}{\ministral{8}}
\DefMacro{THead-pcp-IMP-K-IMP-SOS-accuracy-qwen-coder-by-split-human-written-mistralai-Ministral-3-14B-Instruct-2512-BF16-da}{\ministral{14}}
\DefMacro{THead-pcp-IMP-K-IMP-SOS-accuracy-qwen-coder-by-split-human-written-Qwen-Qwen2.5-Coder-3B-Instruct-da}{\qwenCoder{3}}
\DefMacro{THead-pcp-IMP-K-IMP-SOS-accuracy-qwen-coder-by-split-human-written-Qwen-Qwen2.5-Coder-7B-Instruct-da}{\qwenCoder{7}}
\DefMacro{THead-pcp-IMP-K-IMP-SOS-accuracy-qwen-coder-by-split-human-written-Qwen-Qwen2.5-Coder-14B-Instruct-da}{\qwenCoder{14}}
\DefMacro{THead-pcp-IMP-K-IMP-SOS-accuracy-qwen-coder-by-split-human-written-Qwen-Qwen2.5-Coder-32B-Instruct-da}{\qwenCoder{32}}
\DefMacro{THead-pcp-IMP-K-IMP-SOS-accuracy-qwen-coder-by-split-human-written-deepseek-ai-DeepSeek-R1-Distill-Qwen-14B-da}{\dpskQwen{14}}
\DefMacro{THead-pcp-IMP-K-IMP-SOS-accuracy-qwen-coder-by-split-human-written-deepseek-ai-DeepSeek-R1-Distill-Qwen-32B-da}{\dpskQwen{32}}
\DefMacro{THead-pcp-IMP-K-IMP-SOS-accuracy-qwen-coder-by-split-human-written-mistralai-Ministral-3-3B-Instruct-2512-BF16-cot}{\ministralCoT{3}}
\DefMacro{THead-pcp-IMP-K-IMP-SOS-accuracy-qwen-coder-by-split-human-written-mistralai-Ministral-3-8B-Instruct-2512-BF16-cot}{\ministralCoT{8}}
\DefMacro{THead-pcp-IMP-K-IMP-SOS-accuracy-qwen-coder-by-split-human-written-mistralai-Ministral-3-14B-Instruct-2512-BF16-cot}{\ministralCoT{14}}
\DefMacro{THead-pcp-IMP-K-IMP-SOS-accuracy-qwen-coder-by-split-human-written-Qwen-Qwen2.5-Coder-3B-Instruct-cot}{\qwenCoderCoT{3}}
\DefMacro{THead-pcp-IMP-K-IMP-SOS-accuracy-qwen-coder-by-split-human-written-Qwen-Qwen2.5-Coder-7B-Instruct-cot}{\qwenCoderCoT{7}}
\DefMacro{THead-pcp-IMP-K-IMP-SOS-accuracy-qwen-coder-by-split-human-written-Qwen-Qwen2.5-Coder-14B-Instruct-cot}{\qwenCoderCoT{14}}
\DefMacro{THead-pcp-IMP-K-IMP-SOS-accuracy-qwen-coder-by-split-human-written-Qwen-Qwen2.5-Coder-32B-Instruct-cot}{\qwenCoderCoT{32}}
\DefMacro{res-random-guesser-dataset-distribution-da-pcp-uk-IMP-K-accuracy-human-written}{18}
\DefMacro{res-random-guesser-dataset-distribution-da-pcp-mk-IMP-K-KeywordSwap-accuracy-human-written}{18}
\DefMacro{res-random-guesser-dataset-distribution-da-pcp-mk-IMP-K-KeywordObf-accuracy-human-written}{16}
\DefMacro{res-random-guesser-dataset-distribution-da-pcp-uk-IMP-SOS-accuracy-human-written}{18}
\DefMacro{res-random-guesser-dataset-distribution-da-pcp-mk-IMP-SOS-KeywordSwap-accuracy-human-written}{18}
\DefMacro{res-random-guesser-dataset-distribution-da-pcp-mk-IMP-SOS-KeywordObf-accuracy-human-written}{16}
\DefMacro{res-mistralai-Ministral-3-3B-Instruct-2512-BF16-da-pcp-uk-IMP-K-accuracy-human-written}{61}
\DefMacro{res-mistralai-Ministral-3-3B-Instruct-2512-BF16-da-pcp-mk-IMP-K-KeywordSwap-accuracy-human-written}{47}
\DefMacro{res-mistralai-Ministral-3-3B-Instruct-2512-BF16-da-pcp-mk-IMP-K-KeywordObf-accuracy-human-written}{14}
\DefMacro{res-mistralai-Ministral-3-3B-Instruct-2512-BF16-da-pcp-uk-IMP-SOS-accuracy-human-written}{46}
\DefMacro{res-mistralai-Ministral-3-3B-Instruct-2512-BF16-da-pcp-mk-IMP-SOS-KeywordSwap-accuracy-human-written}{41}
\DefMacro{res-mistralai-Ministral-3-3B-Instruct-2512-BF16-da-pcp-mk-IMP-SOS-KeywordObf-accuracy-human-written}{21}
\DefMacro{res-mistralai-Ministral-3-8B-Instruct-2512-BF16-da-pcp-uk-IMP-K-accuracy-human-written}{77}
\DefMacro{res-mistralai-Ministral-3-8B-Instruct-2512-BF16-da-pcp-mk-IMP-K-KeywordSwap-accuracy-human-written}{66}
\DefMacro{res-mistralai-Ministral-3-8B-Instruct-2512-BF16-da-pcp-mk-IMP-K-KeywordObf-accuracy-human-written}{24}
\DefMacro{res-mistralai-Ministral-3-8B-Instruct-2512-BF16-da-pcp-uk-IMP-SOS-accuracy-human-written}{69}
\DefMacro{res-mistralai-Ministral-3-8B-Instruct-2512-BF16-da-pcp-mk-IMP-SOS-KeywordSwap-accuracy-human-written}{55}
\DefMacro{res-mistralai-Ministral-3-8B-Instruct-2512-BF16-da-pcp-mk-IMP-SOS-KeywordObf-accuracy-human-written}{20}
\DefMacro{res-mistralai-Ministral-3-14B-Instruct-2512-BF16-da-pcp-uk-IMP-K-accuracy-human-written}{90}
\DefMacro{res-mistralai-Ministral-3-14B-Instruct-2512-BF16-da-pcp-mk-IMP-K-KeywordSwap-accuracy-human-written}{75}
\DefMacro{res-mistralai-Ministral-3-14B-Instruct-2512-BF16-da-pcp-mk-IMP-K-KeywordObf-accuracy-human-written}{32}
\DefMacro{res-mistralai-Ministral-3-14B-Instruct-2512-BF16-da-pcp-uk-IMP-SOS-accuracy-human-written}{87}
\DefMacro{res-mistralai-Ministral-3-14B-Instruct-2512-BF16-da-pcp-mk-IMP-SOS-KeywordSwap-accuracy-human-written}{78}
\DefMacro{res-mistralai-Ministral-3-14B-Instruct-2512-BF16-da-pcp-mk-IMP-SOS-KeywordObf-accuracy-human-written}{45}
\DefMacro{res-Qwen-Qwen2.5-Coder-3B-Instruct-da-pcp-uk-IMP-K-accuracy-human-written}{44}
\DefMacro{res-Qwen-Qwen2.5-Coder-3B-Instruct-da-pcp-mk-IMP-K-KeywordSwap-accuracy-human-written}{23}
\DefMacro{res-Qwen-Qwen2.5-Coder-3B-Instruct-da-pcp-mk-IMP-K-KeywordObf-accuracy-human-written}{7}
\DefMacro{res-Qwen-Qwen2.5-Coder-3B-Instruct-da-pcp-uk-IMP-SOS-accuracy-human-written}{23}
\DefMacro{res-Qwen-Qwen2.5-Coder-3B-Instruct-da-pcp-mk-IMP-SOS-KeywordSwap-accuracy-human-written}{28}
\DefMacro{res-Qwen-Qwen2.5-Coder-3B-Instruct-da-pcp-mk-IMP-SOS-KeywordObf-accuracy-human-written}{18}
\DefMacro{res-Qwen-Qwen2.5-Coder-7B-Instruct-da-pcp-uk-IMP-K-accuracy-human-written}{45}
\DefMacro{res-Qwen-Qwen2.5-Coder-7B-Instruct-da-pcp-mk-IMP-K-KeywordSwap-accuracy-human-written}{22}
\DefMacro{res-Qwen-Qwen2.5-Coder-7B-Instruct-da-pcp-mk-IMP-K-KeywordObf-accuracy-human-written}{6}
\DefMacro{res-Qwen-Qwen2.5-Coder-7B-Instruct-da-pcp-uk-IMP-SOS-accuracy-human-written}{55}
\DefMacro{res-Qwen-Qwen2.5-Coder-7B-Instruct-da-pcp-mk-IMP-SOS-KeywordSwap-accuracy-human-written}{35}
\DefMacro{res-Qwen-Qwen2.5-Coder-7B-Instruct-da-pcp-mk-IMP-SOS-KeywordObf-accuracy-human-written}{21}
\DefMacro{res-Qwen-Qwen2.5-Coder-14B-Instruct-da-pcp-uk-IMP-K-accuracy-human-written}{64}
\DefMacro{res-Qwen-Qwen2.5-Coder-14B-Instruct-da-pcp-mk-IMP-K-KeywordSwap-accuracy-human-written}{41}
\DefMacro{res-Qwen-Qwen2.5-Coder-14B-Instruct-da-pcp-mk-IMP-K-KeywordObf-accuracy-human-written}{26}
\DefMacro{res-Qwen-Qwen2.5-Coder-14B-Instruct-da-pcp-uk-IMP-SOS-accuracy-human-written}{83}
\DefMacro{res-Qwen-Qwen2.5-Coder-14B-Instruct-da-pcp-mk-IMP-SOS-KeywordSwap-accuracy-human-written}{74}
\DefMacro{res-Qwen-Qwen2.5-Coder-14B-Instruct-da-pcp-mk-IMP-SOS-KeywordObf-accuracy-human-written}{37}
\DefMacro{res-Qwen-Qwen2.5-Coder-32B-Instruct-da-pcp-uk-IMP-K-accuracy-human-written}{70}
\DefMacro{res-Qwen-Qwen2.5-Coder-32B-Instruct-da-pcp-mk-IMP-K-KeywordSwap-accuracy-human-written}{57}
\DefMacro{res-Qwen-Qwen2.5-Coder-32B-Instruct-da-pcp-mk-IMP-K-KeywordObf-accuracy-human-written}{39}
\DefMacro{res-Qwen-Qwen2.5-Coder-32B-Instruct-da-pcp-uk-IMP-SOS-accuracy-human-written}{92}
\DefMacro{res-Qwen-Qwen2.5-Coder-32B-Instruct-da-pcp-mk-IMP-SOS-KeywordSwap-accuracy-human-written}{71}
\DefMacro{res-Qwen-Qwen2.5-Coder-32B-Instruct-da-pcp-mk-IMP-SOS-KeywordObf-accuracy-human-written}{45}
\DefMacro{res-deepseek-ai-DeepSeek-R1-Distill-Qwen-14B-da-pcp-uk-IMP-K-accuracy-human-written}{91}
\DefMacro{res-deepseek-ai-DeepSeek-R1-Distill-Qwen-14B-da-pcp-mk-IMP-K-KeywordSwap-accuracy-human-written}{69}
\DefMacro{res-deepseek-ai-DeepSeek-R1-Distill-Qwen-14B-da-pcp-mk-IMP-K-KeywordObf-accuracy-human-written}{81}
\DefMacro{res-deepseek-ai-DeepSeek-R1-Distill-Qwen-14B-da-pcp-uk-IMP-SOS-accuracy-human-written}{92}
\DefMacro{res-deepseek-ai-DeepSeek-R1-Distill-Qwen-14B-da-pcp-mk-IMP-SOS-KeywordSwap-accuracy-human-written}{71}
\DefMacro{res-deepseek-ai-DeepSeek-R1-Distill-Qwen-14B-da-pcp-mk-IMP-SOS-KeywordObf-accuracy-human-written}{61}
\DefMacro{res-deepseek-ai-DeepSeek-R1-Distill-Qwen-32B-da-pcp-uk-IMP-K-accuracy-human-written}{99}
\DefMacro{res-deepseek-ai-DeepSeek-R1-Distill-Qwen-32B-da-pcp-mk-IMP-K-KeywordSwap-accuracy-human-written}{77}
\DefMacro{res-deepseek-ai-DeepSeek-R1-Distill-Qwen-32B-da-pcp-mk-IMP-K-KeywordObf-accuracy-human-written}{98}
\DefMacro{res-deepseek-ai-DeepSeek-R1-Distill-Qwen-32B-da-pcp-uk-IMP-SOS-accuracy-human-written}{98}
\DefMacro{res-deepseek-ai-DeepSeek-R1-Distill-Qwen-32B-da-pcp-mk-IMP-SOS-KeywordSwap-accuracy-human-written}{71}
\DefMacro{res-deepseek-ai-DeepSeek-R1-Distill-Qwen-32B-da-pcp-mk-IMP-SOS-KeywordObf-accuracy-human-written}{88}
\DefMacro{res-mistralai-Ministral-3-3B-Instruct-2512-BF16-cot-pcp-uk-IMP-K-accuracy-human-written}{94}
\DefMacro{res-mistralai-Ministral-3-3B-Instruct-2512-BF16-cot-pcp-mk-IMP-K-KeywordSwap-accuracy-human-written}{76}
\DefMacro{res-mistralai-Ministral-3-3B-Instruct-2512-BF16-cot-pcp-mk-IMP-K-KeywordObf-accuracy-human-written}{54}
\DefMacro{res-mistralai-Ministral-3-3B-Instruct-2512-BF16-cot-pcp-uk-IMP-SOS-accuracy-human-written}{89}
\DefMacro{res-mistralai-Ministral-3-3B-Instruct-2512-BF16-cot-pcp-mk-IMP-SOS-KeywordSwap-accuracy-human-written}{75}
\DefMacro{res-mistralai-Ministral-3-3B-Instruct-2512-BF16-cot-pcp-mk-IMP-SOS-KeywordObf-accuracy-human-written}{64}
\DefMacro{res-mistralai-Ministral-3-8B-Instruct-2512-BF16-cot-pcp-uk-IMP-K-accuracy-human-written}{99}
\DefMacro{res-mistralai-Ministral-3-8B-Instruct-2512-BF16-cot-pcp-mk-IMP-K-KeywordSwap-accuracy-human-written}{79}
\DefMacro{res-mistralai-Ministral-3-8B-Instruct-2512-BF16-cot-pcp-mk-IMP-K-KeywordObf-accuracy-human-written}{77}
\DefMacro{res-mistralai-Ministral-3-8B-Instruct-2512-BF16-cot-pcp-uk-IMP-SOS-accuracy-human-written}{97}
\DefMacro{res-mistralai-Ministral-3-8B-Instruct-2512-BF16-cot-pcp-mk-IMP-SOS-KeywordSwap-accuracy-human-written}{78}
\DefMacro{res-mistralai-Ministral-3-8B-Instruct-2512-BF16-cot-pcp-mk-IMP-SOS-KeywordObf-accuracy-human-written}{64}
\DefMacro{res-mistralai-Ministral-3-14B-Instruct-2512-BF16-cot-pcp-uk-IMP-K-accuracy-human-written}{99}
\DefMacro{res-mistralai-Ministral-3-14B-Instruct-2512-BF16-cot-pcp-mk-IMP-K-KeywordSwap-accuracy-human-written}{79}
\DefMacro{res-mistralai-Ministral-3-14B-Instruct-2512-BF16-cot-pcp-mk-IMP-K-KeywordObf-accuracy-human-written}{91}
\DefMacro{res-mistralai-Ministral-3-14B-Instruct-2512-BF16-cot-pcp-uk-IMP-SOS-accuracy-human-written}{99}
\DefMacro{res-mistralai-Ministral-3-14B-Instruct-2512-BF16-cot-pcp-mk-IMP-SOS-KeywordSwap-accuracy-human-written}{78}
\DefMacro{res-mistralai-Ministral-3-14B-Instruct-2512-BF16-cot-pcp-mk-IMP-SOS-KeywordObf-accuracy-human-written}{89}
\DefMacro{res-Qwen-Qwen2.5-Coder-3B-Instruct-cot-pcp-uk-IMP-K-accuracy-human-written}{32}
\DefMacro{res-Qwen-Qwen2.5-Coder-3B-Instruct-cot-pcp-mk-IMP-K-KeywordSwap-accuracy-human-written}{25}
\DefMacro{res-Qwen-Qwen2.5-Coder-3B-Instruct-cot-pcp-mk-IMP-K-KeywordObf-accuracy-human-written}{16}
\DefMacro{res-Qwen-Qwen2.5-Coder-3B-Instruct-cot-pcp-uk-IMP-SOS-accuracy-human-written}{33}
\DefMacro{res-Qwen-Qwen2.5-Coder-3B-Instruct-cot-pcp-mk-IMP-SOS-KeywordSwap-accuracy-human-written}{27}
\DefMacro{res-Qwen-Qwen2.5-Coder-3B-Instruct-cot-pcp-mk-IMP-SOS-KeywordObf-accuracy-human-written}{17}
\DefMacro{res-Qwen-Qwen2.5-Coder-7B-Instruct-cot-pcp-uk-IMP-K-accuracy-human-written}{69}
\DefMacro{res-Qwen-Qwen2.5-Coder-7B-Instruct-cot-pcp-mk-IMP-K-KeywordSwap-accuracy-human-written}{49}
\DefMacro{res-Qwen-Qwen2.5-Coder-7B-Instruct-cot-pcp-mk-IMP-K-KeywordObf-accuracy-human-written}{32}
\DefMacro{res-Qwen-Qwen2.5-Coder-7B-Instruct-cot-pcp-uk-IMP-SOS-accuracy-human-written}{71}
\DefMacro{res-Qwen-Qwen2.5-Coder-7B-Instruct-cot-pcp-mk-IMP-SOS-KeywordSwap-accuracy-human-written}{46}
\DefMacro{res-Qwen-Qwen2.5-Coder-7B-Instruct-cot-pcp-mk-IMP-SOS-KeywordObf-accuracy-human-written}{26}
\DefMacro{res-Qwen-Qwen2.5-Coder-14B-Instruct-cot-pcp-uk-IMP-K-accuracy-human-written}{96}
\DefMacro{res-Qwen-Qwen2.5-Coder-14B-Instruct-cot-pcp-mk-IMP-K-KeywordSwap-accuracy-human-written}{74}
\DefMacro{res-Qwen-Qwen2.5-Coder-14B-Instruct-cot-pcp-mk-IMP-K-KeywordObf-accuracy-human-written}{78}
\DefMacro{res-Qwen-Qwen2.5-Coder-14B-Instruct-cot-pcp-uk-IMP-SOS-accuracy-human-written}{93}
\DefMacro{res-Qwen-Qwen2.5-Coder-14B-Instruct-cot-pcp-mk-IMP-SOS-KeywordSwap-accuracy-human-written}{71}
\DefMacro{res-Qwen-Qwen2.5-Coder-14B-Instruct-cot-pcp-mk-IMP-SOS-KeywordObf-accuracy-human-written}{55}
\DefMacro{res-Qwen-Qwen2.5-Coder-32B-Instruct-cot-pcp-uk-IMP-K-accuracy-human-written}{99}
\DefMacro{res-Qwen-Qwen2.5-Coder-32B-Instruct-cot-pcp-mk-IMP-K-KeywordSwap-accuracy-human-written}{81}
\DefMacro{res-Qwen-Qwen2.5-Coder-32B-Instruct-cot-pcp-mk-IMP-K-KeywordObf-accuracy-human-written}{72}
\DefMacro{res-Qwen-Qwen2.5-Coder-32B-Instruct-cot-pcp-uk-IMP-SOS-accuracy-human-written}{97}
\DefMacro{res-Qwen-Qwen2.5-Coder-32B-Instruct-cot-pcp-mk-IMP-SOS-KeywordSwap-accuracy-human-written}{78}
\DefMacro{res-Qwen-Qwen2.5-Coder-32B-Instruct-cot-pcp-mk-IMP-SOS-KeywordObf-accuracy-human-written}{68}

\DefMacro{TCap-models-pcp-IMP-K-IMP-SOS-accuracy-qwen-coder-by-split-llm-translated}{\pcp accuracy on \llmtrans programs.}
\DefMacro{THead-pcp-IMP-K-IMP-SOS-accuracy-qwen-coder-by-split-llm-translated-models}{Models}
\DefMacro{THead-pcp-IMP-K-IMP-SOS-accuracy-qwen-coder-by-split-llm-translated-pcp-uk-IMP-K-accuracy}{\standardSemCap}
\DefMacro{THead-pcp-IMP-K-IMP-SOS-accuracy-qwen-coder-by-split-llm-translated-pcp-mk-IMP-K-KeywordSwap-accuracy}{\KeywordSwap}
\DefMacro{THead-pcp-IMP-K-IMP-SOS-accuracy-qwen-coder-by-split-llm-translated-pcp-mk-IMP-K-KeywordObf-accuracy}{\KeywordObf}
\DefMacro{THead-pcp-IMP-K-IMP-SOS-accuracy-qwen-coder-by-split-llm-translated-pcp-uk-IMP-SOS-accuracy}{\standardSemCap}
\DefMacro{THead-pcp-IMP-K-IMP-SOS-accuracy-qwen-coder-by-split-llm-translated-pcp-mk-IMP-SOS-KeywordSwap-accuracy}{\KeywordSwap}
\DefMacro{THead-pcp-IMP-K-IMP-SOS-accuracy-qwen-coder-by-split-llm-translated-pcp-mk-IMP-SOS-KeywordObf-accuracy}{\KeywordObf}
\DefMacro{THead-pcp-IMP-K-IMP-SOS-accuracy-qwen-coder-by-split-llm-translated-random-guesser-dataset-distribution-da}{Random}
\DefMacro{THead-pcp-IMP-K-IMP-SOS-accuracy-qwen-coder-by-split-llm-translated-mistralai-Ministral-3-3B-Instruct-2512-BF16-da}{\ministral{3}}
\DefMacro{THead-pcp-IMP-K-IMP-SOS-accuracy-qwen-coder-by-split-llm-translated-mistralai-Ministral-3-8B-Instruct-2512-BF16-da}{\ministral{8}}
\DefMacro{THead-pcp-IMP-K-IMP-SOS-accuracy-qwen-coder-by-split-llm-translated-mistralai-Ministral-3-14B-Instruct-2512-BF16-da}{\ministral{14}}
\DefMacro{THead-pcp-IMP-K-IMP-SOS-accuracy-qwen-coder-by-split-llm-translated-Qwen-Qwen2.5-Coder-3B-Instruct-da}{\qwenCoder{3}}
\DefMacro{THead-pcp-IMP-K-IMP-SOS-accuracy-qwen-coder-by-split-llm-translated-Qwen-Qwen2.5-Coder-7B-Instruct-da}{\qwenCoder{7}}
\DefMacro{THead-pcp-IMP-K-IMP-SOS-accuracy-qwen-coder-by-split-llm-translated-Qwen-Qwen2.5-Coder-14B-Instruct-da}{\qwenCoder{14}}
\DefMacro{THead-pcp-IMP-K-IMP-SOS-accuracy-qwen-coder-by-split-llm-translated-Qwen-Qwen2.5-Coder-32B-Instruct-da}{\qwenCoder{32}}
\DefMacro{THead-pcp-IMP-K-IMP-SOS-accuracy-qwen-coder-by-split-llm-translated-deepseek-ai-DeepSeek-R1-Distill-Qwen-14B-da}{\dpskQwen{14}}
\DefMacro{THead-pcp-IMP-K-IMP-SOS-accuracy-qwen-coder-by-split-llm-translated-deepseek-ai-DeepSeek-R1-Distill-Qwen-32B-da}{\dpskQwen{32}}
\DefMacro{THead-pcp-IMP-K-IMP-SOS-accuracy-qwen-coder-by-split-llm-translated-mistralai-Ministral-3-3B-Instruct-2512-BF16-cot}{\ministralCoT{3}}
\DefMacro{THead-pcp-IMP-K-IMP-SOS-accuracy-qwen-coder-by-split-llm-translated-mistralai-Ministral-3-8B-Instruct-2512-BF16-cot}{\ministralCoT{8}}
\DefMacro{THead-pcp-IMP-K-IMP-SOS-accuracy-qwen-coder-by-split-llm-translated-mistralai-Ministral-3-14B-Instruct-2512-BF16-cot}{\ministralCoT{14}}
\DefMacro{THead-pcp-IMP-K-IMP-SOS-accuracy-qwen-coder-by-split-llm-translated-Qwen-Qwen2.5-Coder-3B-Instruct-cot}{\qwenCoderCoT{3}}
\DefMacro{THead-pcp-IMP-K-IMP-SOS-accuracy-qwen-coder-by-split-llm-translated-Qwen-Qwen2.5-Coder-7B-Instruct-cot}{\qwenCoderCoT{7}}
\DefMacro{THead-pcp-IMP-K-IMP-SOS-accuracy-qwen-coder-by-split-llm-translated-Qwen-Qwen2.5-Coder-14B-Instruct-cot}{\qwenCoderCoT{14}}
\DefMacro{THead-pcp-IMP-K-IMP-SOS-accuracy-qwen-coder-by-split-llm-translated-Qwen-Qwen2.5-Coder-32B-Instruct-cot}{\qwenCoderCoT{32}}
\DefMacro{res-random-guesser-dataset-distribution-da-pcp-uk-IMP-K-accuracy-llm-translated}{16}
\DefMacro{res-random-guesser-dataset-distribution-da-pcp-mk-IMP-K-KeywordSwap-accuracy-llm-translated}{16}
\DefMacro{res-random-guesser-dataset-distribution-da-pcp-mk-IMP-K-KeywordObf-accuracy-llm-translated}{16}
\DefMacro{res-random-guesser-dataset-distribution-da-pcp-uk-IMP-SOS-accuracy-llm-translated}{16}
\DefMacro{res-random-guesser-dataset-distribution-da-pcp-mk-IMP-SOS-KeywordSwap-accuracy-llm-translated}{16}
\DefMacro{res-random-guesser-dataset-distribution-da-pcp-mk-IMP-SOS-KeywordObf-accuracy-llm-translated}{16}
\DefMacro{res-mistralai-Ministral-3-3B-Instruct-2512-BF16-da-pcp-uk-IMP-K-accuracy-llm-translated}{44}
\DefMacro{res-mistralai-Ministral-3-3B-Instruct-2512-BF16-da-pcp-mk-IMP-K-KeywordSwap-accuracy-llm-translated}{33}
\DefMacro{res-mistralai-Ministral-3-3B-Instruct-2512-BF16-da-pcp-mk-IMP-K-KeywordObf-accuracy-llm-translated}{10}
\DefMacro{res-mistralai-Ministral-3-3B-Instruct-2512-BF16-da-pcp-uk-IMP-SOS-accuracy-llm-translated}{41}
\DefMacro{res-mistralai-Ministral-3-3B-Instruct-2512-BF16-da-pcp-mk-IMP-SOS-KeywordSwap-accuracy-llm-translated}{31}
\DefMacro{res-mistralai-Ministral-3-3B-Instruct-2512-BF16-da-pcp-mk-IMP-SOS-KeywordObf-accuracy-llm-translated}{18}
\DefMacro{res-mistralai-Ministral-3-8B-Instruct-2512-BF16-da-pcp-uk-IMP-K-accuracy-llm-translated}{74}
\DefMacro{res-mistralai-Ministral-3-8B-Instruct-2512-BF16-da-pcp-mk-IMP-K-KeywordSwap-accuracy-llm-translated}{59}
\DefMacro{res-mistralai-Ministral-3-8B-Instruct-2512-BF16-da-pcp-mk-IMP-K-KeywordObf-accuracy-llm-translated}{22}
\DefMacro{res-mistralai-Ministral-3-8B-Instruct-2512-BF16-da-pcp-uk-IMP-SOS-accuracy-llm-translated}{62}
\DefMacro{res-mistralai-Ministral-3-8B-Instruct-2512-BF16-da-pcp-mk-IMP-SOS-KeywordSwap-accuracy-llm-translated}{48}
\DefMacro{res-mistralai-Ministral-3-8B-Instruct-2512-BF16-da-pcp-mk-IMP-SOS-KeywordObf-accuracy-llm-translated}{19}
\DefMacro{res-mistralai-Ministral-3-14B-Instruct-2512-BF16-da-pcp-uk-IMP-K-accuracy-llm-translated}{80}
\DefMacro{res-mistralai-Ministral-3-14B-Instruct-2512-BF16-da-pcp-mk-IMP-K-KeywordSwap-accuracy-llm-translated}{69}
\DefMacro{res-mistralai-Ministral-3-14B-Instruct-2512-BF16-da-pcp-mk-IMP-K-KeywordObf-accuracy-llm-translated}{29}
\DefMacro{res-mistralai-Ministral-3-14B-Instruct-2512-BF16-da-pcp-uk-IMP-SOS-accuracy-llm-translated}{76}
\DefMacro{res-mistralai-Ministral-3-14B-Instruct-2512-BF16-da-pcp-mk-IMP-SOS-KeywordSwap-accuracy-llm-translated}{68}
\DefMacro{res-mistralai-Ministral-3-14B-Instruct-2512-BF16-da-pcp-mk-IMP-SOS-KeywordObf-accuracy-llm-translated}{32}
\DefMacro{res-Qwen-Qwen2.5-Coder-3B-Instruct-da-pcp-uk-IMP-K-accuracy-llm-translated}{35}
\DefMacro{res-Qwen-Qwen2.5-Coder-3B-Instruct-da-pcp-mk-IMP-K-KeywordSwap-accuracy-llm-translated}{18}
\DefMacro{res-Qwen-Qwen2.5-Coder-3B-Instruct-da-pcp-mk-IMP-K-KeywordObf-accuracy-llm-translated}{16}
\DefMacro{res-Qwen-Qwen2.5-Coder-3B-Instruct-da-pcp-uk-IMP-SOS-accuracy-llm-translated}{21}
\DefMacro{res-Qwen-Qwen2.5-Coder-3B-Instruct-da-pcp-mk-IMP-SOS-KeywordSwap-accuracy-llm-translated}{23}
\DefMacro{res-Qwen-Qwen2.5-Coder-3B-Instruct-da-pcp-mk-IMP-SOS-KeywordObf-accuracy-llm-translated}{18}
\DefMacro{res-Qwen-Qwen2.5-Coder-7B-Instruct-da-pcp-uk-IMP-K-accuracy-llm-translated}{44}
\DefMacro{res-Qwen-Qwen2.5-Coder-7B-Instruct-da-pcp-mk-IMP-K-KeywordSwap-accuracy-llm-translated}{18}
\DefMacro{res-Qwen-Qwen2.5-Coder-7B-Instruct-da-pcp-mk-IMP-K-KeywordObf-accuracy-llm-translated}{8}
\DefMacro{res-Qwen-Qwen2.5-Coder-7B-Instruct-da-pcp-uk-IMP-SOS-accuracy-llm-translated}{44}
\DefMacro{res-Qwen-Qwen2.5-Coder-7B-Instruct-da-pcp-mk-IMP-SOS-KeywordSwap-accuracy-llm-translated}{25}
\DefMacro{res-Qwen-Qwen2.5-Coder-7B-Instruct-da-pcp-mk-IMP-SOS-KeywordObf-accuracy-llm-translated}{19}
\DefMacro{res-Qwen-Qwen2.5-Coder-14B-Instruct-da-pcp-uk-IMP-K-accuracy-llm-translated}{55}
\DefMacro{res-Qwen-Qwen2.5-Coder-14B-Instruct-da-pcp-mk-IMP-K-KeywordSwap-accuracy-llm-translated}{39}
\DefMacro{res-Qwen-Qwen2.5-Coder-14B-Instruct-da-pcp-mk-IMP-K-KeywordObf-accuracy-llm-translated}{20}
\DefMacro{res-Qwen-Qwen2.5-Coder-14B-Instruct-da-pcp-uk-IMP-SOS-accuracy-llm-translated}{74}
\DefMacro{res-Qwen-Qwen2.5-Coder-14B-Instruct-da-pcp-mk-IMP-SOS-KeywordSwap-accuracy-llm-translated}{69}
\DefMacro{res-Qwen-Qwen2.5-Coder-14B-Instruct-da-pcp-mk-IMP-SOS-KeywordObf-accuracy-llm-translated}{29}
\DefMacro{res-Qwen-Qwen2.5-Coder-32B-Instruct-da-pcp-uk-IMP-K-accuracy-llm-translated}{60}
\DefMacro{res-Qwen-Qwen2.5-Coder-32B-Instruct-da-pcp-mk-IMP-K-KeywordSwap-accuracy-llm-translated}{45}
\DefMacro{res-Qwen-Qwen2.5-Coder-32B-Instruct-da-pcp-mk-IMP-K-KeywordObf-accuracy-llm-translated}{29}
\DefMacro{res-Qwen-Qwen2.5-Coder-32B-Instruct-da-pcp-uk-IMP-SOS-accuracy-llm-translated}{88}
\DefMacro{res-Qwen-Qwen2.5-Coder-32B-Instruct-da-pcp-mk-IMP-SOS-KeywordSwap-accuracy-llm-translated}{65}
\DefMacro{res-Qwen-Qwen2.5-Coder-32B-Instruct-da-pcp-mk-IMP-SOS-KeywordObf-accuracy-llm-translated}{38}
\DefMacro{res-deepseek-ai-DeepSeek-R1-Distill-Qwen-14B-da-pcp-uk-IMP-K-accuracy-llm-translated}{90}
\DefMacro{res-deepseek-ai-DeepSeek-R1-Distill-Qwen-14B-da-pcp-mk-IMP-K-KeywordSwap-accuracy-llm-translated}{71}
\DefMacro{res-deepseek-ai-DeepSeek-R1-Distill-Qwen-14B-da-pcp-mk-IMP-K-KeywordObf-accuracy-llm-translated}{77}
\DefMacro{res-deepseek-ai-DeepSeek-R1-Distill-Qwen-14B-da-pcp-uk-IMP-SOS-accuracy-llm-translated}{88}
\DefMacro{res-deepseek-ai-DeepSeek-R1-Distill-Qwen-14B-da-pcp-mk-IMP-SOS-KeywordSwap-accuracy-llm-translated}{69}
\DefMacro{res-deepseek-ai-DeepSeek-R1-Distill-Qwen-14B-da-pcp-mk-IMP-SOS-KeywordObf-accuracy-llm-translated}{60}
\DefMacro{res-deepseek-ai-DeepSeek-R1-Distill-Qwen-32B-da-pcp-uk-IMP-K-accuracy-llm-translated}{98}
\DefMacro{res-deepseek-ai-DeepSeek-R1-Distill-Qwen-32B-da-pcp-mk-IMP-K-KeywordSwap-accuracy-llm-translated}{75}
\DefMacro{res-deepseek-ai-DeepSeek-R1-Distill-Qwen-32B-da-pcp-mk-IMP-K-KeywordObf-accuracy-llm-translated}{92}
\DefMacro{res-deepseek-ai-DeepSeek-R1-Distill-Qwen-32B-da-pcp-uk-IMP-SOS-accuracy-llm-translated}{97}
\DefMacro{res-deepseek-ai-DeepSeek-R1-Distill-Qwen-32B-da-pcp-mk-IMP-SOS-KeywordSwap-accuracy-llm-translated}{70}
\DefMacro{res-deepseek-ai-DeepSeek-R1-Distill-Qwen-32B-da-pcp-mk-IMP-SOS-KeywordObf-accuracy-llm-translated}{83}
\DefMacro{res-mistralai-Ministral-3-3B-Instruct-2512-BF16-cot-pcp-uk-IMP-K-accuracy-llm-translated}{88}
\DefMacro{res-mistralai-Ministral-3-3B-Instruct-2512-BF16-cot-pcp-mk-IMP-K-KeywordSwap-accuracy-llm-translated}{67}
\DefMacro{res-mistralai-Ministral-3-3B-Instruct-2512-BF16-cot-pcp-mk-IMP-K-KeywordObf-accuracy-llm-translated}{42}
\DefMacro{res-mistralai-Ministral-3-3B-Instruct-2512-BF16-cot-pcp-uk-IMP-SOS-accuracy-llm-translated}{88}
\DefMacro{res-mistralai-Ministral-3-3B-Instruct-2512-BF16-cot-pcp-mk-IMP-SOS-KeywordSwap-accuracy-llm-translated}{69}
\DefMacro{res-mistralai-Ministral-3-3B-Instruct-2512-BF16-cot-pcp-mk-IMP-SOS-KeywordObf-accuracy-llm-translated}{48}
\DefMacro{res-mistralai-Ministral-3-8B-Instruct-2512-BF16-cot-pcp-uk-IMP-K-accuracy-llm-translated}{96}
\DefMacro{res-mistralai-Ministral-3-8B-Instruct-2512-BF16-cot-pcp-mk-IMP-K-KeywordSwap-accuracy-llm-translated}{78}
\DefMacro{res-mistralai-Ministral-3-8B-Instruct-2512-BF16-cot-pcp-mk-IMP-K-KeywordObf-accuracy-llm-translated}{66}
\DefMacro{res-mistralai-Ministral-3-8B-Instruct-2512-BF16-cot-pcp-uk-IMP-SOS-accuracy-llm-translated}{96}
\DefMacro{res-mistralai-Ministral-3-8B-Instruct-2512-BF16-cot-pcp-mk-IMP-SOS-KeywordSwap-accuracy-llm-translated}{78}
\DefMacro{res-mistralai-Ministral-3-8B-Instruct-2512-BF16-cot-pcp-mk-IMP-SOS-KeywordObf-accuracy-llm-translated}{54}
\DefMacro{res-mistralai-Ministral-3-14B-Instruct-2512-BF16-cot-pcp-uk-IMP-K-accuracy-llm-translated}{97}
\DefMacro{res-mistralai-Ministral-3-14B-Instruct-2512-BF16-cot-pcp-mk-IMP-K-KeywordSwap-accuracy-llm-translated}{78}
\DefMacro{res-mistralai-Ministral-3-14B-Instruct-2512-BF16-cot-pcp-mk-IMP-K-KeywordObf-accuracy-llm-translated}{81}
\DefMacro{res-mistralai-Ministral-3-14B-Instruct-2512-BF16-cot-pcp-uk-IMP-SOS-accuracy-llm-translated}{88}
\DefMacro{res-mistralai-Ministral-3-14B-Instruct-2512-BF16-cot-pcp-mk-IMP-SOS-KeywordSwap-accuracy-llm-translated}{71}
\DefMacro{res-mistralai-Ministral-3-14B-Instruct-2512-BF16-cot-pcp-mk-IMP-SOS-KeywordObf-accuracy-llm-translated}{68}
\DefMacro{res-Qwen-Qwen2.5-Coder-3B-Instruct-cot-pcp-uk-IMP-K-accuracy-llm-translated}{29}
\DefMacro{res-Qwen-Qwen2.5-Coder-3B-Instruct-cot-pcp-mk-IMP-K-KeywordSwap-accuracy-llm-translated}{18}
\DefMacro{res-Qwen-Qwen2.5-Coder-3B-Instruct-cot-pcp-mk-IMP-K-KeywordObf-accuracy-llm-translated}{14}
\DefMacro{res-Qwen-Qwen2.5-Coder-3B-Instruct-cot-pcp-uk-IMP-SOS-accuracy-llm-translated}{31}
\DefMacro{res-Qwen-Qwen2.5-Coder-3B-Instruct-cot-pcp-mk-IMP-SOS-KeywordSwap-accuracy-llm-translated}{25}
\DefMacro{res-Qwen-Qwen2.5-Coder-3B-Instruct-cot-pcp-mk-IMP-SOS-KeywordObf-accuracy-llm-translated}{16}
\DefMacro{res-Qwen-Qwen2.5-Coder-7B-Instruct-cot-pcp-uk-IMP-K-accuracy-llm-translated}{58}
\DefMacro{res-Qwen-Qwen2.5-Coder-7B-Instruct-cot-pcp-mk-IMP-K-KeywordSwap-accuracy-llm-translated}{36}
\DefMacro{res-Qwen-Qwen2.5-Coder-7B-Instruct-cot-pcp-mk-IMP-K-KeywordObf-accuracy-llm-translated}{23}
\DefMacro{res-Qwen-Qwen2.5-Coder-7B-Instruct-cot-pcp-uk-IMP-SOS-accuracy-llm-translated}{59}
\DefMacro{res-Qwen-Qwen2.5-Coder-7B-Instruct-cot-pcp-mk-IMP-SOS-KeywordSwap-accuracy-llm-translated}{37}
\DefMacro{res-Qwen-Qwen2.5-Coder-7B-Instruct-cot-pcp-mk-IMP-SOS-KeywordObf-accuracy-llm-translated}{26}
\DefMacro{res-Qwen-Qwen2.5-Coder-14B-Instruct-cot-pcp-uk-IMP-K-accuracy-llm-translated}{86}
\DefMacro{res-Qwen-Qwen2.5-Coder-14B-Instruct-cot-pcp-mk-IMP-K-KeywordSwap-accuracy-llm-translated}{68}
\DefMacro{res-Qwen-Qwen2.5-Coder-14B-Instruct-cot-pcp-mk-IMP-K-KeywordObf-accuracy-llm-translated}{48}
\DefMacro{res-Qwen-Qwen2.5-Coder-14B-Instruct-cot-pcp-uk-IMP-SOS-accuracy-llm-translated}{85}
\DefMacro{res-Qwen-Qwen2.5-Coder-14B-Instruct-cot-pcp-mk-IMP-SOS-KeywordSwap-accuracy-llm-translated}{67}
\DefMacro{res-Qwen-Qwen2.5-Coder-14B-Instruct-cot-pcp-mk-IMP-SOS-KeywordObf-accuracy-llm-translated}{37}
\DefMacro{res-Qwen-Qwen2.5-Coder-32B-Instruct-cot-pcp-uk-IMP-K-accuracy-llm-translated}{95}
\DefMacro{res-Qwen-Qwen2.5-Coder-32B-Instruct-cot-pcp-mk-IMP-K-KeywordSwap-accuracy-llm-translated}{75}
\DefMacro{res-Qwen-Qwen2.5-Coder-32B-Instruct-cot-pcp-mk-IMP-K-KeywordObf-accuracy-llm-translated}{52}
\DefMacro{res-Qwen-Qwen2.5-Coder-32B-Instruct-cot-pcp-uk-IMP-SOS-accuracy-llm-translated}{95}
\DefMacro{res-Qwen-Qwen2.5-Coder-32B-Instruct-cot-pcp-mk-IMP-SOS-KeywordSwap-accuracy-llm-translated}{77}
\DefMacro{res-Qwen-Qwen2.5-Coder-32B-Instruct-cot-pcp-mk-IMP-SOS-KeywordObf-accuracy-llm-translated}{49}

\DefMacro{TCap-models-pcp-IMP-K-IMP-SOS-accuracy-qwen-coder-by-split-fuzzer-generated}{\pcp accuracy on \fuzzgen programs.}
\DefMacro{THead-pcp-IMP-K-IMP-SOS-accuracy-qwen-coder-by-split-fuzzer-generated-models}{Models}
\DefMacro{THead-pcp-IMP-K-IMP-SOS-accuracy-qwen-coder-by-split-fuzzer-generated-pcp-uk-IMP-K-accuracy}{\standardSemCap}
\DefMacro{THead-pcp-IMP-K-IMP-SOS-accuracy-qwen-coder-by-split-fuzzer-generated-pcp-mk-IMP-K-KeywordSwap-accuracy}{\KeywordSwap}
\DefMacro{THead-pcp-IMP-K-IMP-SOS-accuracy-qwen-coder-by-split-fuzzer-generated-pcp-mk-IMP-K-KeywordObf-accuracy}{\KeywordObf}
\DefMacro{THead-pcp-IMP-K-IMP-SOS-accuracy-qwen-coder-by-split-fuzzer-generated-pcp-uk-IMP-SOS-accuracy}{\standardSemCap}
\DefMacro{THead-pcp-IMP-K-IMP-SOS-accuracy-qwen-coder-by-split-fuzzer-generated-pcp-mk-IMP-SOS-KeywordSwap-accuracy}{\KeywordSwap}
\DefMacro{THead-pcp-IMP-K-IMP-SOS-accuracy-qwen-coder-by-split-fuzzer-generated-pcp-mk-IMP-SOS-KeywordObf-accuracy}{\KeywordObf}
\DefMacro{THead-pcp-IMP-K-IMP-SOS-accuracy-qwen-coder-by-split-fuzzer-generated-random-guesser-dataset-distribution-da}{Random}
\DefMacro{THead-pcp-IMP-K-IMP-SOS-accuracy-qwen-coder-by-split-fuzzer-generated-mistralai-Ministral-3-3B-Instruct-2512-BF16-da}{\ministral{3}}
\DefMacro{THead-pcp-IMP-K-IMP-SOS-accuracy-qwen-coder-by-split-fuzzer-generated-mistralai-Ministral-3-8B-Instruct-2512-BF16-da}{\ministral{8}}
\DefMacro{THead-pcp-IMP-K-IMP-SOS-accuracy-qwen-coder-by-split-fuzzer-generated-mistralai-Ministral-3-14B-Instruct-2512-BF16-da}{\ministral{14}}
\DefMacro{THead-pcp-IMP-K-IMP-SOS-accuracy-qwen-coder-by-split-fuzzer-generated-Qwen-Qwen2.5-Coder-3B-Instruct-da}{\qwenCoder{3}}
\DefMacro{THead-pcp-IMP-K-IMP-SOS-accuracy-qwen-coder-by-split-fuzzer-generated-Qwen-Qwen2.5-Coder-7B-Instruct-da}{\qwenCoder{7}}
\DefMacro{THead-pcp-IMP-K-IMP-SOS-accuracy-qwen-coder-by-split-fuzzer-generated-Qwen-Qwen2.5-Coder-14B-Instruct-da}{\qwenCoder{14}}
\DefMacro{THead-pcp-IMP-K-IMP-SOS-accuracy-qwen-coder-by-split-fuzzer-generated-Qwen-Qwen2.5-Coder-32B-Instruct-da}{\qwenCoder{32}}
\DefMacro{THead-pcp-IMP-K-IMP-SOS-accuracy-qwen-coder-by-split-fuzzer-generated-deepseek-ai-DeepSeek-R1-Distill-Qwen-14B-da}{\dpskQwen{14}}
\DefMacro{THead-pcp-IMP-K-IMP-SOS-accuracy-qwen-coder-by-split-fuzzer-generated-deepseek-ai-DeepSeek-R1-Distill-Qwen-32B-da}{\dpskQwen{32}}
\DefMacro{THead-pcp-IMP-K-IMP-SOS-accuracy-qwen-coder-by-split-fuzzer-generated-mistralai-Ministral-3-3B-Instruct-2512-BF16-cot}{\ministralCoT{3}}
\DefMacro{THead-pcp-IMP-K-IMP-SOS-accuracy-qwen-coder-by-split-fuzzer-generated-mistralai-Ministral-3-8B-Instruct-2512-BF16-cot}{\ministralCoT{8}}
\DefMacro{THead-pcp-IMP-K-IMP-SOS-accuracy-qwen-coder-by-split-fuzzer-generated-mistralai-Ministral-3-14B-Instruct-2512-BF16-cot}{\ministralCoT{14}}
\DefMacro{THead-pcp-IMP-K-IMP-SOS-accuracy-qwen-coder-by-split-fuzzer-generated-Qwen-Qwen2.5-Coder-3B-Instruct-cot}{\qwenCoderCoT{3}}
\DefMacro{THead-pcp-IMP-K-IMP-SOS-accuracy-qwen-coder-by-split-fuzzer-generated-Qwen-Qwen2.5-Coder-7B-Instruct-cot}{\qwenCoderCoT{7}}
\DefMacro{THead-pcp-IMP-K-IMP-SOS-accuracy-qwen-coder-by-split-fuzzer-generated-Qwen-Qwen2.5-Coder-14B-Instruct-cot}{\qwenCoderCoT{14}}
\DefMacro{THead-pcp-IMP-K-IMP-SOS-accuracy-qwen-coder-by-split-fuzzer-generated-Qwen-Qwen2.5-Coder-32B-Instruct-cot}{\qwenCoderCoT{32}}
\DefMacro{res-random-guesser-dataset-distribution-da-pcp-uk-IMP-K-accuracy-fuzzer-generated}{17}
\DefMacro{res-random-guesser-dataset-distribution-da-pcp-mk-IMP-K-KeywordSwap-accuracy-fuzzer-generated}{17}
\DefMacro{res-random-guesser-dataset-distribution-da-pcp-mk-IMP-K-KeywordObf-accuracy-fuzzer-generated}{16}
\DefMacro{res-random-guesser-dataset-distribution-da-pcp-uk-IMP-SOS-accuracy-fuzzer-generated}{17}
\DefMacro{res-random-guesser-dataset-distribution-da-pcp-mk-IMP-SOS-KeywordSwap-accuracy-fuzzer-generated}{17}
\DefMacro{res-random-guesser-dataset-distribution-da-pcp-mk-IMP-SOS-KeywordObf-accuracy-fuzzer-generated}{16}
\DefMacro{res-mistralai-Ministral-3-3B-Instruct-2512-BF16-da-pcp-uk-IMP-K-accuracy-fuzzer-generated}{24}
\DefMacro{res-mistralai-Ministral-3-3B-Instruct-2512-BF16-da-pcp-mk-IMP-K-KeywordSwap-accuracy-fuzzer-generated}{26}
\DefMacro{res-mistralai-Ministral-3-3B-Instruct-2512-BF16-da-pcp-mk-IMP-K-KeywordObf-accuracy-fuzzer-generated}{14}
\DefMacro{res-mistralai-Ministral-3-3B-Instruct-2512-BF16-da-pcp-uk-IMP-SOS-accuracy-fuzzer-generated}{29}
\DefMacro{res-mistralai-Ministral-3-3B-Instruct-2512-BF16-da-pcp-mk-IMP-SOS-KeywordSwap-accuracy-fuzzer-generated}{17}
\DefMacro{res-mistralai-Ministral-3-3B-Instruct-2512-BF16-da-pcp-mk-IMP-SOS-KeywordObf-accuracy-fuzzer-generated}{15}
\DefMacro{res-mistralai-Ministral-3-8B-Instruct-2512-BF16-da-pcp-uk-IMP-K-accuracy-fuzzer-generated}{43}
\DefMacro{res-mistralai-Ministral-3-8B-Instruct-2512-BF16-da-pcp-mk-IMP-K-KeywordSwap-accuracy-fuzzer-generated}{30}
\DefMacro{res-mistralai-Ministral-3-8B-Instruct-2512-BF16-da-pcp-mk-IMP-K-KeywordObf-accuracy-fuzzer-generated}{17}
\DefMacro{res-mistralai-Ministral-3-8B-Instruct-2512-BF16-da-pcp-uk-IMP-SOS-accuracy-fuzzer-generated}{35}
\DefMacro{res-mistralai-Ministral-3-8B-Instruct-2512-BF16-da-pcp-mk-IMP-SOS-KeywordSwap-accuracy-fuzzer-generated}{26}
\DefMacro{res-mistralai-Ministral-3-8B-Instruct-2512-BF16-da-pcp-mk-IMP-SOS-KeywordObf-accuracy-fuzzer-generated}{17}
\DefMacro{res-mistralai-Ministral-3-14B-Instruct-2512-BF16-da-pcp-uk-IMP-K-accuracy-fuzzer-generated}{65}
\DefMacro{res-mistralai-Ministral-3-14B-Instruct-2512-BF16-da-pcp-mk-IMP-K-KeywordSwap-accuracy-fuzzer-generated}{37}
\DefMacro{res-mistralai-Ministral-3-14B-Instruct-2512-BF16-da-pcp-mk-IMP-K-KeywordObf-accuracy-fuzzer-generated}{24}
\DefMacro{res-mistralai-Ministral-3-14B-Instruct-2512-BF16-da-pcp-uk-IMP-SOS-accuracy-fuzzer-generated}{71}
\DefMacro{res-mistralai-Ministral-3-14B-Instruct-2512-BF16-da-pcp-mk-IMP-SOS-KeywordSwap-accuracy-fuzzer-generated}{31}
\DefMacro{res-mistralai-Ministral-3-14B-Instruct-2512-BF16-da-pcp-mk-IMP-SOS-KeywordObf-accuracy-fuzzer-generated}{27}
\DefMacro{res-Qwen-Qwen2.5-Coder-3B-Instruct-da-pcp-uk-IMP-K-accuracy-fuzzer-generated}{19}
\DefMacro{res-Qwen-Qwen2.5-Coder-3B-Instruct-da-pcp-mk-IMP-K-KeywordSwap-accuracy-fuzzer-generated}{16}
\DefMacro{res-Qwen-Qwen2.5-Coder-3B-Instruct-da-pcp-mk-IMP-K-KeywordObf-accuracy-fuzzer-generated}{12}
\DefMacro{res-Qwen-Qwen2.5-Coder-3B-Instruct-da-pcp-uk-IMP-SOS-accuracy-fuzzer-generated}{20}
\DefMacro{res-Qwen-Qwen2.5-Coder-3B-Instruct-da-pcp-mk-IMP-SOS-KeywordSwap-accuracy-fuzzer-generated}{20}
\DefMacro{res-Qwen-Qwen2.5-Coder-3B-Instruct-da-pcp-mk-IMP-SOS-KeywordObf-accuracy-fuzzer-generated}{19}
\DefMacro{res-Qwen-Qwen2.5-Coder-7B-Instruct-da-pcp-uk-IMP-K-accuracy-fuzzer-generated}{21}
\DefMacro{res-Qwen-Qwen2.5-Coder-7B-Instruct-da-pcp-mk-IMP-K-KeywordSwap-accuracy-fuzzer-generated}{17}
\DefMacro{res-Qwen-Qwen2.5-Coder-7B-Instruct-da-pcp-mk-IMP-K-KeywordObf-accuracy-fuzzer-generated}{16}
\DefMacro{res-Qwen-Qwen2.5-Coder-7B-Instruct-da-pcp-uk-IMP-SOS-accuracy-fuzzer-generated}{17}
\DefMacro{res-Qwen-Qwen2.5-Coder-7B-Instruct-da-pcp-mk-IMP-SOS-KeywordSwap-accuracy-fuzzer-generated}{17}
\DefMacro{res-Qwen-Qwen2.5-Coder-7B-Instruct-da-pcp-mk-IMP-SOS-KeywordObf-accuracy-fuzzer-generated}{17}
\DefMacro{res-Qwen-Qwen2.5-Coder-14B-Instruct-da-pcp-uk-IMP-K-accuracy-fuzzer-generated}{58}
\DefMacro{res-Qwen-Qwen2.5-Coder-14B-Instruct-da-pcp-mk-IMP-K-KeywordSwap-accuracy-fuzzer-generated}{30}
\DefMacro{res-Qwen-Qwen2.5-Coder-14B-Instruct-da-pcp-mk-IMP-K-KeywordObf-accuracy-fuzzer-generated}{25}
\DefMacro{res-Qwen-Qwen2.5-Coder-14B-Instruct-da-pcp-uk-IMP-SOS-accuracy-fuzzer-generated}{65}
\DefMacro{res-Qwen-Qwen2.5-Coder-14B-Instruct-da-pcp-mk-IMP-SOS-KeywordSwap-accuracy-fuzzer-generated}{39}
\DefMacro{res-Qwen-Qwen2.5-Coder-14B-Instruct-da-pcp-mk-IMP-SOS-KeywordObf-accuracy-fuzzer-generated}{26}
\DefMacro{res-Qwen-Qwen2.5-Coder-32B-Instruct-da-pcp-uk-IMP-K-accuracy-fuzzer-generated}{51}
\DefMacro{res-Qwen-Qwen2.5-Coder-32B-Instruct-da-pcp-mk-IMP-K-KeywordSwap-accuracy-fuzzer-generated}{26}
\DefMacro{res-Qwen-Qwen2.5-Coder-32B-Instruct-da-pcp-mk-IMP-K-KeywordObf-accuracy-fuzzer-generated}{28}
\DefMacro{res-Qwen-Qwen2.5-Coder-32B-Instruct-da-pcp-uk-IMP-SOS-accuracy-fuzzer-generated}{69}
\DefMacro{res-Qwen-Qwen2.5-Coder-32B-Instruct-da-pcp-mk-IMP-SOS-KeywordSwap-accuracy-fuzzer-generated}{36}
\DefMacro{res-Qwen-Qwen2.5-Coder-32B-Instruct-da-pcp-mk-IMP-SOS-KeywordObf-accuracy-fuzzer-generated}{34}
\DefMacro{res-deepseek-ai-DeepSeek-R1-Distill-Qwen-14B-da-pcp-uk-IMP-K-accuracy-fuzzer-generated}{78}
\DefMacro{res-deepseek-ai-DeepSeek-R1-Distill-Qwen-14B-da-pcp-mk-IMP-K-KeywordSwap-accuracy-fuzzer-generated}{43}
\DefMacro{res-deepseek-ai-DeepSeek-R1-Distill-Qwen-14B-da-pcp-mk-IMP-K-KeywordObf-accuracy-fuzzer-generated}{51}
\DefMacro{res-deepseek-ai-DeepSeek-R1-Distill-Qwen-14B-da-pcp-uk-IMP-SOS-accuracy-fuzzer-generated}{78}
\DefMacro{res-deepseek-ai-DeepSeek-R1-Distill-Qwen-14B-da-pcp-mk-IMP-SOS-KeywordSwap-accuracy-fuzzer-generated}{42}
\DefMacro{res-deepseek-ai-DeepSeek-R1-Distill-Qwen-14B-da-pcp-mk-IMP-SOS-KeywordObf-accuracy-fuzzer-generated}{37}
\DefMacro{res-deepseek-ai-DeepSeek-R1-Distill-Qwen-32B-da-pcp-uk-IMP-K-accuracy-fuzzer-generated}{89}
\DefMacro{res-deepseek-ai-DeepSeek-R1-Distill-Qwen-32B-da-pcp-mk-IMP-K-KeywordSwap-accuracy-fuzzer-generated}{58}
\DefMacro{res-deepseek-ai-DeepSeek-R1-Distill-Qwen-32B-da-pcp-mk-IMP-K-KeywordObf-accuracy-fuzzer-generated}{68}
\DefMacro{res-deepseek-ai-DeepSeek-R1-Distill-Qwen-32B-da-pcp-uk-IMP-SOS-accuracy-fuzzer-generated}{90}
\DefMacro{res-deepseek-ai-DeepSeek-R1-Distill-Qwen-32B-da-pcp-mk-IMP-SOS-KeywordSwap-accuracy-fuzzer-generated}{57}
\DefMacro{res-deepseek-ai-DeepSeek-R1-Distill-Qwen-32B-da-pcp-mk-IMP-SOS-KeywordObf-accuracy-fuzzer-generated}{57}
\DefMacro{res-mistralai-Ministral-3-3B-Instruct-2512-BF16-cot-pcp-uk-IMP-K-accuracy-fuzzer-generated}{55}
\DefMacro{res-mistralai-Ministral-3-3B-Instruct-2512-BF16-cot-pcp-mk-IMP-K-KeywordSwap-accuracy-fuzzer-generated}{29}
\DefMacro{res-mistralai-Ministral-3-3B-Instruct-2512-BF16-cot-pcp-mk-IMP-K-KeywordObf-accuracy-fuzzer-generated}{28}
\DefMacro{res-mistralai-Ministral-3-3B-Instruct-2512-BF16-cot-pcp-uk-IMP-SOS-accuracy-fuzzer-generated}{49}
\DefMacro{res-mistralai-Ministral-3-3B-Instruct-2512-BF16-cot-pcp-mk-IMP-SOS-KeywordSwap-accuracy-fuzzer-generated}{30}
\DefMacro{res-mistralai-Ministral-3-3B-Instruct-2512-BF16-cot-pcp-mk-IMP-SOS-KeywordObf-accuracy-fuzzer-generated}{24}
\DefMacro{res-mistralai-Ministral-3-8B-Instruct-2512-BF16-cot-pcp-uk-IMP-K-accuracy-fuzzer-generated}{82}
\DefMacro{res-mistralai-Ministral-3-8B-Instruct-2512-BF16-cot-pcp-mk-IMP-K-KeywordSwap-accuracy-fuzzer-generated}{48}
\DefMacro{res-mistralai-Ministral-3-8B-Instruct-2512-BF16-cot-pcp-mk-IMP-K-KeywordObf-accuracy-fuzzer-generated}{40}
\DefMacro{res-mistralai-Ministral-3-8B-Instruct-2512-BF16-cot-pcp-uk-IMP-SOS-accuracy-fuzzer-generated}{82}
\DefMacro{res-mistralai-Ministral-3-8B-Instruct-2512-BF16-cot-pcp-mk-IMP-SOS-KeywordSwap-accuracy-fuzzer-generated}{45}
\DefMacro{res-mistralai-Ministral-3-8B-Instruct-2512-BF16-cot-pcp-mk-IMP-SOS-KeywordObf-accuracy-fuzzer-generated}{37}
\DefMacro{res-mistralai-Ministral-3-14B-Instruct-2512-BF16-cot-pcp-uk-IMP-K-accuracy-fuzzer-generated}{88}
\DefMacro{res-mistralai-Ministral-3-14B-Instruct-2512-BF16-cot-pcp-mk-IMP-K-KeywordSwap-accuracy-fuzzer-generated}{54}
\DefMacro{res-mistralai-Ministral-3-14B-Instruct-2512-BF16-cot-pcp-mk-IMP-K-KeywordObf-accuracy-fuzzer-generated}{66}
\DefMacro{res-mistralai-Ministral-3-14B-Instruct-2512-BF16-cot-pcp-uk-IMP-SOS-accuracy-fuzzer-generated}{76}
\DefMacro{res-mistralai-Ministral-3-14B-Instruct-2512-BF16-cot-pcp-mk-IMP-SOS-KeywordSwap-accuracy-fuzzer-generated}{51}
\DefMacro{res-mistralai-Ministral-3-14B-Instruct-2512-BF16-cot-pcp-mk-IMP-SOS-KeywordObf-accuracy-fuzzer-generated}{52}
\DefMacro{res-Qwen-Qwen2.5-Coder-3B-Instruct-cot-pcp-uk-IMP-K-accuracy-fuzzer-generated}{20}
\DefMacro{res-Qwen-Qwen2.5-Coder-3B-Instruct-cot-pcp-mk-IMP-K-KeywordSwap-accuracy-fuzzer-generated}{14}
\DefMacro{res-Qwen-Qwen2.5-Coder-3B-Instruct-cot-pcp-mk-IMP-K-KeywordObf-accuracy-fuzzer-generated}{14}
\DefMacro{res-Qwen-Qwen2.5-Coder-3B-Instruct-cot-pcp-uk-IMP-SOS-accuracy-fuzzer-generated}{22}
\DefMacro{res-Qwen-Qwen2.5-Coder-3B-Instruct-cot-pcp-mk-IMP-SOS-KeywordSwap-accuracy-fuzzer-generated}{18}
\DefMacro{res-Qwen-Qwen2.5-Coder-3B-Instruct-cot-pcp-mk-IMP-SOS-KeywordObf-accuracy-fuzzer-generated}{13}
\DefMacro{res-Qwen-Qwen2.5-Coder-7B-Instruct-cot-pcp-uk-IMP-K-accuracy-fuzzer-generated}{47}
\DefMacro{res-Qwen-Qwen2.5-Coder-7B-Instruct-cot-pcp-mk-IMP-K-KeywordSwap-accuracy-fuzzer-generated}{31}
\DefMacro{res-Qwen-Qwen2.5-Coder-7B-Instruct-cot-pcp-mk-IMP-K-KeywordObf-accuracy-fuzzer-generated}{22}
\DefMacro{res-Qwen-Qwen2.5-Coder-7B-Instruct-cot-pcp-uk-IMP-SOS-accuracy-fuzzer-generated}{38}
\DefMacro{res-Qwen-Qwen2.5-Coder-7B-Instruct-cot-pcp-mk-IMP-SOS-KeywordSwap-accuracy-fuzzer-generated}{25}
\DefMacro{res-Qwen-Qwen2.5-Coder-7B-Instruct-cot-pcp-mk-IMP-SOS-KeywordObf-accuracy-fuzzer-generated}{25}
\DefMacro{res-Qwen-Qwen2.5-Coder-14B-Instruct-cot-pcp-uk-IMP-K-accuracy-fuzzer-generated}{68}
\DefMacro{res-Qwen-Qwen2.5-Coder-14B-Instruct-cot-pcp-mk-IMP-K-KeywordSwap-accuracy-fuzzer-generated}{47}
\DefMacro{res-Qwen-Qwen2.5-Coder-14B-Instruct-cot-pcp-mk-IMP-K-KeywordObf-accuracy-fuzzer-generated}{23}
\DefMacro{res-Qwen-Qwen2.5-Coder-14B-Instruct-cot-pcp-uk-IMP-SOS-accuracy-fuzzer-generated}{69}
\DefMacro{res-Qwen-Qwen2.5-Coder-14B-Instruct-cot-pcp-mk-IMP-SOS-KeywordSwap-accuracy-fuzzer-generated}{45}
\DefMacro{res-Qwen-Qwen2.5-Coder-14B-Instruct-cot-pcp-mk-IMP-SOS-KeywordObf-accuracy-fuzzer-generated}{22}
\DefMacro{res-Qwen-Qwen2.5-Coder-32B-Instruct-cot-pcp-uk-IMP-K-accuracy-fuzzer-generated}{73}
\DefMacro{res-Qwen-Qwen2.5-Coder-32B-Instruct-cot-pcp-mk-IMP-K-KeywordSwap-accuracy-fuzzer-generated}{48}
\DefMacro{res-Qwen-Qwen2.5-Coder-32B-Instruct-cot-pcp-mk-IMP-K-KeywordObf-accuracy-fuzzer-generated}{24}
\DefMacro{res-Qwen-Qwen2.5-Coder-32B-Instruct-cot-pcp-uk-IMP-SOS-accuracy-fuzzer-generated}{77}
\DefMacro{res-Qwen-Qwen2.5-Coder-32B-Instruct-cot-pcp-mk-IMP-SOS-KeywordSwap-accuracy-fuzzer-generated}{44}
\DefMacro{res-Qwen-Qwen2.5-Coder-32B-Instruct-cot-pcp-mk-IMP-SOS-KeywordObf-accuracy-fuzzer-generated}{29}

\DefMacro{THead-benchmark-mean}{Mean}
\DefMacro{THead-benchmark-max}{Max}
\DefMacro{THead-imp}{\imp}
\DefMacro{THead-imp-tokens}{\# Tokens}
\DefMacro{THead-imp-vars}{\# Variables}
\DefMacro{THead-imp-loc}{\# LOC}
\DefMacro{THead-srp}{\srp}
\DefMacro{THead-srp-selected-stmts}{\# Chosen Stmt.}
\DefMacro{THead-srp-rules-per-stmt}{\# Rules/Stmt.}
\DefMacro{THead-etp}{\etp}
\DefMacro{THead-etp-trace-length}{Len. exec. trace}

\DefMacro{TCap-program-executability-stats}{Introduced semantic errors and corresponding rules.}
\DefMacro{THead-executability}{Executability}
\DefMacro{THead-semantic-error-type}{Semantic error type}
\DefMacro{THead-corresponding-rule}{Corresponding Rule}
\DefMacro{THead-count}{Count}
\DefMacro{THead-percentage}{Percentage}
\DefMacro{THead-success}{success}
\DefMacro{THead-error}{error}
\DefMacro{THead-none}{---}
\DefMacro{dataset-program-executability-success-rule-k}{---}
\DefMacro{dataset-program-executability-success-rule-s}{---}
\DefMacro{THead-semantic-error-break-outside-loop}{Break outside loop}
\DefMacro{dataset-program-executability-break-outside-loop-rule-k}{34}
\DefMacro{dataset-program-executability-break-outside-loop-rule-s}{73}
\DefMacro{THead-semantic-error-continue-outside-loop}{Continue outside loop}
\DefMacro{dataset-program-executability-continue-outside-loop-rule-k}{31}
\DefMacro{dataset-program-executability-continue-outside-loop-rule-s}{76}
\DefMacro{THead-semantic-error-divide-by-zero}{Divide by zero}
\DefMacro{dataset-program-executability-divide-by-zero-rule-k}{7}
\DefMacro{dataset-program-executability-divide-by-zero-rule-s}{19}
\DefMacro{THead-semantic-error-modulo-zero}{Modulo by zero}
\DefMacro{dataset-program-executability-modulo-zero-rule-k}{9}
\DefMacro{dataset-program-executability-modulo-zero-rule-s}{23}
\DefMacro{THead-semantic-error-var-use-before-declare}{Variable use before declare}
\DefMacro{dataset-program-executability-var-use-before-declare-rule-k}{2}
\DefMacro{dataset-program-executability-var-use-before-declare-rule-s}{2}

\DefMacro{TCap-pcp-fuzzing-examples}{Example fuzzing transformations applied to a shared valid base \IMP program (\CodeIn{addition.imp}). Green highlights mark inserted or modified statements in each invalid variant.}
\DefMacro{THead-pcp-fuzzing-valid-group}{Valid program}
\DefMacro{THead-pcp-fuzzing-invalid-group}{Invalid variants}
\DefMacro{THead-pcp-fuzzing-valid}{Valid program}

\DefMacro{TCap-imp-split-loc-stats}{Program size in lines of code (LOC) and token counts measured by the GPT-4o tokenizer per dataset split.}
\DefMacro{THead-imp-split-loc-split}{Split}
\DefMacro{THead-imp-split-loc-num-programs}{\# programs}
\DefMacro{THead-imp-split-loc-loc}{LOC}
\DefMacro{THead-imp-split-loc-tokens}{Tokens}
\DefMacro{THead-imp-split-loc-min}{Min}
\DefMacro{THead-imp-split-loc-median}{Median}
\DefMacro{THead-imp-split-loc-max}{Max}
\DefMacro{THead-imp-split-loc-human-written}{Human-Written}
\DefMacro{THead-imp-split-loc-synthetic-cpp}{LLM-Translated}
\DefMacro{THead-imp-split-loc-fuzzer-generated}{Fuzzer-Generated}

\DefMacro{TCap-imp-split-complexity-stats}{Median control-flow complexity per dataset split. CC is extended cyclomatic complexity. Max nested if/while report static nesting depth.}
\DefMacro{THead-imp-split-complexity-split}{Split}
\DefMacro{THead-imp-split-complexity-num-programs}{\# programs}
\DefMacro{THead-imp-split-complexity-cc}{CC}
\DefMacro{THead-imp-split-complexity-max-nested-if}{Max nested if}
\DefMacro{THead-imp-split-complexity-max-nested-while}{Max nested while}
\DefMacro{THead-imp-split-complexity-human-written}{Human-Written}
\DefMacro{THead-imp-split-complexity-synthetic-cpp}{LLM-Translated}
\DefMacro{THead-imp-split-complexity-fuzzer-generated}{Fuzzer-Generated}

\setcopyright{cc}
\setcctype{by}
\acmDOI{10.1145/3843750.3843842}
\acmYear{2026}
\copyrightyear{2026}
\acmISBN{979-8-4007-2986-7/2026/10}
\acmConference[LMPL '26]{Proceedings of the 2nd ACM SIGPLAN International Workshop on Language Models and Programming Languages}{October 4--9, 2026}{Oakland, CA, USA}
\acmBooktitle{Proceedings of the 2nd ACM SIGPLAN International Workshop on Language Models and Programming Languages (LMPL '26), October 4--9, 2026, Oakland, CA, USA}
\acmSubmissionID{splashws26lmplmain-p42-p}
\received{2026-07-10}
\received[accepted]{2026-08-14}

\begin{document}

\title{\PaperTitle}

\author{Lara Marinov}
\orcid{0009-0002-6785-4230}
\affiliation{%
  \institution{The University of Texas at Austin}
  \city{Austin}
  \country{USA}
}
\email{marinov@utexas.edu}

\author{Aditya Thimmaiah}
\orcid{0009-0002-1917-7386}
\affiliation{%
  \institution{The University of Texas at Austin}
  \city{Austin}
  \country{USA}
}
\email{auditt@utexas.edu}

\author{Jayanth Srinivasa}
\orcid{0000-0002-7732-8827}
\affiliation{%
  \institution{Cisco Research}
  \city{San Francisco}
  \country{USA}
}
\email{jasriniv@cisco.com}

\author{Junyi Jessy Li}
\orcid{0000-0002-2550-5262}
\affiliation{%
  \institution{The University of Texas at Austin}
  \city{Austin}
  \country{USA}
}
\email{jessy@utexas.edu}

\author{Milos Gligoric}
\orcid{0000-0002-5894-7649}
\affiliation{%
  \institution{The University of Texas at Austin}
  \city{Austin}
  \country{USA}
}
\email{gligoric@utexas.edu}

\begin{abstract}
Large language models (\LLMs) have shown proficiency in various
software engineering tasks, such as code generation and translation.
However, a key limitation in their performance may be their
(lack of) understanding of programming-language semantics.
Even when explicit semantics are given, it remains
unclear whether \LLMs apply those rules or lean on priors learned
during pre-training instead.
We study if LLMs lean on priors or given semantics with a novel
task--Program Executability Prediction (\pcp)--that asks models to
predict whether a program is semantically valid or invalid (and, if
invalid, which formal rule it violates) given the program's syntax and
operational semantics.  Because \pcp requires both valid and invalid
programs, we build a dataset with systematically generated \transfs
derived from valid programs.  We evaluate open-source coding \LLMs
under two semantic formalisms and two semantic shifts across
\humanwrit, \llmtrans, and \fuzzgen program splits.  Our findings show
that \LLMs lean on
pre-training
priors rather than systematically
applying the given rules, performing especially poorly on modified
semantics and degrading further as program complexity increases.
\pcp is available at \url{https://github.com/EngineeringSoftware/prex}.
\end{abstract}



\maketitle

\captionsetup[table]{skip=1mm}
\captionsetup[figure]{skip=2mm}
\captionsetup[subfigure]{skip=2mm}
\setlength{\columnsep}{2mm}%
\setlength{\textfloatsep}{8pt plus 1pt minus 1pt}
\setlength{\intextsep}{6pt plus 1pt minus 1pt}
\setlength{\parskip}{4pt}

\section{Introduction}

\LLMs have become widely used in software engineering, demonstrating remarkable proficiency in tasks like code generation~\cite{nijkampcodegen, longCodercodecompletion, lu-etal-2022-reacccodecompletion}, code translation~\cite{llmcodetranslation, intertranscodetranslation}, bug fixing~\cite{aprbugfixing, liu2024marscodeagentainativeautomated}, and summarizing complex codebases~\cite{11028203codesummarization, 11029737codesummarization, lomshakov-etal-2024-proconsulcodesummarization}.
However, despite their practical success, it remains unclear to what extent these models truly comprehend the underlying semantics of the programs they process.

Much of an \LLM's apparent coding ability seems to come from sophisticated pattern matching and the memorization of vast amounts of training data, rather than an inherent ability to logically simulate program execution~\cite{memorizationinmodels, meeus2026detectingfunctionalmemorizationcode}.
For example, consider a scenario where an \LLM is presented with a standard sorting algorithm.
Because the model has previously encountered this exact structure many times, it can easily summarize its purpose, generate missing lines of code, or predict the final outcome~\cite{storek-etal-2026-sense, euraste2026learnedormemorizedcodellms}.
However, if we introduce a subtle change to the program, such as slightly modifying the loop bounds, obfuscating variable names, or redefining how the comparison operator works, and ask the model to predict the result, it frequently fails~\cite{wang2023naming}.
Instead of systematically tracing the data flow step-by-step as an interpreter would, the model often falls back on its \emph{priors} and simply guesses the standard sorted output~\cite{lamalfa2024codesimulationchallengeslarge, Bernstein_2026}.
Such a situation clearly reveals a gap between rote memorization and genuine, systematic semantic reasoning.

Recent work introduced \dataset~\cite{thimmaiah2026llms} and established a framework for evaluating \LLMses semantic understanding through three tasks: predicting final variable states (PredState), identifying the formal semantic rules applied during execution (PredRule), and generating step-by-step execution traces (PredTrace)~\cite{thimmaiah2026llms}.
Evaluations revealed that current \LLMs struggle significantly across all three of these tasks.
Because \dataset presented a set of complex evaluations where models needed to articulate intermediate steps and output long execution traces, the models' uniformly low performance leaves an open question regarding exactly where their capability threshold is.
To determine whether these failures are in-part due to the natural multi-step complexity of the tasks presented in \dataset or a fundamental inability to reason about semantics, we establish a more foundational baseline that separates trace generation from semantic judgment.
Building on this premise, we present \pcp: a task that asks models, given a program and formal programming language semantics, to predict whether execution succeeds or halts with a semantic error---and, when it fails, which rule was violated.
The required output is deliberately small with just a discrete executability verdict rather than a long trace but the evaluation is not: we pair each valid program with invalid counterparts produced by semantics-aware \transfs, vary program source and structural complexity across three splits, and probe rule-following under alternate semantic formalisms and shifts.
This design lets us ask whether models systematically apply the supplied rules or instead continue to rely on memorized priors.

\MyPara{Dataset}
We build on the valid \C programs from \dataset and extend them with semantically \emph{\transfs} required by \pcp.
Invalid programs are generated from valid programs using a semantics-aware \emph{transformation procedure} that applies one of five errors: \breakloop, \contloop, \divbyzero, \modbyzero, and \varusebdec.
Each valid program has a matched invalid counterpart in every error category.
Programs come from three \emph{splits}---\humanwrit, \llmtrans, and \fuzzgen---with average length and complexity increasing across splits.
Together, these design choices let us test whether semantic reasoning generalizes across program sources and complexity levels, not just memorized coding patterns.

\MyPara{Semantic formalisms}
We evaluate models under two semantic formalisms: small-step operational semantics (\S) and the \KTool (\K).
\S defines execution through finer granularity inference rules, where each rule represents one atomic computation step~\cite{plotkin_small_step_sos}.
\K instead uses coarser rewriting rules that transform program configurations in larger chunks, relying on built-in rewriting to handle intermediate expression reduction~\cite{rosu_kframework}.
We include both to test whether an \LLM's ability to apply provided semantics depends on rule granularity or persists across formalisms.

\MyPara{Semantic shifts}
We also evaluate models under two semantic shifts, \kswap and \kobf, that~alter~the~meaning of selected symbols while keeping execution~defined by the provided formal rules.
\kswap interchanges the meanings of familiar operator symbols (e.g., \texttt{+} denotes subtraction), requiring models to override pre-trained associations and follow the supplied semantics.
\kobf replaces standard operators and keywords with novel single-token symbols, removing recognizable cues from pre-training while preserving the same underlying rules.
Together, these conditions test whether \LLMs can reason from explicitly provided semantics rather than from their prior knowledge.

We evaluate open-source \LLMs designed for coding tasks and enhanced reasoning using both non-chain-of-thought and chain-of-thought (\COT) reasoning.
Our findings indicate that even on the simpler task (\pcp) of predicting program exit codes, \LLMs continue to lean on priors learned during pre-training rather than systematically applying the supplied formal rules.
Performance is especially poor under \kswap and \kobf, where models must override familiar symbol meanings to follow the provided semantics.
Accuracy also degrades significantly on the \llmtrans and \fuzzgen splits compared to \humanwrit, indicating that semantic reasoning does not generalize well as structural complexity increases.

\section{Task Description}

\newenvironment{PEPProgramBox}{%
  \begin{list}{}{%
    \setlength{\labelwidth}{1em}%
    \setlength{\labelsep}{0.45em}%
    \setlength{\leftmargin}{\dimexpr\labelwidth+\labelsep\relax}%
    \setlength{\itemindent}{0pt}%
    \setlength{\listparindent}{0pt}%
    \setlength{\rightmargin}{0pt}%
    \setlength{\topsep}{0pt}%
    \setlength{\partopsep}{0pt}%
    \setlength{\itemsep}{0pt}%
    \setlength{\parsep}{0pt}%
  }%
  \footnotesize
  \ttfamily
  \raggedright
  \renewcommand{\makelabel}[1]{\footnotesize\ttfamily ##1}%
}{%
  \end{list}%
}

\newcommand{\PEPProgramLine}[2]{\item[#1] #2}
\newcommand{\PEPProgramLineIndent}[2]{\item[#1] \hspace{1.5em}#2}

\newcommand{\PEPHighlightError}[1]{%
    \begingroup\setlength{\fboxsep}{0.15ex}\colorbox{red!20}{#1}\endgroup
}

\providecommand{\PEPTaskCodeBoxWidth}{3.8cm}
\providecommand{\PEPTaskCodeBoxHeight}{3.6cm}
\providecommand{\PEPTaskBoxDistance}{0.12cm}

\newlength{\pepStatusIconBugXShift}
\newlength{\pepStatusIconBugYShift}
\setlength{\pepStatusIconBugXShift}{0cm}
\setlength{\pepStatusIconBugYShift}{0.75cm}

\begin{figure}[!t]
  \centering
  \begin{subfigure}[t]{0.49\columnwidth}
    \centering
    \resizebox{\linewidth}{!}{\input{figures/pep-task-example_styles}
\begin{tikzpicture}
\node[pep codecontainervalid] (box1) at (0,0) {%
    \begin{minipage}[t][\PEPTaskCodeBoxHeight][t]{\PEPTaskCodeBoxWidth}%
        \input{figures/pep-task-example-listings/valid-content.tex}%
    \end{minipage}%
};
\node[pep statusiconvalid] at (box1.north east) {\pepStatusIconCheck};
\end{tikzpicture}}%
    \caption{\FigPEPTaskValidCaption}
  \end{subfigure}
  \hfill
  \begin{subfigure}[t]{0.49\columnwidth}
    \centering
    \resizebox{\linewidth}{!}{\input{figures/pep-task-example_styles}
\begin{tikzpicture}
\node[pep codecontainerinvalid] (box1) at (0,0) {%
    \begin{minipage}[t][\PEPTaskCodeBoxHeight][t]{\PEPTaskCodeBoxWidth}%
        \input{figures/pep-task-example-listings/invalid-content.tex}%
    \end{minipage}%
};
\node[pep statusiconinvalid] (pepstatuscross) at (box1.north east) {\pepStatusIconCross};
\node[pep statusiconinvalid] at ([xshift=\pepStatusIconBugXShift, yshift=-\pepStatusIconBugYShift]pepstatuscross.center) {\pepStatusIconBug};
\end{tikzpicture}}%
    \caption{\FigPEPTaskInvalidCaption}
  \end{subfigure}
  \caption{\FigPEPTaskExampleCaption}
\end{figure}

In traditional program interpretation, the semantic analyzer identifies semantic errors in the input program, typically following parsing~\cite{dragon,tiger}.
As a step towards our long term goal--evaluating whether LLMs can function as interpreters--in this paper, we tasked LLMs with detecting semantic errors.

We not only provide the model with a program it is expected to analyze but also supply the formal semantics of the language, including the inference rules that govern valid and invalid executions, thereby encouraging it to reason about semantic correctness.
If an \LLM identifies a program as semantically invalid, it is further instructed to specify the rule that formalizes the error, assessing its understanding of the programming language semantics.
Consequently, we test not only if these models can identify a program as semantically invalid, but also if they lean on priors learned in their training data or on the semantics provided to them.

To illustrate the task, consider the example in Figure~\ref{figure:pep-task-example} of a \C program which calculates the integer square root of the variable $n$.
The program in Figure~\ref{figure:pep-task-valid} is semantically valid and contains no rule violations, so an \LLM successfully performing \pcp should classify it as valid, based on semantic reasoning over the inference rules rather than needing to simulate a full step-by-step execution.
In contrast, Figure~\ref{figure:pep-task-invalid} shows an invalid variant of the same base program.
Because line \PepTaskExampleDivByZeroLine\ of the invalid program contains a division-by-zero error, an \LLM should successfully classify the program as invalid and accurately specify the violated rule.

\section{Methodology}

This section describes our experimental setup for \pcp.
We describe the \C programming language (Section~\ref{subsec:imp-language}), the semantic formalisms used to specify program behavior (\S and \K; Section~\ref{subsec:semantic-formalisms}), semantic shifts that test whether models follow provided rules or pre-training priors (\kswap and \kobf; Section~\ref{subsec:semantic-shifts}), and the full prompt given to models (Section~\ref{subsec:full-prompt}).
The \C language, semantic formalisms, and semantics shifts are introduced in \dataset~\cite{thimmaiah2026llms}, but we recap them here together with the full prompt used for \pcp.

\subsection{\C Language}\label{subsec:imp-language}

We evaluate models on programs written in \C, a small imperative language with C-like syntax.
The complete \ebnf grammar used in our experiments is given in Figure~\ref{figure:imp-syntax}.
We use the same \C language as \dataset.

\begin{figure}[t]
\begin{lstlisting}[language=bnf-grammar-tiny,
    basicstyle=\scriptsize\ttfamily,
    numberstyle=\scriptsize,
    xleftmargin=14pt]
<program>   ::= <stmt_list>
<stmt_list> ::= (<stmt> ';')*
<stmt>      ::= 'int' <id>
            | <id> '=' <aexp> 
            | 'if' '(' <bexp> ')' '{' <stmt_list> '}' 'else' '{' <stmt_list> '}'
            | 'while' '(' <bexp> ')' '{' <stmt_list> '}'
            | 'loop' '(' <bexp> ')' '{' <stmt_list> '}'
            | 'halt'
            | 'continue'
            | 'break'
            | 'LE'
<aexp>      ::= <id>
            | <literal>
            | '(' <aexp>? <mathop> <aexp> ')'
<bexp>      ::= '(' <bool> ')'
            | '(' <aexp> <relop> <aexp> ')'
            | '(' <lognot> <bexp> ')'
            | '(' <bexp> <logicalop> <bexp> ')'
<bool>      ::= 'true' | 'false'
<mathop>    ::= '+' | '-' | '*' | '/' | '%'
<relop>     ::= '<' | '<=' | '>' | '>=' | '==' | '!='
<lognot>    ::= '!'
<logicalop> ::= '&&' | '||'
<id>        ::= <letter> (<letter> | <digit>)*
<literal>   ::= <digit>+
\end{lstlisting}
  \caption{\FigIMPEBNFCaption}
\end{figure}

\subsection{Semantic Formalisms}\label{subsec:semantic-formalisms}

We evaluate the models using two distinct semantic formalisms: Small-step operational semantics (\S) and the \KTool (\K).
We use both formalisms to understand how the granularity and style of formal rules affect an \LLM's ability to correctly reason about the programs it is given and to discover if presenting the semantic rules in a particular formulization maximizes the model's adherence to them.

\MyPara{Small-step operational semantics (\S)}
\S is a fine-grained semantic formalism, where each rule represents one atomic computation~\cite{plotkin_small_step_sos}.
Execution happens through repeated rule applications as expressions are broken down into tiny constituent parts and resolved one by one.
Rules are written in Gentzen-style inference notation~\cite{gentzen_notation}, with premises and side conditions written above the bar and conclusions below.

Table~\ref{table:imp-sos-rules} shows a subset (due to the space reasons) of the \S rules used in our experiments.
To illustrate how the \S formalism handles both valid execution and error states, we highlight several key rules.
Rules 1 and 2 form a complementary pair describing variable lookup: Rule 1 defines successful evaluation when a variable is bound in the current state, while Rule 2 states that attempting to evaluate an undeclared variable results in an error.
Similarly, Rules 18 and 19 describe the division operation, distinguishing between valid integer division and a division-by-zero error.
Rule 9 highlights the fine-grained, small-step nature of the formalism by explicitly defining the addition transition only after both operands have been fully reduced to values.

\MyPara{\KTool}
The \KTool is a coarser, rewriting-based framework~\cite{rosu_kframework,KRepo}.
\KTool semantics evaluate code in slightly larger, higher-level chunks rather than atomic steps.
Table~\ref{table:imp-k-rules} shows a subset of the \KTool rules used in our experiments, paralleling the \S rules.

Like the \S formalism, Rules 1 and 2 from \K form a pair describing variable lookup.
Rule 1 defines successful evaluation when a variable maps to a value in the current program state, and Rule 2 states that attempting to evaluate an undeclared variable stops execution.
Likewise, Rules 6 and 7 (Table~\ref{table:imp-k-rules}) parallel \S Rules 18 and 19 (Table~\ref{table:imp-sos-rules}) to handle division.
Rule 6 defines valid integer division when the right operand is not zero, and Rule 7 defines the error for division by zero.

\begin{table}[t]
  \centering
  \begin{scriptsize}%
  \caption{\FigIMPSOSRulesSubsetCaption}
  \begin{tabularx}{\columnwidth}{@{}>{\raggedright\arraybackslash\hspace{0pt}}p{0.09\columnwidth}>{\centering\arraybackslash\hspace{0pt}}X>{\raggedright\arraybackslash}m{0.32\columnwidth}@{}}
   \toprule%
   \textbf{Rule} & \textbf{Formalization} & \textbf{Description}\\
   \midrule%
Rule 1 & \begin{minipage}[c]{\hsize}\begin{mathpar}\inferrule{\sigma(\texttt{x}) = \texttt{v}}{\langle \texttt{x},\sigma,\chi\rangle \to \texttt{v}}\end{mathpar}\end{minipage} & If a variable is associated with a value in the current program state, then evaluating it yields that value.\\\hline
Rule 2 & \begin{minipage}[c]{\hsize}\begin{mathpar}\inferrule{\sigma(\texttt{x}) = \bot}{\langle \texttt{x},\sigma,\chi\rangle \to \langle \texttt{ERROR},\sigma,\chi\rangle}\end{mathpar}\end{minipage} & If a variable has no associated value in the current program state, then attempting to evaluate it results in an error and the program execution stops immediately.\\\hline
Rule 9 & \begin{minipage}[c]{\hsize}\begin{mathpar}\inferrule{\texttt{v3} = \texttt{v1 + v2}}{\langle \texttt{v1 + v2},\sigma,\chi\rangle \to \texttt{v3}}\end{mathpar}\end{minipage} & When both operands of a `+` expression are values, the result is obtained by adding those two values.\\\hline
Rule 18 & \begin{minipage}[c]{\hsize}\begin{mathpar}\inferrule{\texttt{v2} \neq \texttt{0} \\ \texttt{v3} = \texttt{v1 / v2}}{\langle \texttt{v1 / v2},\sigma,\chi\rangle \to \texttt{v3}}\end{mathpar}\end{minipage} & When both operands of a `/` expression are values and the right operand value is not zero, the result is obtained by integer division of the left operand value by the right operand value.\\\hline
Rule 19 & \begin{minipage}[c]{\hsize}\begin{mathpar}\inferrule{\texttt{v2} = \texttt{0}}{\langle \texttt{v1 / v2},\sigma,\chi\rangle \to \langle \texttt{ERROR},\sigma,\chi\rangle}\end{mathpar}\end{minipage} & When both operands of a `/` expression are values and the right operand value is zero, then an error occurs and the program execution stops immediately.\\
   \bottomrule%
  \end{tabularx}%
  \end{scriptsize}%
\end{table}

\begin{figure}[t]
  \centering
  \resizebox{\linewidth}{!}{%
  \begin{tikzpicture}[
    innerbox/.style={
      draw=black!75,
      rounded corners=1.5pt,
      line width=0.5pt,
      align=left,
      inner sep=3pt,
      text width=2.05cm,
      minimum height=0.9cm,
      font=\tiny\ttfamily,
    },
    innerboxtop/.style={
      innerbox,
      minimum height=1.15cm,
    },
    outerbox/.style={
      draw=black!85,
      rounded corners=2pt,
      line width=0.6pt,
      inner sep=6pt,
    },
    innerlabel/.style={
      font=\tiny\bfseries,
      anchor=south,
      align=center,
    },
    mainlabel/.style={
      font=\scriptsize\bfseries,
      anchor=south,
      align=center,
    },
    arrow/.style={
      -{Stealth[length=2.8mm, width=1.8mm]},
      line width=1.1pt,
      draw=black!50,
    },
  ]
    \def\innerxgap{0.22cm}
    \def\innerygap{0.55cm}
    \def\outputorvspace{0.12cm}
    \def\outputxgap{0.62cm}

    \node[innerboxtop, fill=white, anchor=north west] (prog) at (0,0) {%
      int a;\\
      int b;\\
      a = 3 / b;
    };
    \node[innerlabel] (proglabel) at ([yshift=0.03cm]prog.north) {Program};

    \node[innerboxtop, fill=white, anchor=north west] (ebnf) at ([xshift=\innerxgap]prog.north east) {%
      \ldots\\
      \textless stmt\textgreater{} ::= `int' \textless id\textgreater{}\\
      \textbar{} \textless id\textgreater{} `=' \textless aexp\textgreater{}\\
      \textbar{} `if' `(' \textless bexp\textgreater{} `)' \ldots
    };
    \node[innerlabel] (ebnflabel) at ([yshift=0.03cm]ebnf.north) {\C{} syntax in \ebnf};

    \coordinate (toprow-bottom) at (ebnf.south -| prog.south west);

    \node[innerbox, fill=white, anchor=north west] (sbox) at ([yshift=-\innerygap]toprow-bottom) {%
      \ldots\\
      Rule 19:\\
      \hspace*{0.5em}v2 = 0\\
      \hspace*{0.5em}---------\\
      \hspace*{0.5em}\textless v1/v2,\ldots\textgreater{} $\to$ ERROR \ldots
    };
    \node[innerlabel] (slabel) at ([yshift=0.03cm]sbox.north) {\S};

    \node[innerbox, fill=white, anchor=north west] (kbox) at ([xshift=\innerxgap]sbox.north east) {%
      \ldots\\
      rule [7]:\\
      \hspace*{0.5em}\textless k\textgreater{} I1 / I2 $\Rightarrow$ ERROR\\
      \hspace*{0.5em}requires I2 = 0 \ldots
    };
    \node[innerlabel] (klabel) at ([yshift=0.03cm]kbox.north) {\K};

    \node[font=\tiny\bfseries, anchor=center] (or) at ($(sbox.east)!0.5!(kbox.west)$) {or};

    \begin{scope}[on background layer]
      \node[outerbox, fill=blue!6, fit=(prog) (ebnf) (sbox) (kbox) (or) (proglabel) (ebnflabel) (slabel) (klabel)] (input) {};
    \end{scope}
    \node[mainlabel] at ([yshift=0.04cm]input.north) {Prompt input};

    \coordinate (output-west) at ([xshift=\outputxgap]input.east);
    \coordinate (output-mid) at ([xshift=1.025cm]output-west |- input.center);

    \node[font=\tiny\bfseries, anchor=center] (or-out) at (output-mid) {or};

    \node[innerbox, fill=white, anchor=south west, minimum height=0.55cm] (success) at ([yshift=\outputorvspace]or-out.north -| output-west) {%
      \texttt{\#\#success\#\#}
    };

    \node[innerbox, fill=white, anchor=north west, minimum height=0.75cm] (errorbox) at ([yshift=-\outputorvspace]or-out.south -| output-west) {%
      \texttt{\#\#error\#\#}\\
      + violated rule
    };

    \begin{scope}[on background layer]
      \node[outerbox, fill=olive!10, fit=(success) (errorbox) (or-out)] (output) {};
    \end{scope}
    \node[mainlabel] at ([yshift=0.04cm]output.north) {Model output};

    \draw[arrow] (input.east) -- (output.west |- input.center);
  \end{tikzpicture}%
  }
  \vspace{2mm}
  \caption{\FigPCPPromptOverviewCaption}
  \vspace{3.5mm}
\end{figure}

The primary difference between the two formalisms is how they handle intermediate computations, which is best illustrated by the addition operation.
In the \S formalism, Rule 9 dictates the addition of two values, but its strictly small-step nature means this rule only applies after separate, explicitly defined atomic steps have reduced the operands into discrete values first.
In contrast, the \K semantics uses its built-in rewriting to handle this operand reduction implicitly.
So, the corresponding Rule 3 for \K encapsulates the addition evaluation into a single, coarser rewrite step, evaluating the expression once both operands are values.


\subsection{Semantic Shifts}\label{subsec:semantic-shifts}
We use two semantic shifts, \kswap and \kobf, to test whether models can trace programs using explicitly provided rules rather than relying on their prior knowledge.

\MyPara{\kswap}
\kswap reassigns the semantic meanings of selected operator symbols to their counterparts (Table~\ref{tab:mutation-rules-combined}).
For example, addition (+) and subtraction (-) are interchanged, such that the + symbol dictates subtraction.
To reason correctly, models must strictly condition on the provided rules and override pre-trained semantic associations.

\begin{table}[t]
  \centering
  \begin{scriptsize}%
  \caption{\FigIMPKRulesSubsetCaption}
  \setlength{\tabcolsep}{2pt}%
  \begin{tabularx}{\columnwidth}{@{}>{\raggedright\arraybackslash\hspace{0pt}}p{0.09\columnwidth}>{\raggedright\arraybackslash\hspace{0pt}}X>{\raggedright\arraybackslash}m{0.32\columnwidth}@{}}
   \toprule%
   \textbf{Rule} & \textbf{Formalization} & \textbf{Description}\\
   \midrule%
Rule 1 & \begin{minipage}[c]{\hsize}\raggedright \hspace*{1em}\textless{}k\textgreater{} X:Id $\Rightarrow$ I $\ldots$ \textless{}/k\textgreater{} \newline \hspace*{1em}\textless{}state\textgreater{} $\ldots$ X $\mapsto$ I $\ldots$ \textless{}/state\textgreater{} \newline \hspace*{1em}\textless{}rules\textgreater{} L $\Rightarrow$ L ListItem(``Rule 1'') \textless{}/rules\textgreater{}\end{minipage} & If a variable is associated with a value in the current program state, then evaluating it yields that value.\\\hline
Rule 2 & \begin{minipage}[c]{\hsize}\raggedright \hspace*{1em}\textless{}k\textgreater{} X:Id $\Rightarrow$ \{ERROR\} \textless{}/k\textgreater{} \newline \hspace*{1em}\textless{}state\textgreater{} Rho:Map \textless{}/state\textgreater{} \newline \hspace*{1em}\textless{}rules\textgreater{} L $\Rightarrow$ L ListItem(``Rule 2'') \textless{}/rules\textgreater{} \newline \hspace*{1em}requires notBool (X in\_keys(Rho))\end{minipage} & If a variable has no associated value in the current program state, then attempting to evaluate it results in an error and the program execution stops immediately.\\\hline
Rule 3 & \begin{minipage}[c]{\hsize}\raggedright \hspace*{1em}\textless{}k\textgreater{} I1 \{PLUS\_OP\} I2 $\Rightarrow$ I1 $+\ \mathrm{Int}$ I2 $\ldots$ \textless{}/k\textgreater{} \newline \hspace*{1em}\textless{}rules\textgreater{} L $\Rightarrow$ L ListItem(``Rule 3'') \textless{}/rules\textgreater{}\end{minipage} & When both operands of a `+` expression are values, the result is obtained by adding those two values.\\\hline
Rule 6 & \begin{minipage}[c]{\hsize}\raggedright \hspace*{1em}\textless{}k\textgreater{} I1 \{DIV\_OP\} I2 $\Rightarrow$ I1 $/\ \mathrm{Int}$ I2 $\ldots$ \textless{}/k\textgreater{} \newline \hspace*{1em}\textless{}rules\textgreater{} L $\Rightarrow$ L ListItem(``Rule 6'') \textless{}/rules\textgreater{} \newline \hspace*{1em}requires I2 $\neq \mathrm{Int}$ 0\end{minipage} & When both operands of a `/` expression are values and the right operand value is not zero, the result is obtained by integer division of the left operand value by the right operand value.\\\hline
Rule 7 & \begin{minipage}[c]{\hsize}\raggedright \hspace*{1em}\textless{}k\textgreater{} I1 \{DIV\_OP\} I2 $\Rightarrow$ \{ERROR\} $\ldots$ \textless{}/k\textgreater{} \newline \hspace*{1em}\textless{}rules\textgreater{} L $\Rightarrow$ L ListItem(``Rule 7'') \textless{}/rules\textgreater{} \newline \hspace*{1em}requires I2 $= \mathrm{Int}$ 0\end{minipage} & When both operands of a `/` expression are values and the right operand value is zero, then an error occurs and the program execution stops immediately.\\
   \bottomrule%
  \end{tabularx}%
  \end{scriptsize}%
  \vspace{3mm}
\end{table}

\MyPara{\kobf}
\kobf replaces standard operator and keyword symbols with novel single-token symbols (Table~\ref{tab:mutation-rules-combined}).
For instance, an expression using the \KOADD{} symbol is executed as addition.
The \kobf condition ensures models can apply explicitly defined rules in the absence of recognizable patterns from prior exposure.
Unlike \dataset whose symbols are tokenized into multiple tokens by \LLMs, we use a different set of symbols that are each one token.
Ensuring each novel symbol maps to a single token prevents artificial inflation of the prompt length, reduces overall token overhead, and minimizes the risk of introducing unintended syntactic complexities that might confuse the model.

\subsection{Full Prompt}\label{subsec:full-prompt}
For each program, we provide the model with the \C syntax in \ebnf, the semantic rules (\S or \K), and the program itself.
Figure~\ref{figure:pcp-prompt-overview} summarizes this prompt structure.
The model must predict whether the program is executable (\texttt{\#\#success\#\#}) or semantically invalid (\texttt{\#\#error\#\#}), and when invalid, identify the violated rule.
For non-reasoning models, we evaluate both direct-answer and \COT using only variants with minor changes to this prompt (\COT is told to provide the reasoning).

\clearpage

\begin{strip}
  \begin{center}
  \centering%
  \TableFont%
  \captionof{table}{\TableNonStandardRulesCaption}%
  \begin{threeparttable}
   \setlength{\tabcolsep}{4pt}%
   \renewcommand{\arraystretch}{1.1}
   \begin{tabular}{l c ccccc cccccc ccc ccccc}%
     \toprule%
     \UseMacro{Caucasiantitle1} &
     \multicolumn{1}{c}{\UseMacro{CaucasianAssignment}} &
     \multicolumn{5}{c}{\UseMacro{CaucasianArithmetic}} &
     \multicolumn{6}{c}{\UseMacro{CaucasianRelational}} &
     \multicolumn{3}{c}{\UseMacro{CaucasianLogical}} &
     \multicolumn{5}{c}{\UseMacro{CaucasianKeyword}} \\
     \cmidrule(lr){1-1}
     \cmidrule(lr){2-2}
     \cmidrule(lr){3-7}
     \cmidrule(lr){8-13}
     \cmidrule(lr){14-16}
     \cmidrule(lr){17-21}
     \UseMacro{Caucasiantitle2} &
     \texttt{=} &
     \texttt{+} & 
     \texttt{-} &
     \texttt{*} &
     \texttt{/} &
     \texttt{\%} &

     \texttt{<} &
     \texttt{<=} &
     \texttt{>} &
     \texttt{>=} &
     \texttt{==} &
     \texttt{!=} &

     \texttt{!} &
     \texttt{\&\&} &
     \texttt{||} &

     \texttt{if-else} &
     \texttt{while} &
     \texttt{halt} &
     \texttt{break} &
     \texttt{continue}\\
     \UseMacro{Caucasiantitle4}\tnote{*} &
     \texttt{=} &
     \texttt{-} & 
     \texttt{+} &
     \texttt{/} &
     \texttt{*} &
     \texttt{\%} &

     \texttt{>} &
     \texttt{>=} &
     \texttt{<} &
     \texttt{<=} &
     \texttt{!=} &
     \texttt{==} &

     \texttt{!} &
     \texttt{||} &
     \texttt{\&\&} &

     \texttt{if-else} &
     \texttt{while} &
     \texttt{halt} &
     \texttt{break} &
     \texttt{continue}\\
     \UseMacro{Caucasiantitle3}\tnote{**} &
     \KOASSIGN &
     \KOADD &
     \KOSUB &
     \KOMUL &
     \KODIV &
     \KOMOD &

     \KOLT &
     \KOLTEQ &
     \KOGT &
     \KOGTEQ &
     \KOEQ &
     \KONEQ &

     \KONOT &
     \KOAND &
     \KOOR &

     \KOIF-\KOELSE &
     \KOWHILE &
     \KOHALT &
     \KOBREAK &
     \KOCONTINUE\\
     \bottomrule%
   \end{tabular}
   \begin{tablenotes}
     \scriptsize
     \item * Swaps the semantics of standard operator/keyword symbols; ** Assigns semantics of standard operators/keywords to novel single-token symbols.
   \end{tablenotes}
  \end{threeparttable}
  \vspace{2mm}%
\end{center}

\end{strip}

\section{Dataset Construction}

The set of programs used in our evaluation are comprised of three splits: \humanwrit, \llmtrans, and \fuzzgen.
We use the same valid programs as \dataset and extend their dataset with invalid programs, because \pcp requires both valid and invalid programs.
We obtain these invalid programs by applying semantics-aware transformations to valid programs, so that each invalid counterpart is derived directly from a matched valid one with consistent style and complexity.

\MyPara{\humanwrit}
The \humanwrit split of \C programs are manually adapted from C++ programs sourced from LeetCode, HumanEval, CodeContests, and MBPP~\citep{leetcode,humanEval,codegeex,codecontest,mbpp}.
For each program, a single public test case and its corresponding oracle is selected.

\MyPara{\llmtrans}
The \llmtrans split of \C programs are C++ programs translated to \C by LLMs. The C++ programs are sourced from CodeForces solutions published to HuggingFace~\citep{penedo2025codeforces}.
\qwenCoder{32} is prompted with the \C syntax, semantics, the C++ solution, and one public test case to generate a valid \C program, and answers are filtered for successful test execution using the \KTool.

\MyPara{\fuzzgen}
The \fuzzgen split of \C programs are constructed with a depth-controlled, semantics-aware, grammar-based fuzzer~\citep{YangCsmith,Han2019CodeAlchemist}.
The fuzzer picks statements from assign, if-else, while, break, continue, halt using depth-tapered probabilites to reduce the chance of generating new nested \texttt{if/while} blocks.

\begin{table}[t]
  \centering

\TableFont
\caption{\UseMacro{TCap-imp-split-loc-stats}}
\label{tab:imp-split-loc-stats}
\setlength{\tabcolsep}{3pt}
\renewcommand{\arraystretch}{1.1}
\begin{tabular}{l r r r r r r r}
\toprule
 & & \multicolumn{3}{c}{\textbf{\UseMacro{THead-imp-split-loc-loc}}} & \multicolumn{3}{c}{\textbf{\UseMacro{THead-imp-split-loc-tokens}}}
\\
\cmidrule(lr){3-5}
\cmidrule(lr){6-8}
\textbf{\UseMacro{THead-imp-split-loc-split}} & \textbf{\UseMacro{THead-imp-split-loc-num-programs}} & \textbf{\UseMacro{THead-imp-split-loc-min}} & \textbf{\UseMacro{THead-imp-split-loc-median}} & \textbf{\UseMacro{THead-imp-split-loc-max}} & \textbf{\UseMacro{THead-imp-split-loc-min}} & \textbf{\UseMacro{THead-imp-split-loc-median}} & \textbf{\UseMacro{THead-imp-split-loc-max}}
\\
\midrule
\UseMacro{THead-imp-split-loc-human-written}
 & \UseMacro{dataset-imp-split-loc-human-written-num-programs} & \UseMacro{dataset-imp-split-loc-human-written-min-loc} & \UseMacro{dataset-imp-split-loc-human-written-median-loc} & \UseMacro{dataset-imp-split-loc-human-written-max-loc} & \UseMacro{dataset-imp-split-loc-human-written-min-tokens} & \UseMacro{dataset-imp-split-loc-human-written-median-tokens} & \UseMacro{dataset-imp-split-loc-human-written-max-tokens}
\\
\UseMacro{THead-imp-split-loc-synthetic-cpp}
 & \UseMacro{dataset-imp-split-loc-synthetic-cpp-num-programs} & \UseMacro{dataset-imp-split-loc-synthetic-cpp-min-loc} & \UseMacro{dataset-imp-split-loc-synthetic-cpp-median-loc} & \UseMacro{dataset-imp-split-loc-synthetic-cpp-max-loc} & \UseMacro{dataset-imp-split-loc-synthetic-cpp-min-tokens} & \UseMacro{dataset-imp-split-loc-synthetic-cpp-median-tokens} & \UseMacro{dataset-imp-split-loc-synthetic-cpp-max-tokens}
\\
\UseMacro{THead-imp-split-loc-fuzzer-generated}
 & \UseMacro{dataset-imp-split-loc-fuzzer-generated-num-programs} & \UseMacro{dataset-imp-split-loc-fuzzer-generated-min-loc} & \UseMacro{dataset-imp-split-loc-fuzzer-generated-median-loc} & \UseMacro{dataset-imp-split-loc-fuzzer-generated-max-loc} & \UseMacro{dataset-imp-split-loc-fuzzer-generated-min-tokens} & \UseMacro{dataset-imp-split-loc-fuzzer-generated-median-tokens} & \UseMacro{dataset-imp-split-loc-fuzzer-generated-max-tokens}
\\
\bottomrule
\end{tabular}
  \vspace{2mm}
\end{table}

Table~\ref{tab:imp-split-loc-stats} shows program length across the three
splits. Median lines of code (LOC) increases from \humanwrit to \llmtrans to \fuzzgen. Even
the shortest \fuzzgen program is longer than the longest \humanwrit program.
%
%
%
We start with \BaseIMPDataset valid programs\footnotemark[8]. These programs are transformed to create \InvalidIMPDataset semantically invalid programs.
The \transformationprocess uses an ANTLR-based parser-visitor to apply the five predefined transformations to each valid program, corresponding to the five semantic error rules under \S and \K formalizations.
These transformations introduce specific semantic errors: 1)~inserting a break outside a loop, 2)~inserting a continue outside a loop, 3)~division by zero, 4)~modulo by zero, and 5)~using an undeclared variable.
Each invalid program is generated by applying one random \transformation rule to a valid \C program, ensuring each violates exactly one semantic error rule.
For each valid program, we generate one corresponding semantically invalid program in each of the five error categories.
Figure~\ref{figure:valid-invalid-example} illustrates these transformations on a shared valid program.

\begin{table}[t]
  \centering

\TableFont
\caption{\UseMacro{TCap-program-executability-stats}}
\label{tab:program-executability-stats}
\setlength{\tabcolsep}{2pt}
\resizebox{\linewidth}{!}{%
\begin{tabular}{l l >{\centering\arraybackslash}p{0.5in} >{\centering\arraybackslash}p{0.5in} r r}
\toprule
\multirow{2}{*}{\textbf{\UseMacro{THead-executability}}} & \multirow{2}{*}{\textbf{\UseMacro{THead-semantic-error-type}}} & \multicolumn{2}{c}{\textbf{\UseMacro{THead-corresponding-rule}}} & \multirow{2}{*}{\textbf{\UseMacro{THead-count}}} & \multirow{2}{*}{\textbf{\UseMacro{THead-percentage}}}
\\
\cmidrule(lr){3-4}
 &  & $\mathbb{K}$ & $\mathbb{S}$ &  & 
\\
\midrule
\UseMacro{THead-success}
 & \UseMacro{THead-none}
 & \UseMacro{dataset-program-executability-success-rule-k} & \UseMacro{dataset-program-executability-success-rule-s} & \UseMacro{dataset-program-executability-success-count} & \UseMacro{dataset-program-executability-success-percent}
\\
\UseMacro{THead-error}
 & \UseMacro{THead-semantic-error-break-outside-loop}
 & \UseMacro{dataset-program-executability-break-outside-loop-rule-k} & \UseMacro{dataset-program-executability-break-outside-loop-rule-s} & \UseMacro{dataset-program-executability-break-outside-loop-count} & \UseMacro{dataset-program-executability-break-outside-loop-percent}
\\
\UseMacro{THead-error}
 & \UseMacro{THead-semantic-error-continue-outside-loop}
 & \UseMacro{dataset-program-executability-continue-outside-loop-rule-k} & \UseMacro{dataset-program-executability-continue-outside-loop-rule-s} & \UseMacro{dataset-program-executability-continue-outside-loop-count} & \UseMacro{dataset-program-executability-continue-outside-loop-percent}
\\
\UseMacro{THead-error}
 & \UseMacro{THead-semantic-error-divide-by-zero}
 & \UseMacro{dataset-program-executability-divide-by-zero-rule-k} & \UseMacro{dataset-program-executability-divide-by-zero-rule-s} & \UseMacro{dataset-program-executability-divide-by-zero-count} & \UseMacro{dataset-program-executability-divide-by-zero-percent}
\\
\UseMacro{THead-error}
 & \UseMacro{THead-semantic-error-modulo-zero}
 & \UseMacro{dataset-program-executability-modulo-zero-rule-k} & \UseMacro{dataset-program-executability-modulo-zero-rule-s} & \UseMacro{dataset-program-executability-modulo-zero-count} & \UseMacro{dataset-program-executability-modulo-zero-percent}
\\
\UseMacro{THead-error}
 & \UseMacro{THead-semantic-error-var-use-before-declare}
 & \UseMacro{dataset-program-executability-var-use-before-declare-rule-k} & \UseMacro{dataset-program-executability-var-use-before-declare-rule-s} & \UseMacro{dataset-program-executability-var-use-before-declare-count} & \UseMacro{dataset-program-executability-var-use-before-declare-percent}
\\
\bottomrule
\end{tabular}%
}
  \vspace{2mm}
\end{table}

\footnotetext[8]{The original \dataset set had \OriginalValidProgs valid programs, but we removed one \fuzzgen program because its full \pcp prompt (\C syntax, semantic rules, and program) requires $\sim$37K tokens and exceeds the 32K context window of our \qwenCoderGeneral models, leaving \BaseIMPDataset semantically valid programs.}

\providecommand{\PCPFuzzListing}{%
  \lstset{language=imp-pretty-no-lines,basicstyle=\tiny\ttfamily,breaklines=true}%
}

\providecommand{\PCPFuzzFigureListing}{%
  \lstset{%
    language=imp-pretty-no-lines,%
    basicstyle=\Large\ttfamily,%
    breaklines=true,%
    columns=fullflexible,%
    aboveskip=0pt,%
    belowskip=0pt,%
    xleftmargin=0pt,%
    xrightmargin=0pt,%
    framesep=0pt,%
  }%
}

\providecommand{\PCPFuzzCodeBoxWidth}{4.8cm}
\providecommand{\PCPFuzzCodeBoxHeight}{4.6cm}

\providecommand{\PCPFuzzTitleTopShift}{8pt}        
\providecommand{\PCPFuzzTitleToCodeGap}{6pt}       
\providecommand{\PCPFuzzInnerBoxTopAdjust}{10pt}    

\providecommand{\PCPFuzzInvalidBoxDistance}{0.10cm}

\providecommand{\PCPFuzzLstInput}[1]{%
  \PCPFuzzListing
  \lstinputlisting{#1}%
}

\PCPFuzzFigureListing

\newsavebox{\PCPFuzzTitleHeightBox}
\newlength{\PCPFuzzInnerBoxTopOffset}
\begin{lrbox}{\PCPFuzzTitleHeightBox}
{\Large\bfseries Ay}
\end{lrbox}
\settoheight{\PCPFuzzInnerBoxTopOffset}{\usebox{\PCPFuzzTitleHeightBox}}
\addtolength{\PCPFuzzInnerBoxTopOffset}{\dimexpr\PCPFuzzTitleTopShift+\PCPFuzzTitleToCodeGap+\PCPFuzzInnerBoxTopAdjust\relax}

\newsavebox{\PCPFuzzBoxValid}
\begin{lrbox}{\PCPFuzzBoxValid}
\begin{lstlisting}
int a;
int b;
int ans;
a = 3045;
b = 1078;
ans = (a + b);
\end{lstlisting}
\end{lrbox}

\newsavebox{\PCPFuzzBoxBreakOutsideLoop}
\begin{lrbox}{\PCPFuzzBoxBreakOutsideLoop}
\begin{lstlisting}
int a;
int b;
int ans;
a = 3045;
b = 1078;
+(*@\colorbox{green!20}{break;}@*)
ans = (a + b);
\end{lstlisting}
\end{lrbox}

\newsavebox{\PCPFuzzBoxContinueOutsideLoop}
\begin{lrbox}{\PCPFuzzBoxContinueOutsideLoop}
\begin{lstlisting}
int a;
int b;
int ans;
+(*@\colorbox{green!20}{continue;}@*)
a = 3045;
b = 1078;
ans = (a + b);
\end{lstlisting}
\end{lrbox}

\newsavebox{\PCPFuzzBoxDivideByZero}
\begin{lrbox}{\PCPFuzzBoxDivideByZero}
\begin{lstlisting}
int a;
int b;
int ans;
-(*@\colorbox{red!20}{a = 3045;}@*)
+(*@\colorbox{green!20}{a = (3045 / 0);}@*)
b = 1078;
ans = (a + b);
\end{lstlisting}
\end{lrbox}

\newsavebox{\PCPFuzzBoxModuloZero}
\begin{lrbox}{\PCPFuzzBoxModuloZero}
\begin{lstlisting}
int a;
int b;
int ans;
+(*@\colorbox{green!20}{ans = 0;}@*)
-(*@\colorbox{red!20}{a = 3045;}@*)
+(*@\colorbox{green!20}{a = (3045 \% ans);}@*)
b = 1078;
ans = (a + b);
\end{lstlisting}
\end{lrbox}

\newsavebox{\PCPFuzzBoxVarUseBeforeDeclare}
\begin{lrbox}{\PCPFuzzBoxVarUseBeforeDeclare}
\begin{lstlisting}
int a;
int b;
int ans;
a = 3045;
b = 1078;
-(*@\colorbox{red!20}{ans = (a + b);}@*)
+(*@\colorbox{green!20}{ans = (y + b);}@*)
\end{lstlisting}
\end{lrbox}

\newcommand{\createprogrambox}[2]{%
    \ifnum#1=1\relax
        \node[programboxvalid] (box#1) at (0,0) {};
    \else
        \ifnum#1=2\relax
            \node[programboxinvalid, right=of box1] (box#1) {};
        \else
            \node[programboxinvalid, right=\PCPFuzzInvalidBoxDistance of box\the\numexpr#1-1\relax] (box#1) {};
        \fi
    \fi
    \node[title, text width=\PCPFuzzCodeBoxWidth] (box#1-title) at ([yshift=-\PCPFuzzTitleTopShift]box#1.north) {#2};
    \node[codecontainer, anchor=north] (box#1-inner) at ([yshift=-\PCPFuzzInnerBoxTopOffset]box#1.north -| box#1.center) {%
        \begin{minipage}[t][\PCPFuzzCodeBoxHeight][t]{\PCPFuzzCodeBoxWidth}%
            \ifnum#1=1\relax \usebox{\PCPFuzzBoxValid}\fi
            \ifnum#1=2\relax \usebox{\PCPFuzzBoxBreakOutsideLoop}\fi
            \ifnum#1=3\relax \usebox{\PCPFuzzBoxContinueOutsideLoop}\fi
            \ifnum#1=4\relax \usebox{\PCPFuzzBoxDivideByZero}\fi
            \ifnum#1=5\relax \usebox{\PCPFuzzBoxModuloZero}\fi
            \ifnum#1=6\relax \usebox{\PCPFuzzBoxVarUseBeforeDeclare}\fi
        \end{minipage}%
    };
}

\begin{figure*}
  \centering
  \resizebox{\textwidth}{!}{\begin{tikzpicture}[
    node distance=0.25cm,
    programbox/.style={
        rounded corners=6pt,
        inner sep=10pt,
        minimum height=6.3cm,
        minimum width=5.5cm,
        anchor=north west
    },
    programboxvalid/.style={
        programbox,
        fill=green!10
    },
    programboxinvalid/.style={
        programbox,
        fill=red!10
    },
    title/.style={
        font=\Large\bfseries,
        text=black,
        align=center,
        anchor=north
    },
    codecontainer/.style={
        draw=gray!100,
        line width=0.8pt,
        fill=white,
        inner sep=5pt,
        rounded corners=4pt,
        anchor=north
    }
]

\createprogrambox{1}{\UseMacro{THead-pcp-fuzzing-valid}}
\createprogrambox{2}{\UseMacro{THead-semantic-error-break-outside-loop}}
\createprogrambox{3}{\UseMacro{THead-semantic-error-continue-outside-loop}}
\createprogrambox{4}{\UseMacro{THead-semantic-error-divide-by-zero}}
\createprogrambox{5}{\UseMacro{THead-semantic-error-modulo-zero}}
\createprogrambox{6}{Var use before declare}

\usetikzlibrary{backgrounds}
\begin{scope}[on background layer]
\node[
    fill=red!40!white,
    draw=none,
    rounded corners=6pt,
    inner sep=2pt,
    fit=(box2) (box3) (box4) (box5) (box6),
] (invalid-variant-group) {};
\end{scope}

\end{tikzpicture}}%
  \caption{\FigValidInvalidExampleCaption}
\end{figure*}

Table~\ref{tab:program-executability-stats} shows the type of semantic errors
introduced, their corresponding rules under \S and \K formalizations, 
and the number of programs in the dataset that violate each rule. The 
percentages are computed over the full set of \IMPPCPDataset programs.

\section{Experiments and Results}
\label{sec:results}

We benchmark models under three settings by providing: the \standardSem semantics and the unmodified program,
\keywordMut semantics and the transformed program, and
\keywordObf semantics and the transformed program.
Ground truths labels are derived from the \transformation procedure. Valid programs are labeled \#\#success\#\#, and each semantically invalid variant is labeled \#\#error\#\# with the corresponding violated rule.
We evaluate open-source \LLMs designed for coding tasks and enhanced reasoning ability on \pcp, including models in the Qwen~\citep{hui2024qwen2}, DeepSeek~\citep{guo2025deepseek}, and Ministral~\citep{liu2026ministral3} families.
For non-reasoning models, we experiment with both \COT and non-\COT prompting.
%
We run each model three times per configuration and report the mean accuracy across runs.
Configurations are
\begin{align*}
  \{\S, \K\} &\times \{\standardSemCap, \keywordMut, \keywordObf\} \\
  &\times \{\text{non-}\COT, \COT\}.
\end{align*}
Because reasoning models generate chain-of-thought traces by default, we evaluate them under one prompting condition only and do not run separate \COT and non-\COT variants.


\begin{table*}[t]
\begin{center}
\TableFont
\caption{\UseMacro{TCap-models-pcp-IMP-K-IMP-SOS-accuracy-qwen-coder-by-split-human-written}\label{tab:pcp-IMP-K-IMP-SOS-accuracy-qwen-coder-by-split-human-written}}
\begin{tabular}{l  c  c  c  c  c  c  c }
\toprule
\multirow{3}{*}{\textbf{\UseMacro{THead-pcp-IMP-K-IMP-SOS-accuracy-qwen-coder-by-split-human-written-models}}}
& \multicolumn{3}{c}{\textbf{\Kos-semantics}}
& \multicolumn{3}{c}{\textbf{\Sos}}
\\
\cmidrule(lr){2-4}
\cmidrule(lr){5-7}
 & \textbf{\UseMacro{THead-pcp-IMP-K-IMP-SOS-accuracy-qwen-coder-by-split-human-written-pcp-uk-IMP-K-accuracy}} & \textbf{\UseMacro{THead-pcp-IMP-K-IMP-SOS-accuracy-qwen-coder-by-split-human-written-pcp-mk-IMP-K-KeywordSwap-accuracy}} & \textbf{\UseMacro{THead-pcp-IMP-K-IMP-SOS-accuracy-qwen-coder-by-split-human-written-pcp-mk-IMP-K-KeywordObf-accuracy}} & \textbf{\UseMacro{THead-pcp-IMP-K-IMP-SOS-accuracy-qwen-coder-by-split-human-written-pcp-uk-IMP-SOS-accuracy}} & \textbf{\UseMacro{THead-pcp-IMP-K-IMP-SOS-accuracy-qwen-coder-by-split-human-written-pcp-mk-IMP-SOS-KeywordSwap-accuracy}} & \textbf{\UseMacro{THead-pcp-IMP-K-IMP-SOS-accuracy-qwen-coder-by-split-human-written-pcp-mk-IMP-SOS-KeywordObf-accuracy}} \\
 &  & \scriptsize(pp) & \scriptsize(pp) &  & \scriptsize(pp) & \scriptsize(pp) \\
\bottomrule
\end{tabular}
\end{center}
\end{table*}

Tables~\ref{tab:pcp-IMP-K-IMP-SOS-accuracy-qwen-coder-by-split-human-written}--\ref{tab:pcp-IMP-K-IMP-SOS-accuracy-qwen-coder-by-split-fuzzer-generated} summarize model performance on \pcp across the three dataset splits.

\subsection{Performance Across \datspts}
Across all models, performance consistently drops between \humanwrit and the other two splits, especially \fuzzgen.

Each of Tables~\ref{tab:pcp-IMP-K-IMP-SOS-accuracy-qwen-coder-by-split-human-written}, \ref{tab:pcp-IMP-K-IMP-SOS-accuracy-qwen-coder-by-split-llm-translated}, and \ref{tab:pcp-IMP-K-IMP-SOS-accuracy-qwen-coder-by-split-fuzzer-generated} reports \pcp accuracy on one dataset split (\humanwrit, \llmtrans, and \fuzzgen, respectively).
Within each table, results are organized by semantics formalism: \K semantics and \S semantics (two groups of columns).
For each formalism, we show accuracy under the three semantic shift configurations: \standardSemCap, \keywordMut, and \keywordObf. \standardSemCap uses the unmodified programs; \keywordMut and \keywordObf use the corresponding modified programs with swapped symbols and new symbols, respectively.  For \keywordMut and \keywordObf, we show, in parenthesis, percentage drop/increase compared to the \standardSemCap shift.

\dpskQwen{32}, \ministralCoT{14}, and \qwenCoderCoT{32} achieve the strongest \humanwrit accuracy among all evaluated models (Table~\ref{tab:pcp-IMP-K-IMP-SOS-accuracy-qwen-coder-by-split-human-written}), but all three still degrade on other splits.
On \llmtrans programs (Table~\ref{tab:pcp-IMP-K-IMP-SOS-accuracy-qwen-coder-by-split-llm-translated}), mean accuracy---averaged across the six \K/\S and semantic-shift columns in each model's row---falls by \UseMacro{pcp-split-drop-dpsk32-mean-llmtrans-pp}, \UseMacro{pcp-split-drop-min14cot-mean-llmtrans-pp}, and \UseMacro{pcp-split-drop-qwen32cot-mean-llmtrans-pp} percentage points (pp) respectively when comparing  to \humanwrit (Table~\ref{tab:pcp-IMP-K-IMP-SOS-accuracy-qwen-coder-by-split-human-written}).
On \fuzzgen programs (Table~\ref{tab:pcp-IMP-K-IMP-SOS-accuracy-qwen-coder-by-split-fuzzer-generated}) the decline is even sharper, with mean drops of \UseMacro{pcp-split-drop-dpsk32-mean-fuzzgen-pp}pp, \UseMacro{pcp-split-drop-min14cot-mean-fuzzgen-pp}pp, and \UseMacro{pcp-split-drop-qwen32cot-mean-fuzzgen-pp}pp computed the same way across Tables~\ref{tab:pcp-IMP-K-IMP-SOS-accuracy-qwen-coder-by-split-fuzzer-generated} and~\ref{tab:pcp-IMP-K-IMP-SOS-accuracy-qwen-coder-by-split-human-written}.
The largest single-configuration drop reaches \UseMacro{pcp-split-drop-overall-max-fuzzgen-pp}pp for \qwenCoderCoT{14} on \KeywordObf semantics under \K where it falls from \UseMacro{res-Qwen-Qwen2.5-Coder-14B-Instruct-cot-pcp-mk-IMP-K-KeywordObf-accuracy-human-written}\% to \UseMacro{res-Qwen-Qwen2.5-Coder-14B-Instruct-cot-pcp-mk-IMP-K-KeywordObf-accuracy-fuzzer-generated}\% (\qwenCoderCoT{14} row and \KeywordObf column under \K semantics in Tables~\ref{tab:pcp-IMP-K-IMP-SOS-accuracy-qwen-coder-by-split-human-written} and~\ref{tab:pcp-IMP-K-IMP-SOS-accuracy-qwen-coder-by-split-fuzzer-generated}).

This performance drop extends beyond the best three performing models: every ``capable model'' (mean \humanwrit accuracy $\geq$ \UseMacro{pcp-split-drop-capable-hw-threshold}\%) loses substantial accuracy on \fuzzgen programs, with mean drops ranging from \UseMacro{pcp-split-drop-capable-mean-fuzzgen-min-pp}pp to \UseMacro{pcp-split-drop-capable-mean-fuzzgen-max-pp}pp (median \UseMacro{pcp-split-drop-capable-mean-fuzzgen-median-pp}pp).
\qwenCoder{14} has a smaller absolute drop under non-\COT which drops by only \UseMacro{pcp-split-drop-qwen14da-mean-fuzzgen-pp}pp on average (the \qwenCoder{14} row averaged across all six \K/\S and semantic-shift columns when comparing Tables~\ref{tab:pcp-IMP-K-IMP-SOS-accuracy-qwen-coder-by-split-fuzzer-generated} and~\ref{tab:pcp-IMP-K-IMP-SOS-accuracy-qwen-coder-by-split-human-written}), but its \humanwrit accuracy is much lower (mean \UseMacro{pcp-split-drop-qwen14da-mean-hw}\%, the same six-column average in Table~\ref{tab:pcp-IMP-K-IMP-SOS-accuracy-qwen-coder-by-split-human-written}).
Somewhat differently, \dpskQwen{14} does not perform as well on \humanwrit (mean \UseMacro{pcp-split-drop-dpsk14-mean-hw}\% vs.\ \UseMacro{pcp-split-drop-dpsk32-mean-hw}\%--\UseMacro{pcp-split-drop-qwen32cot-mean-hw}\%; each the six-column row average in Table~\ref{tab:pcp-IMP-K-IMP-SOS-accuracy-qwen-coder-by-split-human-written}) yet its mean \fuzzgen drop is \UseMacro{pcp-split-drop-dpsk14-vs-top3-mean-fuzzgen-gap-pp}pp below the top-three average (again averaged over the six columns in Tables~\ref{tab:pcp-IMP-K-IMP-SOS-accuracy-qwen-coder-by-split-fuzzer-generated} and~\ref{tab:pcp-IMP-K-IMP-SOS-accuracy-qwen-coder-by-split-human-written}), and no configuration falls more than \UseMacro{pcp-split-drop-dpsk14-max-fuzzgen-pp}pp (its largest drop is at Table~\ref{tab:pcp-IMP-K-IMP-SOS-accuracy-qwen-coder-by-split-human-written}/\dpskQwen{14}/\K/\KeywordObf, from \UseMacro{res-deepseek-ai-DeepSeek-R1-Distill-Qwen-14B-da-pcp-mk-IMP-K-KeywordObf-accuracy-human-written}\% to \UseMacro{res-deepseek-ai-DeepSeek-R1-Distill-Qwen-14B-da-pcp-mk-IMP-K-KeywordObf-accuracy-fuzzer-generated}\% in Table~\ref{tab:pcp-IMP-K-IMP-SOS-accuracy-qwen-coder-by-split-fuzzer-generated}, vs.\ up to \UseMacro{pcp-split-drop-overall-max-fuzzgen-pp}pp at Table~\ref{tab:pcp-IMP-K-IMP-SOS-accuracy-qwen-coder-by-split-human-written}/\qwenCoderCoT{14}/\K/\KeywordObf above).

\subsection{Performance on \keywordMut and \keywordObf}
Moving from \standardSemCap to \keywordMut or \keywordObf semantics substantially reduces accuracy across models.
On \humanwrit programs (Table~\ref{tab:pcp-IMP-K-IMP-SOS-accuracy-qwen-coder-by-split-human-written}), \keywordObf is consistently harder than \keywordMut: across all models, median accuracy falls by \UseMacro{pcp-sem-drop-all-swap-median-pp}pp under \keywordMut versus \UseMacro{pcp-sem-drop-all-obf-median-pp}pp under \keywordObf (each median is the drop from \standardSemCap to \keywordMut or \keywordObf, taken over all model rows and \K/\S formalisms in Table~\ref{tab:pcp-IMP-K-IMP-SOS-accuracy-qwen-coder-by-split-human-written}).

The top three \humanwrit models (\dpskQwen{32}, \ministralCoT{14}, and \qwenCoderCoT{32}) incur mean semantic-shift drops of \UseMacro{pcp-sem-drop-dpsk32-mean-hw-pp}pp, \UseMacro{pcp-sem-drop-min14cot-mean-hw-pp}pp, and \UseMacro{pcp-sem-drop-qwen32cot-mean-hw-pp}pp, respectively (each averaged over \{\K{}, \S{}\} $\times$ \{\keywordMut{}, \keywordObf{}\} in Table~\ref{tab:pcp-IMP-K-IMP-SOS-accuracy-qwen-coder-by-split-human-written}). 

The largest \humanwrit drop
reaches \UseMacro{pcp-sem-drop-max-hw-pp}pp (\ministral{14} under \K, \KeywordObf falls from \UseMacro{res-mistralai-Ministral-3-14B-Instruct-2512-BF16-da-pcp-uk-IMP-K-accuracy-human-written}\% to \UseMacro{res-mistralai-Ministral-3-14B-Instruct-2512-BF16-da-pcp-mk-IMP-K-KeywordObf-accuracy-human-written}\%, Table~\ref{tab:pcp-IMP-K-IMP-SOS-accuracy-qwen-coder-by-split-human-written}), and the largest drop among the top three is \qwenCoderCoT{32} on \S, \KeywordObf, which loses \UseMacro{pcp-sem-drop-qwen32cot-sos-obf-hw-pp}pp (from \UseMacro{res-Qwen-Qwen2.5-Coder-32B-Instruct-cot-pcp-uk-IMP-SOS-accuracy-human-written}\% to \UseMacro{res-Qwen-Qwen2.5-Coder-32B-Instruct-cot-pcp-mk-IMP-SOS-KeywordObf-accuracy-human-written}\%, Table~\ref{tab:pcp-IMP-K-IMP-SOS-accuracy-qwen-coder-by-split-human-written}).
Interestingly, \dpskQwen{32} is an outlier under \KeywordObf on \K, dropping only \UseMacro{pcp-sem-drop-dpsk32-k-obf-hw-pp}pp (from \UseMacro{res-deepseek-ai-DeepSeek-R1-Distill-Qwen-32B-da-pcp-uk-IMP-K-accuracy-human-written} to \UseMacro{res-deepseek-ai-DeepSeek-R1-Distill-Qwen-32B-da-pcp-mk-IMP-K-KeywordObf-accuracy-human-written}, Table~\ref{tab:pcp-IMP-K-IMP-SOS-accuracy-qwen-coder-by-split-human-written}) while its \S drop is \UseMacro{pcp-sem-drop-dpsk32-sos-obf-hw-pp}pp (Table~\ref{tab:pcp-IMP-K-IMP-SOS-accuracy-qwen-coder-by-split-human-written}).

On \fuzzgen programs (results shown in Table~\ref{tab:pcp-IMP-K-IMP-SOS-accuracy-qwen-coder-by-split-fuzzer-generated}), absolute semantic-shift drops remain large for the strongest models (e.g., \qwenCoderCoT{32} loses \UseMacro{pcp-sem-drop-qwen32cot-mean-fuzzgen-pp}pp on average), but the \keywordObf versus \keywordMut gap narrows to \UseMacro{pcp-sem-drop-fuzzgen-obf-minus-swap-mean-pp}pp on average (means of \UseMacro{pcp-sem-drop-fuzzgen-obf-mean-pp}pp vs.\ \UseMacro{pcp-sem-drop-fuzzgen-swap-mean-pp}pp) because the baseline \standardSem accuracy is already much lower.

\subsection{Rule Identification Performance by Error Type}
Figure~\ref{figure:pcp-error-type-radar} breaks down per-error-type accuracy on invalid \C programs under \S for all evaluated models.

The top row shows \humanwrit programs and the bottom row shows \fuzzgen programs; polygons shrink substantially on the bottom row, reflecting the split-wise accuracy drop in Table~\ref{tab:pcp-IMP-K-IMP-SOS-accuracy-qwen-coder-by-split-fuzzer-generated}.
Each radar axis is one model; a colored polygon that stays near the outer ring indicates reliable identification of that semantic error type across models.
Across configurations, keyword-dependent errors (\contloop and \breakloop) shrink inward more under \keywordObf than arithmetic and scoping errors (\divbyzero, \modbyzero, and \varusebdec).
Under \keywordMut, \divbyzero accuracy drops relative to \standardSemCap semantics, suggesting \pretraining bias toward standard division.

\subsection{Things That the Models Get Right}
On \humanwrit programs (Table~\ref{tab:pcp-IMP-K-IMP-SOS-accuracy-qwen-coder-by-split-human-written}), all three models are already strong under \standardSemCap (Table~\ref{tab:pcp-IMP-K-IMP-SOS-accuracy-qwen-coder-by-split-human-written}), and our inspection confirms that most remaining failures are subtle rather than a significant misunderstanding of the programs.
When a model does fail on \humanwrit, it typically still recognizes the relevant error category (e.g., that a break statement is illegal outside a loop), even if it cites the wrong rule number within that category.

\subsection{Main Failure Modes by Dataset Split}
The main failure modes differ by dataset split in ways that explain the accuracy drops reported in the previous subsections.
Across all \UseMacro{pcp-qual-config-count} configurations, \humanwrit failures are relatively rare (\UseMacro{pcp-qual-hw-failures-min}-\UseMacro{pcp-qual-hw-failures-max} incorrect predictions per model out of \UseMacro{pcp-qual-hw-programs-total} programs) and are split between false successes (\UseMacro{pcp-qual-hw-false-success-min}--\UseMacro{pcp-qual-hw-false-success-max}) and wrong-rule confusions (\UseMacro{pcp-qual-hw-wrong-rule-min}--\UseMacro{pcp-qual-hw-wrong-rule-max}).
In contrast, \fuzzgen and \llmtrans failures are both more frequent and structurally different.


\begin{table*}[t]
\begin{center}
\TableFont
\caption{\UseMacro{TCap-models-pcp-IMP-K-IMP-SOS-accuracy-qwen-coder-by-split-llm-translated}\label{tab:pcp-IMP-K-IMP-SOS-accuracy-qwen-coder-by-split-llm-translated}}
\begin{tabular}{l  c  c  c  c  c  c  c }
\toprule
\multirow{3}{*}{\textbf{\UseMacro{THead-pcp-IMP-K-IMP-SOS-accuracy-qwen-coder-by-split-llm-translated-models}}}
& \multicolumn{3}{c}{\textbf{\Kos-semantics}}
& \multicolumn{3}{c}{\textbf{\Sos}}
\\
\cmidrule(lr){2-4}
\cmidrule(lr){5-7}
 & \textbf{\UseMacro{THead-pcp-IMP-K-IMP-SOS-accuracy-qwen-coder-by-split-llm-translated-pcp-uk-IMP-K-accuracy}} & \textbf{\UseMacro{THead-pcp-IMP-K-IMP-SOS-accuracy-qwen-coder-by-split-llm-translated-pcp-mk-IMP-K-KeywordSwap-accuracy}} & \textbf{\UseMacro{THead-pcp-IMP-K-IMP-SOS-accuracy-qwen-coder-by-split-llm-translated-pcp-mk-IMP-K-KeywordObf-accuracy}} & \textbf{\UseMacro{THead-pcp-IMP-K-IMP-SOS-accuracy-qwen-coder-by-split-llm-translated-pcp-uk-IMP-SOS-accuracy}} & \textbf{\UseMacro{THead-pcp-IMP-K-IMP-SOS-accuracy-qwen-coder-by-split-llm-translated-pcp-mk-IMP-SOS-KeywordSwap-accuracy}} & \textbf{\UseMacro{THead-pcp-IMP-K-IMP-SOS-accuracy-qwen-coder-by-split-llm-translated-pcp-mk-IMP-SOS-KeywordObf-accuracy}} \\
 &  & \scriptsize(pp) & \scriptsize(pp) &  & \scriptsize(pp) & \scriptsize(pp) \\
\bottomrule
\end{tabular}
\end{center}
\end{table*}


\begin{table*}[t]
\begin{center}
\TableFont
\caption{\UseMacro{TCap-models-pcp-IMP-K-IMP-SOS-accuracy-qwen-coder-by-split-fuzzer-generated}\label{tab:pcp-IMP-K-IMP-SOS-accuracy-qwen-coder-by-split-fuzzer-generated}}
\begin{tabular}{l  c  c  c  c  c  c  c }
\toprule
\multirow{3}{*}{\textbf{\UseMacro{THead-pcp-IMP-K-IMP-SOS-accuracy-qwen-coder-by-split-fuzzer-generated-models}}}
& \multicolumn{3}{c}{\textbf{\Kos-semantics}}
& \multicolumn{3}{c}{\textbf{\Sos}}
\\
\cmidrule(lr){2-4}
\cmidrule(lr){5-7}
 & \textbf{\UseMacro{THead-pcp-IMP-K-IMP-SOS-accuracy-qwen-coder-by-split-fuzzer-generated-pcp-uk-IMP-K-accuracy}} & \textbf{\UseMacro{THead-pcp-IMP-K-IMP-SOS-accuracy-qwen-coder-by-split-fuzzer-generated-pcp-mk-IMP-K-KeywordSwap-accuracy}} & \textbf{\UseMacro{THead-pcp-IMP-K-IMP-SOS-accuracy-qwen-coder-by-split-fuzzer-generated-pcp-mk-IMP-K-KeywordObf-accuracy}} & \textbf{\UseMacro{THead-pcp-IMP-K-IMP-SOS-accuracy-qwen-coder-by-split-fuzzer-generated-pcp-uk-IMP-SOS-accuracy}} & \textbf{\UseMacro{THead-pcp-IMP-K-IMP-SOS-accuracy-qwen-coder-by-split-fuzzer-generated-pcp-mk-IMP-SOS-KeywordSwap-accuracy}} & \textbf{\UseMacro{THead-pcp-IMP-K-IMP-SOS-accuracy-qwen-coder-by-split-fuzzer-generated-pcp-mk-IMP-SOS-KeywordObf-accuracy}} \\
 &  & \scriptsize(pp) & \scriptsize(pp) &  & \scriptsize(pp) & \scriptsize(pp) \\
\bottomrule
\end{tabular}
\end{center}
\end{table*}

\begin{figure*}[t]
  \centering
  \includegraphics[width=\textwidth]{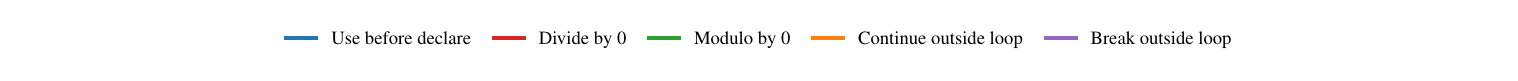}
  \\[-4mm]
  \subcaptionbox{%
    \humanwrit, \FigPCPSOSUKRadarCaption%
    \label{figure:pcp-error-type-radar-human-written-uk-sos}}[0.20\linewidth]{%
    \includegraphics[width=\linewidth]{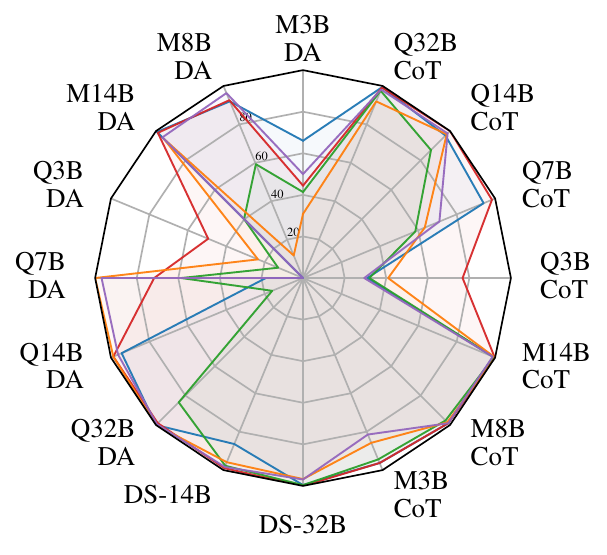}
  }%
  \hspace{0.03\linewidth}
  \subcaptionbox{%
    \humanwrit, \FigPCPSOSSwapRadarCaption%
    \label{figure:pcp-error-type-radar-human-written-mk-sos-swap}}[0.20\linewidth]{%
    \includegraphics[width=\linewidth]{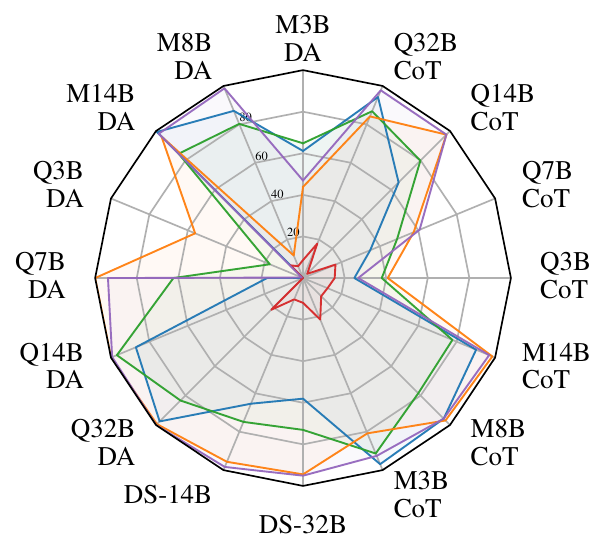}
  }%
  \hspace{0.03\linewidth}
  \subcaptionbox{%
    \humanwrit, \FigPCPSOSObfRadarCaption%
    \label{figure:pcp-error-type-radar-human-written-mk-sos-obf}}[0.20\linewidth]{%
    \includegraphics[width=\linewidth]{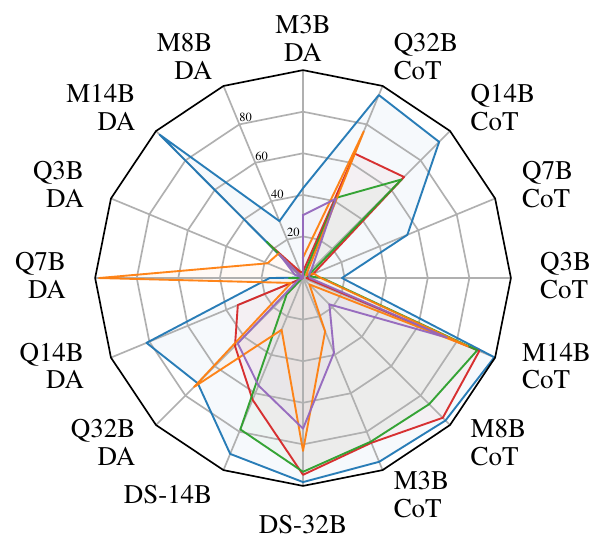}
  }%
  \\[1mm]
  \subcaptionbox{%
    \fuzzgen, \FigPCPSOSUKRadarCaption%
    \label{figure:pcp-error-type-radar-fuzzer-generated-uk-sos}}[0.20\linewidth]{%
    \includegraphics[width=\linewidth]{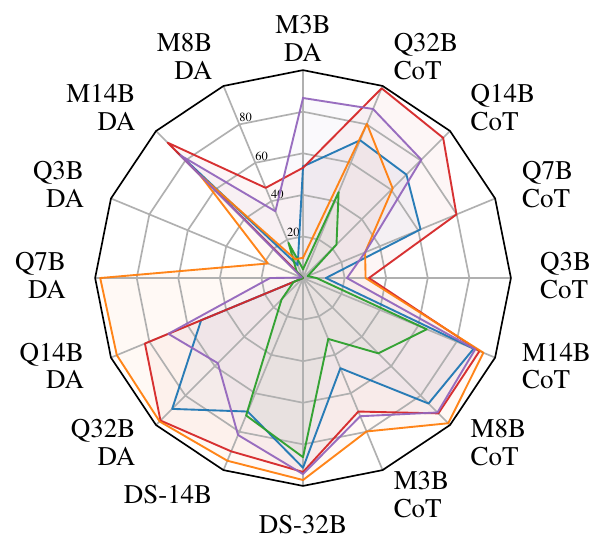}
  }%
  \hspace{0.03\linewidth}
  \subcaptionbox{%
    \fuzzgen, \FigPCPSOSSwapRadarCaption%
    \label{figure:pcp-error-type-radar-fuzzer-generated-mk-sos-swap}}[0.20\linewidth]{%
    \includegraphics[width=\linewidth]{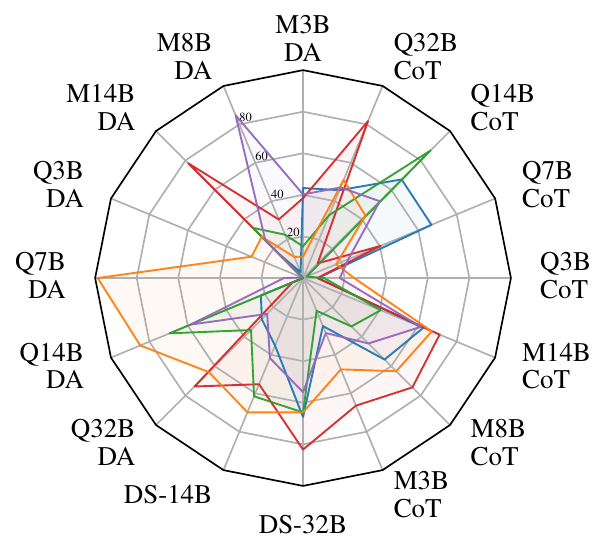}
  }%
  \hspace{0.03\linewidth}
  \subcaptionbox{%
    \fuzzgen, \FigPCPSOSObfRadarCaption%
    \label{figure:pcp-error-type-radar-fuzzer-generated-mk-sos-obf}}[0.20\linewidth]{%
    \includegraphics[width=\linewidth]{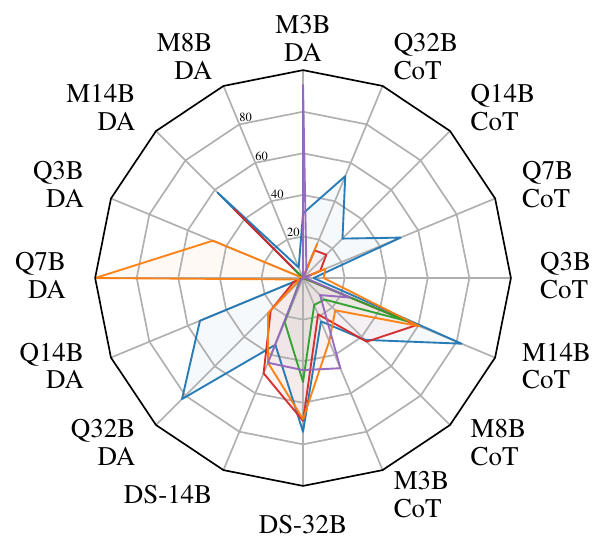}
  }%
  \caption{\FigPCPErrorTypeRadarCaption}
  \vspace{3mm}
\end{figure*}

\MyPara{\humanwrit}
False success is the most common failure mode for \dpskQwen{32} and \qwenCoderCoT{32} on \humanwrit (\UseMacro{pcp-qual-hw-false-success-dpsk32} and \UseMacro{pcp-qual-hw-false-success-qwen32cot} cases respectively, aggregated across configurations).
Wrong-rule errors are less frequent but when they do occur, models confuse nearby rules within the same error family, especially \contloop{} vs.\ \breakloop{} (Rules~\UseMacro{pcp-qual-sos-rule-continue} vs.~\UseMacro{pcp-qual-sos-rule-break} under \S) and adjacent arithmetic-error rules.

\MyPara{\llmtrans and \fuzzgen}
On programs from the \llmtrans split (Table~\ref{tab:pcp-IMP-K-IMP-SOS-accuracy-qwen-coder-by-split-llm-translated}), false success is very frequent.
For example, \dpskQwen{32} produces \UseMacro{pcp-qual-llmtrans-false-success-dpsk32} false-success errors out of \UseMacro{pcp-qual-llmtrans-failures-dpsk32} total failures across configurations, suggesting that models often treat translated C-like code as executable even when it violates the supplied \C semantics.
On \fuzzgen programs (Table~\ref{tab:pcp-IMP-K-IMP-SOS-accuracy-qwen-coder-by-split-fuzzer-generated}), false success remains common (\UseMacro{pcp-qual-fuzzgen-false-success-min}--\UseMacro{pcp-qual-fuzzgen-false-success-max} cases per model), but wrong-rule (\UseMacro{pcp-qual-fuzzgen-wrong-rule-min}--\UseMacro{pcp-qual-fuzzgen-wrong-rule-max}) and false-error (\UseMacro{pcp-qual-fuzzgen-false-error-min}--\UseMacro{pcp-qual-fuzzgen-false-error-max}) predictions become much more frequent as well.
The most common missed rules are \modbyzero, \breakloop, and \varusebdec.
Malformed outputs are also more frequent on \llmtrans and \fuzzgen, especially for \ministralCoT{14}.

Overall, \pcp accuracy is highest on short \humanwrit programs under \standardSemCap semantics but drops sharply on \fuzzgen and \llmtrans splits, under \keywordMut and \keywordObf shifts, and on keyword-based error types such as \contloop{} and \breakloop{}.
The split-wise and semantic-shift declines in Tables~\ref{tab:pcp-IMP-K-IMP-SOS-accuracy-qwen-coder-by-split-human-written}--\ref{tab:pcp-IMP-K-IMP-SOS-accuracy-qwen-coder-by-split-fuzzer-generated} and Figure~\ref{figure:pcp-error-type-radar} coincide with systematic failure modes.
On \humanwrit, remaining errors are relatively rare and tend toward wrong-rule confusions within the same error family; on \fuzzgen and \llmtrans, false success is more common, with models frequently declaring invalid programs executable despite the supplied \C semantics.
Taken together, these results suggest that current \LLMs lean on pre-training priors rather than systematically applying the provided formal semantics
even for execution prediction.





\section{Qualitative Study}
\label{sec:qualitative}

\providecommand{\PCPFuzzListing}{%
  \lstset{language=imp-pretty-no-lines,basicstyle=\tiny\ttfamily,breaklines=true}%
}

\providecommand{\PCPFuzzFigureListing}{%
  \lstset{%
    language=imp-pretty-no-lines,%
    basicstyle=\Large\ttfamily,%
    breaklines=true,%
    columns=fullflexible,%
    aboveskip=0pt,%
    belowskip=0pt,%
    xleftmargin=0pt,%
    xrightmargin=0pt,%
    framesep=0pt,%
  }%
}

\providecommand{\PCPFuzzCodeBoxWidth}{4.8cm}
\providecommand{\PCPFuzzCodeBoxHeight}{4.6cm}

\providecommand{\PCPFuzzTitleTopShift}{8pt}        
\providecommand{\PCPFuzzTitleToCodeGap}{6pt}       
\providecommand{\PCPFuzzInnerBoxTopAdjust}{10pt}    

\providecommand{\PCPFuzzInvalidBoxDistance}{0.10cm}

\providecommand{\PCPFuzzLstInput}[1]{%
  \PCPFuzzListing
  \lstinputlisting{#1}%
}

\begin{table}[t]
  \centering
  \caption{\FigPCPQualitativeExamplesCaption}
  \label{tab:pcp-qualitative-examples}
  \small
  \setlength{\tabcolsep}{1pt}
  \begin{tabular}{@{}p{0.26\linewidth}>{\raggedright\arraybackslash}p{0.17\linewidth}>{\raggedright\arraybackslash}p{0.22\linewidth}>{\raggedright\arraybackslash}p{0.30\linewidth}@{}}
    \toprule
    \textbf{Program} &
    \textbf{Expected} &
    \textbf{Model} &
    \textbf{Predicted} \\
    \midrule
    \begin{minipage}[t]{\linewidth}
      \vspace{0pt}
      \PCPFuzzListing
\begin{lstlisting}
int ans;
int a;
+(*@\colorbox{green!20}{continue;}@*)
a = 156;
ans = (a * a);
\end{lstlisting}
    \end{minipage} &
    Invalid; \mbox{Rule~\UseMacro{pcp-qual-sos-rule-continue}} (\contloop) &
    \qwenCoderCoT{32} &
    Predicted invalid but cited Rule~\UseMacro{pcp-qual-sos-rule-break} (\breakloop) instead of Rule~\UseMacro{pcp-qual-sos-rule-continue} (\emph{wrong rule}) \\
    \midrule
    \begin{minipage}[t]{\linewidth}
      \vspace{0pt}
      \PCPFuzzListing
\begin{lstlisting}
int ans;
int a;
+(*@\colorbox{green!20}{ans = 0;}@*)
+(*@\colorbox{green!20}{a = (156 \% ans);}@*)
ans = (a * a);
\end{lstlisting}
    \end{minipage} &
    Invalid; \mbox{Rule~\UseMacro{pcp-qual-sos-rule-modbyzero}} (\modbyzero) &
    \dpskQwen{32} &
    Predicted invalid by \modbyzero, but cited Rule~\UseMacro{pcp-qual-sos-rule-modbyzero-neighbor} instead of Rule~\UseMacro{pcp-qual-sos-rule-modbyzero} (\emph{wrong rule}) \\
    \bottomrule
  \end{tabular}
  \vspace{2mm}
\end{table}

The quantitative results in Section~\ref{sec:results} report rule-level accuracy aggregated over all programs and runs.
To understand \emph{how} models fail, we complement those aggregates with a qualitative pass over the three strongest \pcp models (\dpskQwen{32}, \qwenCoderCoT{32}, and \ministralCoT{14}).
We only consider programs in the \humanwrit split.
We classify every misprediction into one of \UseMacro{pcp-qual-failure-mode-count} failure modes:
\begin{itemize}[leftmargin=*,itemsep=2pt,topsep=2pt]
  \item \textbf{False success:} predicts \CodeIn{\#\#success\#\#} for invalid program.
  \item \textbf{False error:} predicts \CodeIn{\#\#error\#\#} for valid program.
  \item \textbf{Wrong rule:} correctly predicts invalidity but cites a different violated rule than the ground truth.
  \item \textbf{Malformed output:} the model response does not contain a parseable \CodeIn{<ans>} tag.
\end{itemize}
This taxonomy uses the same rule-level criterion as the main accuracy tables: a prediction counts as correct only when both executability \emph{and} the violated rule (on invalid programs) match the ground truth.

For each of the \UseMacro{pcp-qual-config-count} \pcp configurations
(\K/\S $\times$ \standardSemCap/\allowbreak\keywordMut/\allowbreak\keywordObf),
we select the \UseMacro{pcp-qual-top-n-failures} smallest (by line
count) non-empty failing programs for each top model and inspect the
full model responses.  The goal is not to cherry-pick hard cases, but
to examine minimal counterexamples of programs where even short,
explicit code still triggers systematic errors.
Table~\ref{tab:pcp-qualitative-examples} shows
\UseMacro{pcp-qual-example-count} of the shortest \humanwrit failures
under \S-semantics (\UseMacro{pcp-qual-example-loc} lines each).  Both
programs are semantically invalid for a single, localized reason;
neither requires deep control-flow reasoning.

In the first example, \CodeIn{continue} appears at the top level with
an empty control stack, violating
Rule~\UseMacro{pcp-qual-sos-rule-continue}.  \dpskQwen{32} and
\ministralCoT{14} identify this correctly, but \qwenCoderCoT{32}
predicts invalid while citing Rule~\UseMacro{pcp-qual-sos-rule-break}
(\breakloop) instead.  This matches the per-error-type pattern in
Figure~\ref{figure:pcp-error-type-radar}, where \contloop{} and
\breakloop{} identification degrades more than arithmetic errors.

In the second example, \CodeIn{ans} is initialized to 0 and then used
as a modulo divisor.  All three models predict invalid and the correct
error type (\modbyzero), but \dpskQwen{32} cites
Rule~\UseMacro{pcp-qual-sos-rule-modbyzero-neighbor} rather than
Rule~\UseMacro{pcp-qual-sos-rule-modbyzero}
(Rule~\UseMacro{pcp-qual-sos-rule-modbyzero-neighbor} is an unrelated
unary expression rule).  The model therefore applies the right
semantic reasoning (\modbyzero) but maps it to a neighboring rule in
the \S specification.  Such off-by-one rule citations still count as
incorrect under our metric even when the natural-language explanation
is essentially right.


\section{Related Work}


\MyPara{LLM-Based predictive execution}
A growing body of work asks whether \LLMs can simulate program
execution without running code.  Lyu et
al.~\cite{lyu2024largelanguagemodelscode} treat \LLMs as direct code
executors, feeding snippets to the model and evaluating the returned
outputs on LeetCode programs.  Ni et al.~\cite{nextllmreasoning}
introduced NExT, a self-training approach that teaches \LLMs to
inspect execution traces and reason about run-time behavior via \COT
rationales, improving program repair on MBPP and HumanEval.  Li et
al.~\cite{blendedanalysis} proposed PredEx, a predictive executor for
Python that combines program analysis with \LLM prompting to predict
full execution traces and static runtime errors.  Patel et
al.~\cite{orca} enabled \LLMs to act as predictive interpreters by
autonomously navigating a program's control-flow graph and tracking
variable states at branching points to simulate execution and
statically detect runtime errors.  Le et al.~\cite{codeflow2025}
proposed CodeFlow, a learned CFG-based model that combines static
control-flow structure with dynamic dependencies extracted from
execution traces to predict code coverage and localize runtime errors.
These methods improve predictive execution through specialized
training, program analysis, or learned graph models, but they evaluate
implicit language semantics on real-world programs rather than testing
whether models follow explicitly supplied formal rules.

\MyPara{Evaluating execution and runtime behavior}
Complementary work evaluates code execution reasoning through
benchmarks and empirical studies on real-world programs.  Gu et
al.~\cite{Cruxeval} introduced CRUXEval, a benchmark of short Python
functions with paired input--output examples and two tasks: predicting
the output given an input or the input given an output.  Chen et
al.~\cite{reval2025} proposed REval, a framework that extends this
line of evaluation to intermediate runtime behavior (code coverage,
program state, execution path, and final output) and to incremental
consistency across these dependent sub-tasks.
Hora~\cite{hora2024predictingtestresults} studied a related applied
setting, asking GPT-4 to predict pass/fail outcomes for Python
Standard Library test cases without execution.  These evaluations
measure reasoning under familiar Python semantics; none supplies an
explicit operational rule set in the prompt.  Our \pcp task is simpler
than REval's multi-step runtime behavior reasoning, but adds matched
valid/invalid \C programs, user-provided \S{} and \K semantics, and
\kswap/\kobf shifts to diagnose whether predictions follow the given
rules or priors.

\MyPara{Semantics, identifiers, and formal properties}
Other work examines whether \LLMs reason from program logic or from
surface cues and formal semantic properties.  Wang et
al.~\cite{wang2023naming} showed that replacing variable, method, and
function names with nonsense or misleading identifiers substantially
degrades CodeBERT performance on code analysis tasks, indicating that
models rely heavily on identifier semantics rather than logic alone.
Sultan et al.~\cite{sultan2026llmshalting} evaluated frontier \LLMs on
SV-Comp termination tasks, finding that models can approach
specialized verifiers in termination classification yet frequently
fail to produce machine-valid witness proofs.  Chen et
al.~\cite{chen2025dce} applied \LLMs to dead code elimination, using a
small classifier to locate suspect lines and a fine-tuned model to
judge, explain, and patch unreachable or unused code.  The naming
results motivate our semantic-shift conditions: \kswap and \kobf
perturb familiar symbol meanings to test whether models condition on
provided rules rather than pre-trained associations.  Together with
the termination and dead-code studies, they highlight that \LLM
``understanding'' of code semantics remains fragile; \pcp{} offers a
controlled executability baseline in which success requires applying
supplied formal rules to predict whether execution succeeds or
violates a specific rule.

\section{Conclusion}

We introduced \PEPTask (\pcp), a task that asks models, when given a
program and formal programming language semantics, to predict whether
execution succeeds or halts on a semantic error and, when it fails,
which rule was violated.  Through \pcp, we can evaluate whether \LLMs
apply explicitly provided programming-language semantics or fall back
on pre-training priors.  Building on valid \C programs from \dataset,
we extended the benchmark with invalid programs produced by five
semantics-aware \transfs, and evaluated open-source coding \LLMs under
\S and \K formalisms, \kswap and \kobf semantic shifts, and three
program data splits (\humanwrit, \llmtrans, and \fuzzgen).
We find that current models struggle to follow the supplied semantics.
Accuracy is highest on short \humanwrit programs under \standardSem
semantics, but drops sharply under semantic shifts and on longer,
structurally complex \llmtrans and \fuzzgen programs.
Our results show that \LLMs do not yet reliably reason from supplied
formal semantics and still heavily rely on patterns learned in their
pre-training.

\section*{Acknowledgments}

We thank Cheng Ding, Ivan Grigorik, Yan Levin, Tong-Nong Lin,
Karl Palmskog, Samuel Yuan, Linghan Zhong, and the anonymous reviewers
for helpful feedback and discussions. 
Computational resources were provided by the Texas Advanced 
Computing Center (TACC\footnotemark[2]) at the University of Texas at Austin, and 
AMD (University Program AI \& HPC Cluster).
This work was supported in part by the U.S. National Science
Foundation (NSF) Nos. CCF-2217696, CCF-2313027, CCF-2403036;
the NSF–Simons AI Institute for Cosmic Origins (CosmicAI\footnotemark[3]) funded by NSF award AST-2421782;
the Simons Foundation (MPS-AI-00010515);
and a sponsored research award from Cisco.
Any opinions, findings, conclusions or recommendations expressed in this material
are those of the authors and do not necessarily reflect the views of the sponsoring entities.

\footnotetext[2]{https://www.tacc.utexas.edu/}
\footnotetext[3]{https://www.cosmicai.org/}

\bibliographystyle{ACM-Reference-Format}
\bibliography{bib}

\end{document}